\documentclass[12pt,a4paper]{article}
\usepackage{jheppub}

\graphicspath{{Img/}}

\usepackage{color}
\usepackage[dvipsnames, svgnames, x11names]{xcolor}
\usepackage{lmodern}

\usepackage[utf8]{inputenc}
\usepackage{float}
\usepackage{graphicx}
\usepackage{setspace}
\usepackage{subfigure}
\usepackage{bm}
\usepackage{bbm}
\usepackage{amsmath}
\usepackage{amssymb}
\usepackage{amsbsy}
\usepackage{amsthm}
\usepackage{amsfonts}
\usepackage{braket}
\usepackage{multirow}
\usepackage{slashed}
\usepackage{babel}
 \hypersetup{
	colorlinks=true,
	linkcolor=cyan,
	urlcolor=blue,
	citecolor=red,
}

\def\nn{\nonumber}
\def\Tr{\text{Tr}}
\def\sgn{\text{sgn}}
\def\J{\mathcal{J}}

\def\G{\mathcal{G}}
\def\O{\mathcal{O}}

\def\L{\mathcal{L}}

\def\H{\mathcal{H}}

\def\P{\mathcal{P}}
\def\X{\mathcal{X}}

\def\I{\mathbb{I}}

\def\avg#1{\left\langle#1\right\rangle}
\def\bra#1{\left\langle#1\right|}
\def\ket#1{\left|#1\right\rangle}

\def\abs#1{\left|#1\right|}
\def\kc#1{\left(#1\right)}
\def\kd#1{\left[#1\right]}
\def\ke#1{\left\{#1\right\}}
\def\Re{{\rm Re}}
\def\Im{{\rm Im}}
\def\sgn{{\rm sgn}}

\def\be{\begin{equation}}       \def\ee{\end{equation}}
\def\bea{\begin{eqnarray}}      \def\eea{\end{eqnarray}}
\def\ba{\begin{array}}
	\def\ea{\end{array}}
\def\bnum{\begin{enumerate} }
	\def\enum{\end{enumerate}}

\def\nn{\nonumber}
\def\pa{\partial}
\def\=>{\Rightarrow}
\def\>{\rightarrow}

\def\eye2{Fathbb{I}}

\def\Tr{\mathrm{Tr}}

\def\tsq{\mathrm{sq}}

\def\dd{{\rm d}}

\DeclareMathOperator\csch{csch}
\DeclareMathOperator\sech{sech}

\DeclareMathOperator\arcsinh{arcsinh}

\title{Operator size growth in Lindbladian SYK}

\author[a]{Jiasheng Liu,}
\emailAdd{jiasheng.liu@physik.uni-muenchen.de}

\author[b]{Ren\'e Meyer,}
\emailAdd{rene.meyer@physik.uni-wuerzburg.de}

\author[b,*]{and Zhuo-Yu Xian\note[*]{Corresponding author.}}
\emailAdd{zhuo-yu.xian@physik.uni-wuerzburg.de}

\affiliation[a]{Faculty for Physik, Ludwig-Maximilians-Universit\"at M\"unchen
Schellingstraße 4, 80799, Munich, Germany}
\affiliation[b]{Institute for Theoretical Physics and Astrophysics and W\"urzburg-Dresden Cluster of Excel-lence ct.qmat, Julius-Maximilians-Universit\"at W\"urzburg, 97074 W\"urzburg, Germany}

\abstract{
We investigate the growth of operator size in the Lindbladian Sachdev-Ye-Kitaev model with $q$-body interaction terms and linear jump terms at finite dissipation strength. We compute the operator size as well as its distribution numerically at finite $q$ and analytically at large $q$. 
With dissipative (productive) jump terms, the size converges to a value smaller (larger) than half the number of Majorana fermions.
At weak dissipation, the evolution of operator size displays a quadratic-exponential-plateau behavior. The plateau value is determined by the ratios between the coupling of the interaction and the linear jump term in the large $q$ limit. The operator size distribution remains localized in the finite size region even at late times, contrasting with the unitary case. Moreover, we also derived the time-independent orthogonal basis for operator expansion which exhibits the operator size concentration at finite dissipation.
Finally, we observe that the uncertainty relation for operator size growth is saturated at large $q$, leading to classical dynamics of the operator size growth with dissipation.
}

\arxivnumber{}

\begin{document}
	
\maketitle

\section{Introduction}

The scrambling of quantum information describes how a local operator spreads out and eventually affects the degree of freedom of many-body system under evolution \cite{Lieb2004,Hayden:2007cs,Sekino:2008scramblers}, as measured by the out-of-time-ordering correlator (OTOC) \cite{larkin:1969quasiclassical,Shenker:2013yza,Roberts:2014localized,PRXQuantum.5.010201,Roberts:2016wdl,Shenker:2013black,Hosur:2015channels,Roberts:2014ifa,Stanford:2015owe}. Scrambling can be easily understood as the growth of the size of operators \cite{Roberts:2018operator,Qi:2018quantum}, as a result of the cumulative commutation between the operator and the Hamiltonian during time evolution. Operator size is determined by the distribution of the Heisenberg operator on a local operator basis and is linearly related to the OTOC between initially local operators \cite{Roberts:2014localized,Qi:2018quantum,Qi:2019rpi}. Recently, Krylov complexity has also been employed to describe operator growth \cite{Parker:2018a,Barbon:2019on,Avdoshkin:2019trj,Jian:2020qpp,Rabinovici:2020operator,Avdoshkin:2019trj,Dymarsky:2019quantum,Carrega:2020unveiling,Kar:2021nbm,Caputa:2021sib,Dymarsky:2021bjq,Rabinovici:2021qqt,Muck:2022xfc,Rabinovici:2022beu,Bhattacharjee:2022vlt,Bhattacharjee:2022ave,Avdoshkin:2022xuw,Alishahiha:2022anw,Huh:2023jxt,Tang:2023ocr,Aguilar-Gutierrez:2023nyk,Kundu:2023hbk,Beetar:2023mfn,Bhattacharyya:2023dhp,Loc:2024oen,Malvimat:2024vhr,Bhattacharya:2024uxx}
\footnote{
The Krylov complexity on the state version characterizes the spreading of the wave function in the Hilbert space \cite{Balasubramanian:2022tpr,Balasubramanian:2022dnj,Erdmenger:2023wjg,Balasubramanian:2023kwd,Camargo:2023eev,Bhattacharyya:2023grv,Caputa:2024vrn}.
}
, although the Krylov basis for operator expansion typically is not a set of local operators.

Scrambling is enhanced by strong coupling and exhibits universal behavior in chaotic systems, such as the Sachdev-Ye-Kitaev (SYK) model \cite{Kitaev:2015a,Maldacena:2016remarks,Roberts:2018operator,Qi:2018quantum,Lin:2023trc}, random matrix theory \cite{Roberts:2016design,Cotler:2017jue}, random circuits \cite{Hosur:2015channels,Nahum:2017operator,vonKeyserlingk:2017operator,Xu:2018locality}, and black holes \cite{Shenker:2013black,Roberts:2014localized,Maldacena:2016conformal,Lensky:2020size,Mousatov:2019operator,Lin:2019symmetries}. In particular, in these models, the operator size was found to exhibit exponential growth, whose exponent saturates the chaos bound \cite{Maldacena:2015waa}. In the SYK model, the scrambling time for the size of local operators scales as $\ln N$, with $N$ being the number of fermions \cite{Sekino:2008scramblers}. 
Scrambling is suppressed by localization \cite{Fan:2016ean} and dissipation \cite{Chen:2017dbb,Zhang:2018oop}.

Scrambling can be measured via forward and backward evolution \cite{Li:2016xhw,Garttner:2016mqj,Sanchez:2020LoschmidtEcho,Sanchez:2022LoschmidtEchoNER,Dominguez:2021decoherence,Dominguez:2021keq,Mi:2021gdf,Cotler:2022fin,Swingle:2016var}, entangled double-copy systems \cite{Islam:2015mom,Landsman:2018jpm,Blok:2020may}, or randomized measurements \cite{Brydges:2019probing,Joshi:2020PRL,Blocher:2023hvk}. For a realistic experimental setup, the dynamics of an open system is inevitably affected by the environment and becomes fundamentally non-unitary. When the system is weakly coupled to a Markovian reservoir, its dynamics is described by the Lindbladian master equation \cite{breuer2002theory,lidar2019lecture,manzano2020short}, which is equivalent to the dynamics on the double-copy Hilbert space governed by a non-Hermitian Hamiltonian \cite{Denisov:2018nif}. For Markov processes, such as the Lindbladian spin chain \cite{Schuster:2022bot} or the Lindbladian SYK model \cite{Kulkarni:2021gtt,Kawabata:2022osw,Garcia-Garcia:2022adg,Sa:2021tdr,Bhattacharjee:2023uwx} at weak dissipation, the size for local operators exhibits growth-plateau behavior, which is determined by the competition between the unitary interaction and non-unitary dissipation. Similar plateau behaviors were observed in Krylov complexity \cite{Liu:2022god,Bhattacharjee:2022lzy,Bhattacharya:2022gbz,Bhattacharjee:2023uwx,Bhattacharya:2023zqt,Bhattacharya:2023yec}. For a non-Markov process, such as the Brownian SYK coupled to a bath \cite{Zhang:2023BrownianSize}, the size can decay to zero for strong system-bath couplings. 

In this work, our aim is to comprehensively understand the growth of operator size in the Lindbladian SYK, particularly through analytical methods based on the path integral and the large $q$ limit. By employing these analytical solutions, we explicitly compute quantities such as the Loschmidt echo fidelity, operator size, size distribution, and size variances, and determine their time scales across all parameter regions. Additionally, We also prove the operator size concentration of Krylov basis from the path integral perspective. Finally, we elucidate on the emergence of classical dynamical equations governing operator size growth \cite{Qi:2018quantum}, which are frequently utilized in estimating operator size growth in open systems \cite{Schuster:2022bot}.

In this paper, we numerically and analytically calculate the operator sizes and size distributions of Heisenberg operators in a Lindbladian SYK model with linear jump operators. In Sec.~\ref{sec:Lindbladian}, we define operator size, express its generating function as a path integral, and study the symmetries of the two-point function. In Sec.~\ref{sec:Numeric}, we adopt exact diagonalization (ED) at finite $N$ and $q$, and numerically solve Schwinger-Dyson (SD) equations at infinite $N$ and finite $q$. In Sec.~\ref{sec:Analytic}, we analytically solve the Liouville equations at large $q$ and obtain complete information about the operator size growth in this limit. In Sec.~\ref{sec:Epidemic}, we derive a classical equation for operator size growth by noticing the saturation of a relative uncertainty relation. In Sec.~\ref{sec:conclusion}, we conclude and give an outlook on future directions.

\section{Lindbladian SYK}\label{sec:Lindbladian}
In this section, we first discuss the Linbladian SYK model and apply it to the Heisenberg evolution of operators. We further map the Linbladian superoperator onto a Liouvillian operator on the double copy Hilbert space, and define the operator size and its generating function. Later, we write the corresponding partition function in path integral along the contour in the double copy Hilbert space. Following known methods for the SYK model, we obtain the effective action and the Schwinger-Dyson equation. In the last part, we find the symmetries of the two-point functions based on the symmetries of the model.

\subsection{Lindbladian}

The Lindblad master equation, or Lindbladian, for a density matrix in the Schrödinger picture is
\begin{equation}\label{eq:LindbState}
    \partial_t\rho=\L_S[\rho]=-i[H,\rho]+\nu\sum_j \kd{L_j\rho L_j^\dagger -\frac12\ke{L_j^\dagger L_j,\rho}},
\end{equation}
which is interpreted as a super-operator $\L_S$. 
The Lindbladian with \(\nu \geq 0\) can effectively describe the microscopic model for a system coupled to an environment (bath) in certain limits. On one hand, the Lindbladian can be derived in the weak coupling limit, where the system-bath coupling is weak compared to the characteristic energy scales of both the system and the bath. More precisely, the weak coupling limit encompasses the Born approximation, the Markov approximation, and the rotating wave approximation. On the other hand, the Lindbladian can also be derived in the singular coupling limit, where the bath Hamiltonian dominates over the system Hamiltonian and the system-bath interaction \cite{breuer2002theory,lidar2019lecture,manzano2020short}. Therefore, we will not assume any hierarchy between the strengths of the two terms in the Lindbladian \eqref{eq:LindbState}.

One can also write down the Lindbladian for an operator in the Heisenberg picture
\begin{equation}\label{eq:LindbOperator}
    \partial_t O=\L_H[O]=\L_U[O]+\L_D[O]=i[H,O]+\nu\sum_j \kd{(-1)^\eta L_j^\dagger O L_j -\frac12\ke{L_j^\dagger L_j,O}},
\end{equation}
where $\eta=0$ when any one of $O$ and $L_j$ is bosonic and $\eta=1$ when both are fermionic \cite{Liu:2022god}. In \cite{Liu:2022god}, the Lindbladian for an operator is derived by tracing out the Markovian bath and imposing the white noise approximation on the bath correlation in an integral equation of the unitary Heisenberg equation of the operator.

The Lindbladian itself is a consistent theory for any real value of $\nu$, including $\nu < 0$, mathematically. We will provide an interpretation of the case where $\nu < 0$ later in our Lindbladian SYK model. In this paper, we will use the Lindbladian in \eqref{eq:LindbOperator} as our starting point and will not consider its microscopic origin in the system-plus-bath composite system.

Both Lindbladians preserve the trace and the maximally mixed state, which is proportional to the identity, namely $\L_S[\I]=\L_H[\I]=0$.
When the Lindblad operators are Hermitian, $L_j^\dagger=L_j$, the two equations \eqref{eq:LindbState} and \eqref{eq:LindbOperator} are identical by mapping $H\to -H$. It is exactly this case we consider in this paper, and we will adopt the Lindbladian in the Heisenberg picture \eqref{eq:LindbOperator} throughout out this paper. The first term $\L_U[O]=i[H,O]$ describes the unitary evolution and the second term $\L_D[O]$ introduces the effect of physical non-unitary. The coefficient $\nu$ could be interpreted as the error rate of quantum gates in quantum circuits \cite{Schuster:2022bot}. 

In this paper, we consider an even number $N$ Majorana fermions $\ke{\psi_j\,|\, j=1,2,\cdots N}$, with anti-commutation relation $\{\psi_j,\psi_k\}=2\delta_{jk}$. The Hamiltonian $H$ is taken to be the Sachdev-Ye-Kitaev (SYK) model \cite{Kitaev:2015a,Maldacena:2016remarks}
\begin{equation}\label{eq:SYK}
    H=i^{q/2}\sum_{1\leq j_1<\cdots<j_q\leq N} J_{j_1\cdots j_q} \psi_{j_1}\cdots\psi_{j_q},\quad \avg{J_{j_1\cdots j_q}^2}
    =\frac{J^2}{\binom{N-1}{q-1}}
    =\frac{\J^2}{2q\binom{N-1}{q-1}},
\end{equation}
with even number $q\geq2$ and $J\geq 0$. The Lindblad operators $L_j$ are taken as the linear jump operators \cite{Kulkarni:2021gtt}
\begin{equation}\label{eq:linearjump}
    L_j=\psi_j/\sqrt2,\quad j=1,2,\cdots,N.
\end{equation}
The dimension of Hilbert space is $D=2^{N/2}$. To derive the Lindbladian with linear jump operators, one could couple the SYK system with a Markovia bath, where the bath correlation follows white noise approximation $\avg{\chi_j(t)\chi_k}=\delta(t)\delta_{jk}$, the hopping term is $H_I=i\sum_{j=1}^N \sqrt{\nu}\psi_j\chi_j$ and $\chi_j$ is the Majorana fermion of the bath \cite{Liu:2022god}.
Last but not least, the Lindbladian SYK model we consider in this paper is theoretically well-defined completely positive trace preserving map for all the parameters $\J>0,\, \nu\in\mathbb R$, although is still debatable whether one can find the original from a microscopic dynamics.

Following \cite{Qi:2018quantum,Kulkarni:2021gtt}, we will use the Choi-Jamiolkowski isomorphism with the double-copy Hilbert space $\H_{LR}=\H_L\otimes \H_R$ and $2N$ Majorana fermions satisfying $\ke{\psi_j^a,\psi_k^b}=2\delta_{jk}\delta_{ab}$ with $a,b=L,R$. Following \eqref{eq:SYK}, we use $\psi_j^L$ to construct $H^L$ and use $\psi_j^R$ to construct $H^R$. Given a bosonic (fermionic) operator $O$ acting on $\H$, we can construct two bosinic (fermionic) operators $O^L=O\otimes \I,\, O^R=\I\otimes O$ ($O^L=O\otimes \I,\, O^R=S\otimes O$) acting on $\H_L\otimes \H_R$ where Hermitian operator $S=i^{N(N-1)/2} \psi_1\psi_2\cdots \psi_N$. One can check that $\kd{O,S}=0$ ($\ke{O,S}=0$) for any bosonic (fermionic) linear operator $O$.

We define a maximally entangled state $\ket0$ in $\H\otimes\H_R$ as 
\begin{equation}\label{eq:MES}
    \kc{\psi_j^L+i\psi_j^R} \ket0=0, \ \forall j,\quad \avg{0|0}=1.
\end{equation}
One can check that $H^L\ket0=i^qH^R\ket0$. The maximally entangled state $\ket0$ induces a map from a linear operator acting on the single Hilbert space $\H$ to a state in the double-copy Hilbert space $\H_{LR}$ via
\begin{equation}\label{eq:Mapping}
    O\mapsto O^L\ket0=\ket{O}.
\end{equation}
Then the identity $\I$ acting on $\H$ is mapped to $\ket0$. The operator trace in $\H$ is mapped to the inner product in $\H_{LR}$ via $\Tr[O_1^\dagger O_2]=D\avg{O_1|O_2}$.

The Lindbladian $\L_H$ is mapped to a non-Hermitian Liouvillian
\begin{equation}\label{eq:Liouvillian}
    \L=i\kc{H^L-i^qH^R}-\nu n=i\P+\X
\end{equation}
acting on $\H_{LR}$ via $\L_H[O]\mapsto \L_H[O]\otimes\I\ket0=\L\ket{O}$, where 
\begin{equation}\label{eq:SizeOperator}
    n=\frac12\sum_{j=1}^N\kc{1+i\psi_j^L\psi_j^R}
\end{equation}
is called the size operator, whose meaning will be explained in the next subsection. The factor $(-1)^\eta$ in \eqref{eq:LindbOperator} is canceled out in \eqref{eq:Liouvillian} when $O^L$ and $\psi_j^R$ are exchanged. At the last step of \eqref{eq:Liouvillian}, we decompose the Liouvillian into anti-Hermitian part $i\P=i\kc{H^L-i^qH^R}=iP^\dagger$ and Hermitian part $\X=-\nu n=\X^\dagger$. Thus,
$
\L^\dagger=-i\kc{H^L-i^qH^R}-\nu n=-i\P+\X.
$
One can check that $\P\ket0=\X\ket0=\L\ket0=\L^\dagger\ket0=0$. 
The Liouvillian $\L$ is reminiscent of the Hamiltonian for eternal traversable wormhole \cite{Gao:2016bin,Maldacena:2018lmt,Milekhin:2022bzx} with the same second term, but here the Hamiltonians on each side have the opposite signs.

\subsection{Operator size}

To study the information scrambling in the Lindbladian SYK model, we introduce the operator size following \cite{Qi:2018quantum}. First, we define a local, complete, operator basis acting on $\H$, consisting of Majorana strings, namely
\begin{equation}
    \ke{\Gamma_I=\Gamma_{j_1j_2\cdots j_k}=i^{k(k-1)/2}\psi_{j_1}\psi_{j_2}\cdots \psi_{j_k}, \ | \ 1\leq j_1<j_2<\cdots<j_k\leq N}
\end{equation}
with notation $\abs{I}=k$, where the factor $i^{k(k-1)/2}$ is introduced to make $\Gamma_I$ Hermitian. The basis is orthogonal and normalized under the inner product $\frac1D\Tr[\Gamma_I\Gamma_J]=\avg{\Gamma_I|\Gamma_J}=\delta_{IJ}$. The basis is complete since the number of elements $2^N$ is equal to the square of the dimension of Hilbert space $D^2$, even in the $N\rightarrow \infty$ limit. Second, we define the size superoperator $n[\cdots]$ so that it is diagonal on the Majorana strings basis with the eigenvalues measuring the numbers of Majorana fermions
\begin{equation}\label{eq:AssignSize}
    n[\Gamma_I]=\abs{I},
\end{equation}
which defines the size of all the linear operators acting on $\H$. Obviously, the identity $\I$ has smallest size $0$ and $\Gamma_{12\cdots N}=S$ has largest size $N$. 

Based on the mapping \eqref{eq:Mapping} to the double-copy Hilbert space $H_{LR}$, the assignment \eqref{eq:AssignSize} simply corresponds to the size operator defined in \eqref{eq:SizeOperator}, as one can check from the eigensystem $n\ket{\Gamma_I}=\abs{I}\ket{\Gamma_I}$ and diagonalization $n=\sum_I\ket{\Gamma_I}|I|\bra{\Gamma_I}$. Obviously, $n\ket0=0,\, n\ket{\Gamma_{1\cdots N}}=N\ket{\Gamma_{1\cdots N}}$ and we denote $\ket N\equiv\ket{\Gamma_{1\cdots N}}$. We define the size subspace and its projection operator 
\begin{equation}\label{eq:sizesubspace}
    \H_n=\text{span}\ke{\ket{\Gamma_I}|\abs{I}=n,\forall I},\quad 
    \pi_n=\sum_{\abs{I}=n}\ket{\Gamma_I}\bra{\Gamma_I},\quad n=0,1,\cdots,N.
\end{equation}
The dimension of $\H_n$ is $\binom{N}{n}$. The size operator can be expanded as $n=\sum_n n\pi_n$.
The Lindblad term $-\nu n$ in \eqref{eq:Liouvillian} with $\nu>0$ ($\nu<0$) suppresses (enhances) eigenstates with large sizes. 
So, we refer to the effect induced by the Lindblad term with $\nu>0$ as dissipation, and that with $\nu<0$ as production. One can measure the size of any operator $O$ acting on $\H$ by the size operator
\begin{equation}\label{eq:size}
    n[O]=\frac{\bra O n \ket O}{\avg{O|O}}=-\partial_\mu \ln \G_\mu[O] |_{\mu=0}
\end{equation}
with $n$ the size operator in \eqref{eq:SizeOperator} and
\begin{equation}\label{eq:Generating}
    \G_\mu[O]=\bra O e^{-\mu n}\ket O
\end{equation}
the generating function of the operator size \cite{Qi:2018quantum}. 
The generating function also gives the size distribution via
\begin{equation}\label{eq:Distribution}
    \G_\mu[O]=\avg{O|O}\sum_{n=0}^N e^{-\mu n}P_n[O],\quad  P_n[O]=\frac{\bra{O}\pi_n\ket{O}}{\avg{O|O}},
\end{equation}
where the distribution $P_n[O]$ is normalized as $\sum_n P_n[O]=1$.

For the sake of analytic solvability, in the main text of this paper, we focus on the case of $O=\psi_1(t)$ in the Heisenberg picture and calculate the size $n[\psi_1(t)]$ and the generating function
\begin{equation}\label{eq:Generatring}
    \G_\mu(t)\equiv\G_\mu[\psi_1(t)]=\bra{\psi_1(t)}e^{-\mu n} \ket{\psi_1(t)}. 
\end{equation}
Obviously, $P_n[\psi_1(t)]=0$ for all even $n$. So $1\leq n[\psi_1(t)]\leq N-1$, in contrast to the case of a non-Markovian reservoir \cite{Zhang:2023BrownianSize}. In the App.~\ref{sec:Temperature}, we consider two initial operators $e^{-\beta H/2}$ and $\psi_1e^{-\beta H/2}$ and numerically study their sizes under the Lindbladian evolution.
For later convenience, we further introduce the double-time generating function
\begin{equation}\label{eq:DGeneratring}
    \G_\mu(t_1,t_2)\equiv \bra{\psi_1(t_1)}e^{-\mu n}\ket{\psi_1(t_2)}.
\end{equation}
It is the transition amplitude between the states at two different times $t_1$ and $t_2$ under the influence of the insertion $e^{-\mu n}$. Obviously, $\G_\mu(t)=\G_\mu(t,t)$. 

The operator size $n[\psi_1(t)]$ for small enough $\nu>0$ was studied in \cite{Schuster:2022bot} semi-classically, and also calculated in the same limit in \cite{Bhattacharjee:2023uwx} from the Krylov complexity perspective. Here, we are able to solve this problem for any value of $\nu$ numerically at finite $q$ and analytically at large $q$, so that we can investigate the operator size and distribution dynamics in the Lindbladian SYK model at finite dissipation or production and observe fast saturation of the size and the distribution.

\subsection{Path integral}\label{sec:PathIntegral}

\begin{figure}
    \centering
    \includegraphics{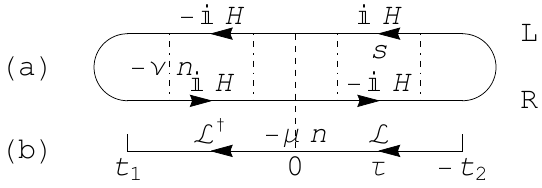}
    \caption{(a) The Keldysh contour used in App.~\ref{sec:Keldysh} and (b) the double-copy contour used in Subsec.~\ref{sec:PathIntegral}.}
    \label{fig:Contour}
\end{figure}

We will utilize the path integral to calculate the generating function of size \eqref{eq:Generatring}. Let us consider the partition function
\begin{equation}\label{eq:PatitionFunction}
        Z_\mu(t_1,t_2)=\bra 0 e^{t_1\L^\dagger} e^{-\mu n} e^{t_2\L}\ket 0=\int \prod_j\prod_{a=L,R} {\mathcal{D}}\psi_j^a e^{-S} .
\end{equation}
Since $\L\ket0=n\ket0=0$, we know that $Z_\mu(t_1,t_2)=1$ for any $t_1,t_2,\mu$. So we are free to choose any $t_1$ and $t_2$ in \eqref{eq:PatitionFunction}. Furthermore, the two-point function should be independent of the choice of $t_1,t_2$ since $\bra0e^{t_1\L^\dagger}\psi_j^a=\bra0\psi_j^a$ and $\psi_j^ae^{t_2\L}\ket0=\psi_j^a\ket0$. On the r.h.s. of \eqref{eq:PatitionFunction}, we write the partition function as the path integral of two real Grassmann fields $\psi_j^L(\tau),\psi_j^R(\tau)$ along the double-copy contour in Fig.~\ref{fig:Contour} with $t_1,t_2\geq0$ \cite{Garcia-Garcia:2022adg,Kawabata:2022osw}. The action is
\begin{align}\label{eq:PathIntegral}
    -S=\int_{-t_2}^{t_1}{\rm d}\tau \Big[&
  -\frac14\sum_{j,a}\psi_j^a(\tau)\partial_\tau\psi_j^a(\tau)  \nn 
  -i \sgn(\tau)\kc{H^L(\tau)-i^qH^R(\tau)} \\
  &-\kc{\nu+\mu \delta(\tau)} n(\tau) \Big],
\end{align}
where $H^L(\tau),\, H^R(\tau)$ and $n(\tau)$ are functions of real Grassmann variables $\psi_j^a(\tau)$ in the forms of \eqref{eq:SYK} and \eqref{eq:SizeOperator}. Notice that the unusual sign function before the Hamiltonians originates from the piecewise evolution $e^{t_1\L^\dagger}e^{t_2\L}$ along the contour. The $\delta(\tau)$ term corresponds to the insertion of $e^{-\mu n}$ at $\tau=0$. Due to the initial and final state \eqref{eq:MES}, the path integral is subjected to the boundary conditions
\begin{equation}\label{eq:BCpsi}
    \psi_j^L(-t_2)+i\psi_j^R(-t_2)=0,\quad \psi_j^L(t_1)-i\psi_j^R(t_1)=0.
\end{equation}

As in the pure SYK model \cite{Maldacena:2016remarks}, we take the disorder average, keep the dominating replica diagonal part and obtain the action
\begin{equation}\label{eq:Actionpsi}
\begin{split}
    -S=&~\int_{-t_2}^{t_1}\dd \tau \sum_j \Big[
    -\frac14\sum_{a}\psi_j^a(\tau)\partial_\tau\psi_j^a(\tau)  
    -\frac12\kc{\nu+\mu\delta(\tau)} (i\psi_j^L(\tau)\psi_j^R(\tau)+1) \Big]    \\  
    &~ -\frac{i^qJ^2}{2qN^{q-1}}\int_{-t_2}^{t_1}\dd\tau_1\dd\tau_2  \sum_{j_1\cdots j_N,ab} s_{ab}(\tau_1,\tau_2) \psi_{j_1}^a(\tau_1)\cdots\psi_{j_q}^a(\tau_1)\psi_{j_1}^b(\tau_2)\cdots\psi_{j_q}^b(\tau_2),
\end{split}
\end{equation}
where $s_{ab}(\tau_1,\tau_2)$ comes from the disorder average between two different locations and is defined as
\begin{equation}
    s_{ab}(\tau_1,\tau_2)\equiv s_{a}(\tau_1)s_b(\tau_2), \quad
    s^L(\tau)=-i^qs_R(\tau)=\sgn(\tau).
\end{equation}
We introduce the bi-local fields
\begin{equation}
	G_{ab}(\tau_1,\tau_2)=\frac{1}{N}\sum_j\psi_j^a(\tau_1)\psi_j^b(\tau_2),\quad 
 a,b=L,R,
\end{equation}
and $\Sigma_{ab}(\tau_1,\tau_2)$ via the Lagrange multiplier method 
\begin{equation}\label{eq:multiplier}
    1=\int {\rm D}\Sigma {\rm D}G \exp\kd{-\frac12 \int_{-t_2}^{t_1} \dd\tau_1\dd\tau_2 \sum_{ab}\Sigma_{ab}(\tau_1,\tau_2)\kc{G_{ab}(\tau_1,\tau_2)-\frac1N\sum_j\psi_j^a(\tau_1)\psi_j^b(\tau_2)}}.
\end{equation}
By integrating out the Grassmann variables $\psi_j^a(\tau)$, we obtain the effective action for bi-local fields
\begin{align}
    -S/N=&~\frac12\log\det\kc{\frac12\delta_{ab}\partial_\tau-\Sigma_{ab}} \nn\\
  &~-\frac{1}{2}\int_{-t_2}^{t_1}\dd\tau_1\dd\tau_2 \sum_{ab}\kc{\Sigma_{ab}(\tau_1,\tau_2)G_{ab}(\tau_1,\tau_2)
  +s_{ab}(\tau_1,\tau_2)\frac{J^2}{q}G_{ab}(\tau_1,\tau_2)^q} \nn\\
		&~-\frac{1}{4}\int_{-t_2}^{t_1}\dd\tau\kc{\nu+\mu\delta(\tau)}\left(iG_{LR}(\tau,\tau)-iG_{RL}(\tau,\tau)+2\right). \label{eq:effaction}
\end{align}
The boundary condition \eqref{eq:BCpsi} becomes
\begin{equation}\label{eq:BCG}
\begin{split}
    &G_{aL}(\tau,-t_2)+iG_{aR}(\tau,-t_2)=G_{La}(-t_2,\tau)+iG_{Ra}(-t_2,\tau)=0,\\
    &G_{aL}(\tau,t_1)-iG_{aR}(\tau,t_1)=G_{La}(t_1,\tau)-iG_{Ra}(t_1,\tau)=0.
\end{split}
\end{equation}

In the large $N$ limit, by taking the variation on the bi-local fields, we obtain the Schwinger-Dyson (SD) equation
\begin{subequations}\label{eq:SDGSigma}
\begin{align}
    	&\frac{1}{2}\partial_{\tau_1}G_{ab}(\tau_1,\tau_2)-\int_{-t_2}^{t_1} \dd\tau_3\sum_c\Sigma_{ac}(\tau_1,\tau_3)G_{cb}(\tau_3,\tau_2)=\delta_{ab}\delta(\tau_1-\tau_2), \label{eq:sdeqo1}\\
	&\Sigma_{ab}(\tau_1,\tau_2)=-J^2s_{ab}(\tau_1,\tau_2)G_{ab}(\tau_1,\tau_2)^{q-1}-\frac i2\epsilon_{ab}\kc{\nu+\mu\delta(\tau_1)}\delta(\tau_1-\tau_2),\label{eq:sdeqo2}
\end{align}
\end{subequations}
where $\epsilon_{LL}=\epsilon_{RR}=0$ and $\epsilon_{LR}=-\epsilon_{RL}=1$. The on-shell solution should be independent of $t_1$ and $t_2$, such that the boundary condition \eqref{eq:BCG} actually holds for any $t$. This fact leads to the following simplification:
\begin{equation}\label{eq:GlobalG}
\begin{split}
    G_{LL}(\tau_1,\tau_2)
    =\sgn(\tau_{21})iG_{LR}(\tau_1,\tau_2)
    =\sgn(\tau_{12})iG_{RL}(\tau_1,\tau_2)
    =G_{RR}(\tau_1,\tau_2),
\end{split}\end{equation}
where $\tau_{12}=\tau_1-\tau_2$. To establish similar relations between the components of the self energy, we need to isolate the contribution from the anti-symmetric $\epsilon_{ab}$ term in \eqref{eq:sdeqo2} by defining $\tilde \Sigma_{ab}(\tau_1,\tau_2)=-J^2s_{ab}(\tau_1,\tau_2)G_{ab}(\tau_1,\tau_2)^{q-1}$, which follows the same relations as  \eqref{eq:GlobalG}, namely
\begin{equation}\label{eq:GlobalSigma}
        \tilde\Sigma_{LL}(\tau_1,\tau_2)
    =\sgn(\tau_{21})i\tilde\Sigma_{LR}(\tau_1,\tau_2)
    =\sgn(\tau_{12})i\tilde\Sigma_{RL}(\tau_1,\tau_2)
    =\tilde\Sigma_{RR}(\tau_1,\tau_2).
\end{equation}
So, only one of the four components in $G_{ab}$ and $\tilde\Sigma_{ab}$ are independent. We choose $G_{LL}$ and $\tilde\Sigma_{LL}$, and simplify the SD equations as
\begin{subequations}\label{eq:SDGSigmaReduce}
\begin{align}
    \delta(\tau_1-\tau_2)=&~\frac{1}{2}\partial_{\tau_1}G_{LL}(\tau_1,\tau_2)-2\sgn(\tau_{12})\int_{\tau_2}^{\tau_1} \dd\tau_3\tilde\Sigma_{LL}(\tau_1,\tau_3)G_{LL}(\tau_3,\tau_2) \nn\\
   & +\frac12\sgn(\tau_{12})\kc{\nu+\mu\delta(\tau_1)}G_{LL}(\tau_{12}), \label{eq:sdeq1}\\
	\tilde\Sigma_{LL}(\tau_1,\tau_2)=&~-J^2\sgn(\tau_1)\sgn(\tau_2)G_{LL}(\tau_1,\tau_2)^{q-1},
\end{align}
\end{subequations}
such that $t_1$ and $t_2$ disappear.
Now the equation is real, so we expect real solutions of $G_{LL}$ and $\tilde\Sigma_{LL}$. Other components could be reconstructed via \eqref{eq:GlobalG} and \eqref{eq:GlobalSigma}. 

The two-point function $G_{ab}(\tau_1,\tau_2)$ derived from the path integral, or from solving the SD equation \eqref{eq:SDGSigma} in the large $N$ limit, can be expressed as the following expectation value
\begin{equation}\label{eq:TwoPtFunction}
	G_{ab}(\tau_1,\tau_2)=\frac1{N Z_\mu(t_1,t_2)}\sum_j\bra0 \mathcal T\kd{e^{-\mu n(0)}\psi_j^a(\tau_1)\psi_j^b(\tau_2)} \ket0,
\end{equation}
where the operator evolution is defined as
\begin{equation}
    O(\tau)=\begin{cases}
        e^{-\tau\L}Oe^{\tau\L},& \tau\leq 0\\
        e^{-\tau\L^\dagger}Oe^{\tau\L^\dagger},& \tau>0\\
    \end{cases}
\end{equation}
and the time ordering $\mathcal T$ compatible with the path integral \eqref{eq:PatitionFunction} and \eqref{eq:PathIntegral} is
\begin{equation}
    \mathcal T \kd{ O_1(\tau_1)O_2(\tau_2)}= \begin{cases}
        O_1(\tau_1)O_2(\tau_2),& \tau_1>\tau_2\\
        (-1)^\eta O_2(\tau_2)O_1(\tau_1),& \tau_1<\tau_2\\
    \end{cases},
\end{equation}
with $\eta=1$ if both operators are fermionic, and $\eta=0$ in other cases. Under the disorder average, the generating functions \eqref{eq:Generatring}\eqref{eq:DGeneratring}  will not depend on the choice of the Majorana index and equals the specific two-point function
\begin{equation}\label{eq:GeneratingG}
    \G_\mu(t)=Z_\mu(t,t)G_{LL}(t,-t),\quad  \G_\mu(t_1,t_2)=Z_\mu(t_1,t_2)G_{LL}(t_1,-t_2).
\end{equation}
where $Z_\mu(t_1,t_2)=1$.

\subsection{Symmetries}\label{sec:symmetry}

The $\sgn(\tau)$ factor in the path integral \eqref{eq:PathIntegral} divides the time domain $[-t_2,t_1]$ into two parts $[-t_2,0)$ and $(0,t_1]$. So, the double-time argument in $G_{ab}(\tau_1,\tau_2)$ has $4$ distinct cases. Besides the relationship \eqref{eq:GlobalG} between the components of $G_{ab}$, we should further identify the independent time domain in $(\tau_1,\tau_2)$ based on other symmetries.

We will review some transformations and symmetries in the $PT$-symmetric Linbladian \cite{Prosen:2012sn,Garcia-Garcia:2023yet}. They will provide us with the symmetries of the spectrum and the symmetries of the two-point function, which connect the two-point function in different time domains.

\begin{itemize}
    \item Anti-commutation relation of fermions. Exchanging the $\psi_j^a(\tau_1)$ and $\psi_j^b(\tau_2)$ in \eqref{eq:TwoPtFunction} leads to the relation
\begin{equation}
    G_{ab}(\tau_1,\tau_2)=-G_{ba}(\tau_2,\tau_1).
\end{equation}

\item Hermitian conjugation $\dagger$. Since $\kc{e^{\tau_1\L^\dagger}e^{-\mu n}e^{\tau_2\L}}^\dagger=e^{\tau_2\L^\dagger}e^{-\mu n}e^{\tau_1\L}$, the conjugate of the two-point function will change the time argument,
\begin{equation}
    G_{ab}(\tau_1,\tau_2)=G_{ba}^*(-\tau_2,-\tau_1).
\end{equation}

\item Parity $P$ and time reversal $T$. 
They are defined as 
\begin{equation}
    P
    =e^{-\frac\pi4\sum_j\psi_j^L\psi_j^R}
    =\kc{e^{-i\frac\pi4 \sigma_z}}^{\otimes N },
    \quad TOT=O^*,
\end{equation} 
where we have expressed the transformation acting on the representation of Majorana fermions of the Jordan–Wigner (JW) transformation,
\begin{equation}\label{eq:JW}
	\psi^L_j=(-1)^{j-1}\sigma_z^{\otimes(j-1)}\otimes\sigma_x\otimes1^{\otimes(N-j)},\quad
 \psi^R_j=(-1)^{j-1}\sigma_z^{\otimes(j-1)}\otimes\sigma_y\otimes1^{\otimes(N-i)},
\end{equation} 
in this representation $T$ performs the complex conjugation of a matrix. The action of $P,\, T$ transformations are listed in Tab.~\ref{tab:Transfromation}.
One can check that $TPT=P^{-1}$ and the Liouvillian has $PT$ symmetry 
\begin{equation}\label{eq:PTsym}
    PT \L PT=\L.
\end{equation}
The $PT$ symmetry ensures that the spectrum of $\L$ is invariant under conjugation \cite{Bender:1998PT,Mostafazadeh:2001jk,Mostafazadeh:2001nr,Mostafazadeh:2002id,Zhang:2019gyc,Prosen:2012sn}.
Combining the $PT$ transformation, we have the relation
\begin{equation}
	G_{ab}(\tau_1,\tau_2) =G_{\bar{a}\bar{b}}^*(\tau_1,\tau_2),
\end{equation}
where $\bar L=R,\, \bar R=L$.

\begin{table}
	\centering
	\begin{tabular}{|c|c|c|c|}		
		\hline $O$ & $POP^{-1}$ & $TOT$ & $S^LOS^L$ \\
            \hline	
            \hline $i$ & $i$ & $-i$ & $i$ \\
            \hline $\psi_j^L$ & $\psi_j^R$&  $\psi_j^L$ & $-\psi_j^L$\\
		\hline $\psi_j^R$ & $-\psi_j^L$&  $-\psi_j^R$ & $\psi_j^R$\\
		\hline $n$&   $n$ & $n$  & $N-n$ \\
            \hline $H^L$ & $H^R$ & $i^qH^L$ & $H^L$  \\
            \hline $H^R$ & $H^L$ & $i^qH^R$ & $H^R$ \\
		\hline
	\end{tabular}\label{tab:Transfromation}
	\caption{The transformation of some operators under $P,\, T$ and $S^L$.}
\end{table}

\item $S^L$ transformation. It is defined as \cite{Garcia-Garcia:2023yet}
\begin{equation}
    S^L=i^{N(N-1)/2} \psi_1^L\psi_2^L\cdots \psi_N^L=\Gamma_{1\cdots N}^L
\end{equation}
with $S^LS^L=1$ and $(S^L)^\dagger=S^L$. Some results of $S^L$ transformation are listed in Tab.~\ref{tab:Transfromation}. 
The particle-hole conjugation is $TS^L$ \cite{Cotler:2016fpe}. 
We can check that
\begin{equation}\label{eq:Csym}
    S^L(\L+\frac12\nu N) S^L=(\L+\frac12\nu N)|_{\nu\to-\nu}=-(\L+\frac12\nu N)^\dagger, \quad S^L\ket0=\ket N.
\end{equation}
So the spectrum of $(\L+\frac12\nu N)$ is symmetric respective to the origin. Since $S^L$ does not preserve the state $\ket0$, it can not be a new symmetry of the two-point function. However, it can relate the two-point function defined on state $\ket0$ \eqref{eq:TwoPtFunction} and the following two-point function defined on state $\ket N$
\begin{equation}
    G_{ab}(\tau_1,\tau_2)=\frac1{N\avg{N|N}}\sum_j\bra N \mathcal T\kd{e^{-\mu(N-n(0))}\psi_j^a(\tau_1)\psi_j^b(\tau_2)} \ket N\Big |_{\nu\to-\nu},
\end{equation}
where $\bra N e^{-\mu n} \ket N=e^{-\mu N}$ is used. The relation between two-point functions results in a simple relation between two sizes
\begin{equation}
    n[\psi_1(t)]=N-n[\Gamma_{2\cdots N}(t)]|_{\nu\to-\nu}.
\end{equation}
Similarly, for the other $\Gamma_I$ operators, we will have $n[\Gamma_I(t)]=N-n[\Gamma_{I^\complement}(t)]|_{\nu\to-\nu}$, where $I^\complement$ is the complement of $I$. This relation aligns with our intuition: a jump term in the Linbladian with strength $|\nu|$, which suppresses the size of $\Gamma_I(t)$, has the same effect as the jump term with $-|\nu|$ that enhances the size of $\Gamma_{I^\complement}(t)$. In this sense, the Lindbladian with $\nu<0$ is also meaningful for the system coupled to a Markovian reservoir.

\end{itemize}

\begin{figure}
    \centering
    \includegraphics[width=0.4\textwidth]{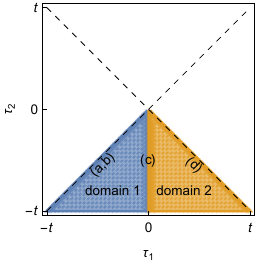}
    \caption{The independent time domain ($ -\tau_2\geq\tau_1\geq\tau_2$) is colored, which consists of domain 1 ($0\geq \tau_1\geq\tau_2$) (blue triangle) and domain 2 ($-\tau_2\geq\tau_1\geq0$) (orange triangle). The locations of the boundary conditions \eqref{eq:LiouvBC} are labeled by the letters in brackets. We have chosen $t_1=t_2=t$.}
    \label{fig:domains}
\end{figure}

In summary, the spectrum of $(\L+\frac12\nu N)$ is symmetric with respect to both the real axis and the imaginary axis. The two-point function has the symmetries:
\begin{equation}\label{eq:sym}
	G_{ab}(\tau_1,\tau_2)=-G_{ba}(\tau_2,\tau_1)=G_{ba}^*(-\tau_2,-\tau_1)=G_{\bar a\bar b}^*(\tau_1,\tau_2).
\end{equation}
Combining them with the simplification \eqref{eq:GlobalG}, we can reduce the two-point function to a single component in an independent time domain. We choose
\begin{equation}\label{eq:triangle}
    G_{LL}(\tau_1,\tau_2)\,\text{ with }  -\tau_2\geq\tau_1\geq\tau_2.
\end{equation}
The time domain forms a triangle, as shown in Fig.~\ref{fig:domains}. With this analysis on our hands, we will now numerically and analytically solve the problem in the following two sections.

\section{Numerical operator size at finite $q$}\label{sec:Numeric}

\subsection{Exact diagonalization at finite $N$}\label{sec:ED}

\begin{figure}
    \centering
    \includegraphics[width=1\textwidth]{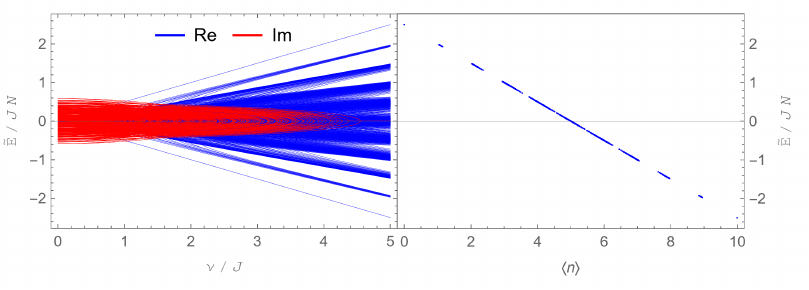}
    \caption{(Left) The spectrum $\ke{\tilde E_k}$ of a realization of shifted Liouvillian $\L+\frac12 \nu N$ as a function of $\nu/J$, where $N=10,\, q=4$. (Right) The shifted spectrum $\ke{\tilde E_k}$ versus the expectation value of size $\avg{n}=\bra{R_k}n\ket{R_k}/\avg{R_k|R_k}$ at $\nu/J=5$. Comparing the left and right panel, we find that the bands in the spectrum at large $\nu/J$, e.g. $\nu/J=5$, could be labeled by the size $\avg{n}$.}
    \label{fig:spectrum}
\end{figure}

\begin{figure}
    \centering
    \includegraphics[height=0.29\textwidth]{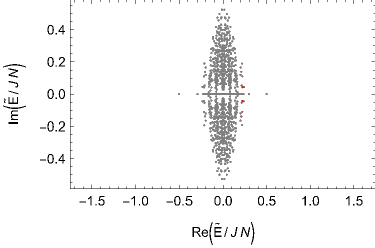}
    \includegraphics[height=0.29\textwidth]{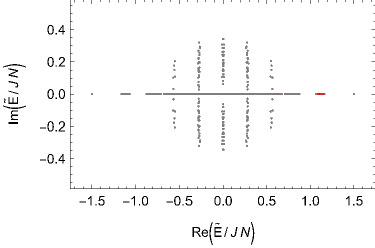}
    \includegraphics[height=0.29\textwidth]{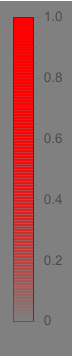}
    \caption{The spectrum of one realization of Liouvillian $\L$ is shown on the complex plane of shifted energy $\tilde E=E+\frac12\nu N$. The shade of red of each dot with energy $\tilde E_k$ denotes the probability $\abs{\avg{L_k|\psi_1}}^2$. The probability on the eigenstate with the energy labeled by a gray dot is negligible. The parameters are $N=10,\,q=4$ and $\nu/J=1$ (left), $3$ (right).}
    \label{fig:SpectrumDistrubition}
\end{figure}

We will numerically diagonalize the Liouvillian \eqref{eq:Liouvillian} at $q=4$ and $N=10$ and then study the evolution of the operator size and size distribution. We observe that, when $\nu>0$ ($\nu<0$), the operator size growth is suppressed (enhanced) and the size reaches a stable value smaller (bigger) than $N/2$ at the late time. We skip the case of $q=2$, since the size is trivially $n[\psi_1(t)]=1$.

We use the Jordan-Wigner transformation \eqref{eq:JW} to construct the Liouvillian \eqref{eq:Liouvillian} and numerically diagonalize it to find its spectrum $\ke{E_k|k=1,2,\cdots, D}$, see also \cite{Kulkarni:2021gtt,Zhou:2021yyw}. We plot the shifted spectrum $\ke{\tilde E_k=E_k+\frac12\nu N}$, because it is symmetric with respect to both the real axis and the imaginary axis, according to the symmetry analysis in Subsec.~\ref{sec:symmetry}. The spectrum of one realization at positive $\nu$ is shown in the left panel of Fig.~\ref{fig:spectrum}, which is identical to the spectrum with $-\nu$. The spectrum of Liouvillian is quite different from the pure SYK spectrum \cite{Cotler:2016fpe,Garcia-Garcia:2016mno,you2017sachdev}. When $\nu/J\gg1$, the spectrum is real and exhibits energy bands and gaps, where the bands are approximately labeled by their sizes, as shown in the right panel of Fig.~\ref{fig:spectrum}. When $\nu/J\sim 1$, some real energy pairs collide with each other $E_k\to E_{k'}$ and then move into the complex plane. The points where such collision happens are called exceptional points \cite{Bender:1998PT}. When $\nu/J\ll1$, most but not all the energies become complex.

We can further find the biorthogonal eigenbasis $\ke{\kc{\ket{R_k},\bra{L_k}}}$ 
with $\L\ket{R_k}=E_k\ket{R_k},\, \bra{L_k}\L=E_k\bra{L_k}$ and $\avg{L_k|R_{k'}}=\delta_{kk'}$ \cite{Brody2013BiorthogonalQM}, except at the exceptional points, where two eigenstates become linearly dependent, $\ket{R_k}\to \ket{R_{k'}}$. Before their collision, the two eigenstate have $PT$ symmetry, namely $PT\ket{R_k}\propto\ket{R_k}$ and $PT\ket{R_{k'}}\propto\ket{R_{k'}}$. After the collision, $PT$ symmetry is spontaneously  broken into $PT\ket{R_k}\propto\ket{R_{k'}}$. The real and imaginary parts of the spectrum are related to the decomposition of Liouvillian \eqref{eq:Liouvillian} via
\begin{equation}
    \Re E_k = \frac{\bra{R_k}\X\ket{R_k}}{\avg{R_k|R_k}},\quad \Im E_k = \frac{\bra{R_k}\P\ket{R_k}}{\avg{R_k|R_k}},
\end{equation}
or, equivalently, with $R_k\to L_k$ being replaced.

The spectrum $\ke{E_k}$ and the probability distribution $\abs{\avg{L_k|\psi_1}}^2$ are plotted in Fig.~\ref{fig:SpectrumDistrubition}. For a large $\nu/J$, the energies exhibiting imaginary parts still form bands and gaps. The state $\ket{\psi_1}$ mainly distributes around the $n=1$ band. When $\nu/J$ decreases, these bands collapse into a cluster. The state $\ket{\psi_1}$ mainly distributes at the edge of the cluster with large $\Re E_k$. Based on the diagonalization and the probability distribution, we can calculate the size and size distribution
\begin{align}
    n[\psi_1(t)]=&~\sum_{kk'}\avg{\psi_1|L_k}e^{t E_k^*}\bra{R_k}n\ket{R_{k'}}e^{t E_{k'}}\avg{L_{k'}|\psi_1},\\
    P_n[\psi_1(t)]=&~ \frac1A\sum_{\abs{I}=n}\abs{\avg{\Gamma_I|R_{k'}}e^{t E_{k'}}\avg{L_{k'}|\psi_1}}^2,
\end{align}
where the normalization factor $A$ is determined by $\sum_n P_n=1$.

\begin{figure}
    \centering
    \includegraphics[width=0.7\textwidth]{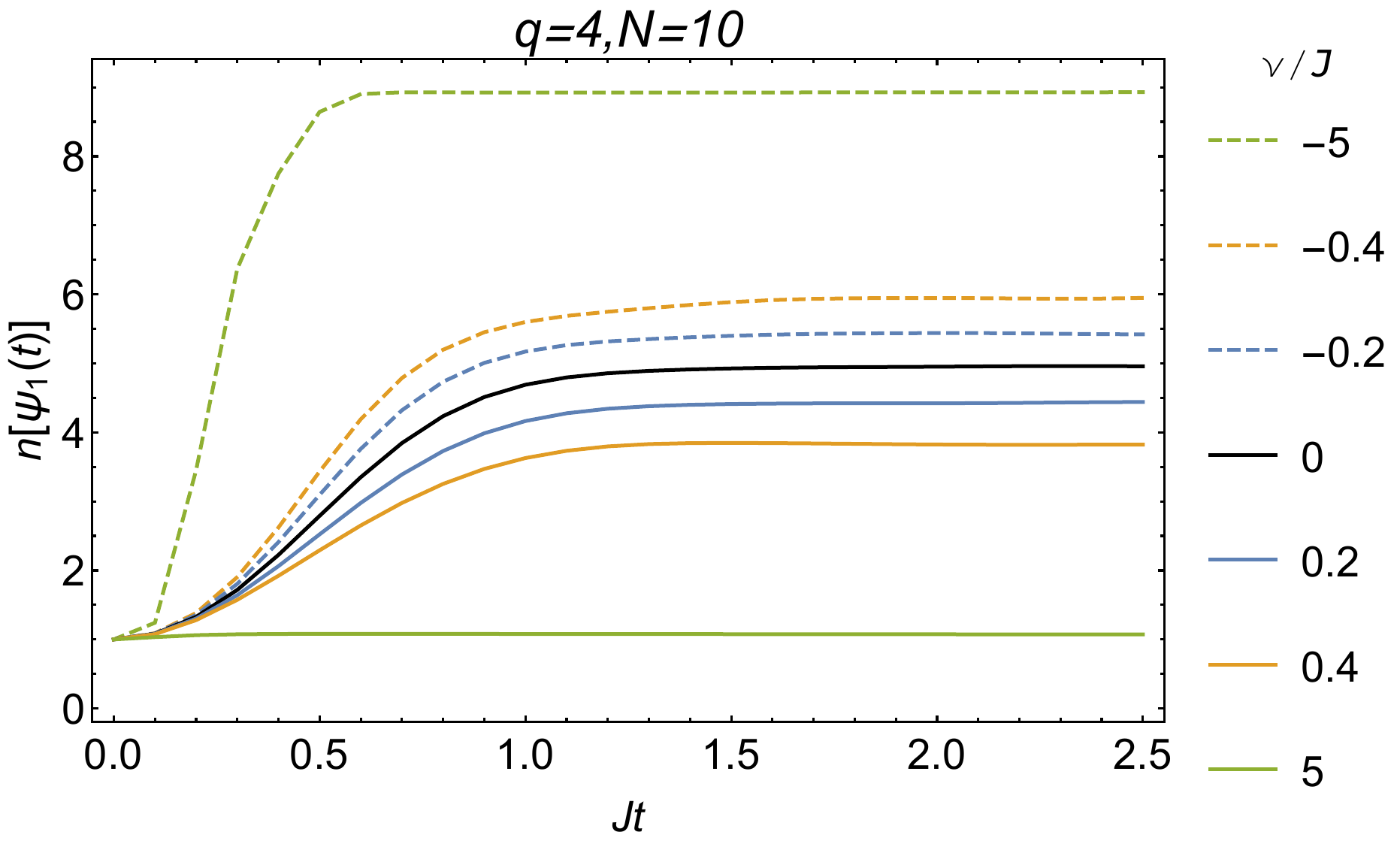}
    \caption{Operator size growth $n[\psi_1(t)]$ in the average of $20$ samples, where $q=4$ and $N=10$.}
    \label{fig:EDSize}
\end{figure}

\begin{figure}
    \centering
    \includegraphics[height=0.29\textwidth]{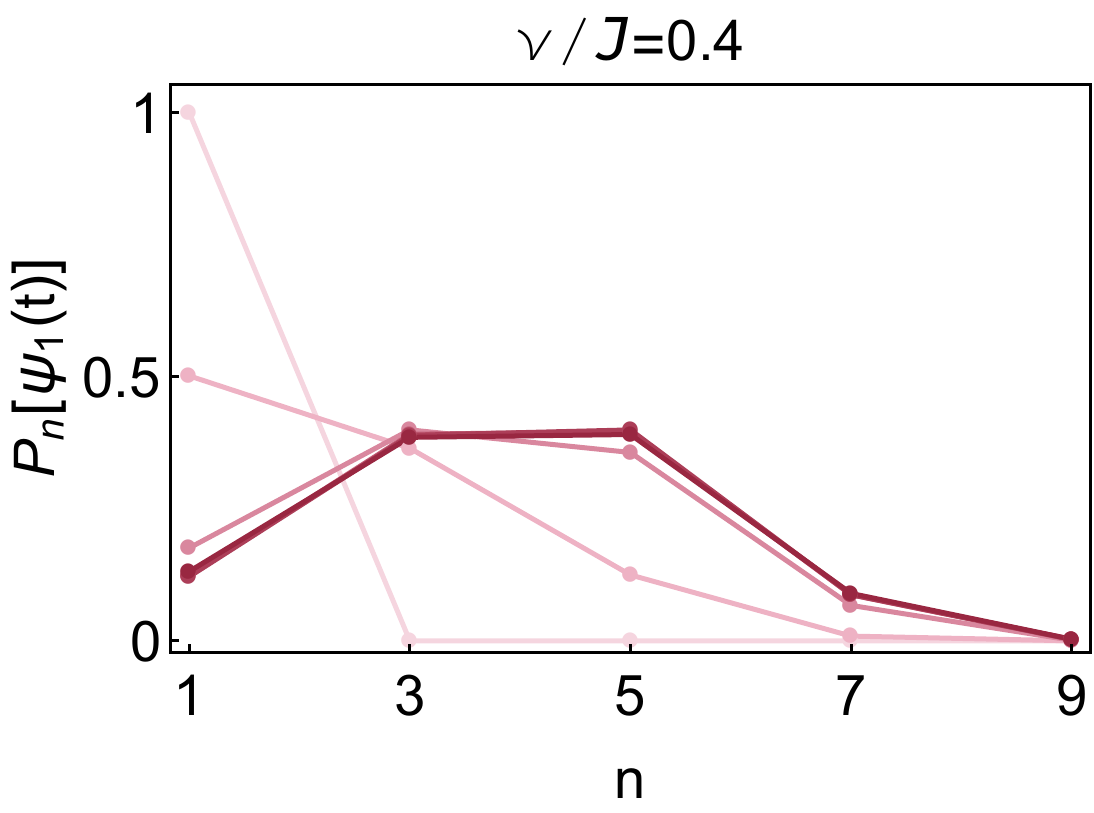}
    \includegraphics[height=0.29\textwidth]{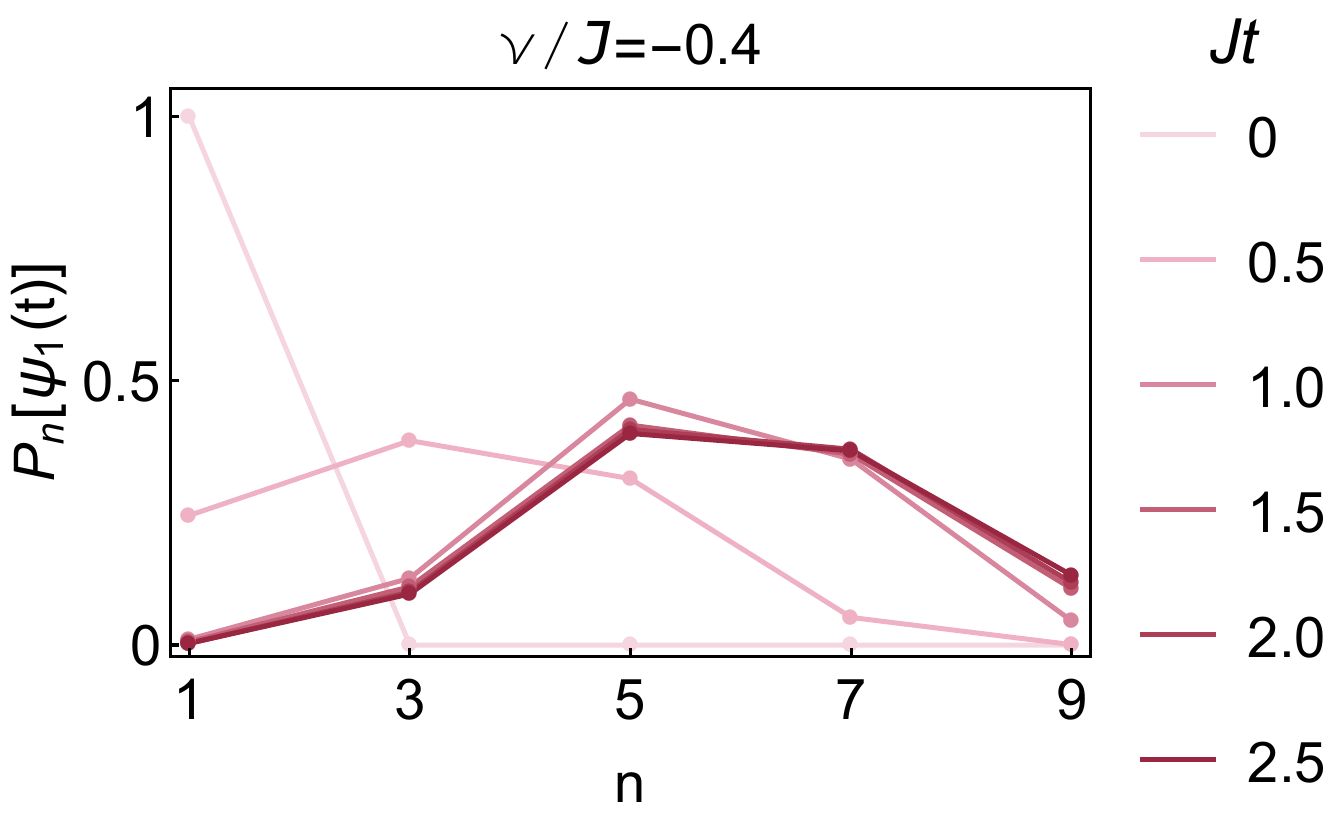}
    \caption{Some snapshots of the operator size distribution $P_n[\psi_1(t)]$ for $\nu/J=0.4$ (left) and $\nu/J=-0.4$ (right), where $q=4$ and $N=10$. Only the distribution for odd $n$ is depicted.}
    \label{fig:EDDist}
\end{figure}

The growth of operator size $n[\psi_1(t)]$ from ED is shown in Fig.~\ref{fig:EDSize}. The larger the $\nu/J$, the slower the operator size grows, and the lower the plateau it reaches. But the reasons for the different sizes plateau of different $\nu$ varies. For $\nu/J=0$, the plateau is given by $N/2$ due to full scrambling. With strong dissipation $\nu/J\gg1$, the plateau is determined by the competition between the interaction $J$ and the dissipation $\nu$. With strong production $-\nu/J\gg1$, the plateau is pushed to a value a small distance away from the maximally possible operator size $N-1$, determined by the ratio between the interaction $J$ and the production $-\nu$. In the limit $-\nu/J \rightarrow \infty$, the operator size will approach $N-1$. In Fig.~\ref{fig:EDDist}, we further show the size distribution $P_n[\psi_1(t)]$ for $\nu>0$ or $\nu<0$. We can see that dissipation $\nu>0$ (production $\nu<0$) suppresses (enhances) the probability of large sizes. Notice that all the probability on even sizes vanish because each commutation with the $q=4$ SYK Hamiltonian can only change the size by even numbers.

In App.~\ref{sec:Temperature}, we calculate the size and size distribution of the pure SYK thermal state $(e^{-\beta H/2})(t)$ and the thermal fermion $(e^{-\beta H/2}\psi_1)(t)$ in the same way. For low temperature, these operators have larger initial sizes. Their sizes decrease or increase, depending on the strength of dissipation or production.

\begin{figure}
	\centering 
	\includegraphics[width=0.45\textwidth]{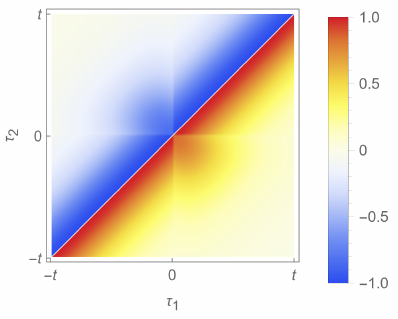}
	\caption{Numerical solution $G_{LL}(\tau_1,\tau_2)$ of the SD equation \eqref{eq:SDGSigma}, where $q=6,\,\hat\nu=q\nu=0.1\J,\,\hat\mu=q\mu=1\J$ and $t_1=t_2=t=6\J$.}
	\label{fig:numSD}
\end{figure}

\subsection{Numerical Schwinger-Dyson equation at infinite $N$}\label{sec:SD}

\begin{figure}
    \centering 
	\includegraphics[height=0.29\textwidth]{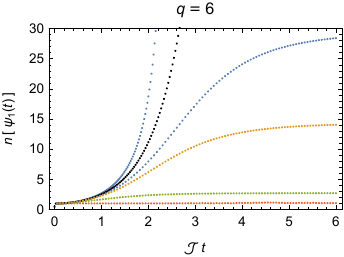}
	\includegraphics[height=0.29\textwidth]{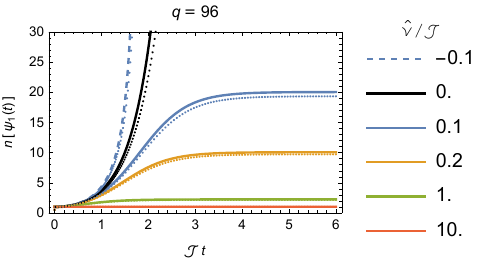}
    \caption{Operator sizes growth $n[\psi_1(t)]$ from the numeric SD equation (dots) at $q=6,96$ and the analytical result \eqref{eq:largeqsize} at infinite $q$ (curves), where $\hat\nu/\J=q\nu/\J=-0.1,0,0.1,0.2,1,10$.}
    \label{fig:SDlargeqSize}
\end{figure}

We also calculated the two-point function by numerically solving the SD equation \eqref{eq:SDGSigma}. We plot the configurations of the numerical solution $G_{LL}(\tau_1,\tau_2)$ at finite $\mu$ in Fig.~\ref{fig:numSD}. Discontinuities appear across the interfaces $\tau_1=0$ and $\tau_2=0$ due to the insertion $e^{-\mu n}$ there. To extract the operator size according to \eqref{eq:size}, we vary $\mu$ around $0$ and calculate the difference of generating functions. We have chosen a long time $t_1=t_2=t$ so that, as shown in Fig.~\ref{fig:SDlargeqSize}, the overall growth-plateau behavior of the operator size is covered. For dissipation $\nu>0$, the growth rate is suppressed and a plateau of operator size emerges at finite time. For production $\nu<0$, the growth rate is enhanced and the numerical method breaks down in finite time, indicating the divergence of operator size and necessitating the regularization by finite $N$ effects. In the next section, we will find the analytical operator size growth at large $q$, which nicely matches the numerical result for any $\nu$ at large but finite $q$, as already demonstrated in Fig.~\ref{fig:SDlargeqSize}.

\section{Analytical operator size at large $q$}\label{sec:Analytic}

In this section, we will solve the Lindbladian SYK in the large $q$ limit, where the SD equations, represented as integral equations, reduce to the Liouville equations, which are differential equations. Following \cite{Maldacena:2016remarks,Maldacena:2018lmt,Qi:2018quantum,Streicher:2019wek,Kulkarni:2021gtt}, when considering the $1\ll q^2\ll N$ limit, we will keep the following parameters fixed
\begin{equation}
    t,\quad \J=\sqrt{2q}J,\quad \hat\nu=q\nu,\quad \hat\mu=q\mu.
\end{equation}
We will analytically solve the Liouville equations with boundary conditions and calculate the generating functions, Loschmidt echo fidelity, operator size, size distribution, and the variances of the size operator and the Liouvillian.

\subsection{Liouville equation}\label{sec:Liouville}

Owing to the symmetries \eqref{eq:sym}, we use the following large $q$ ansatz for the independent parts of two-point function \eqref{eq:triangle}
\begin{equation}\label{eq:ansatz}
	G_{LL}(\tau_1,\tau_2)=\sgn(\tau_1-\tau_2)e^{g(\tau_1,\tau_2)/q},
\end{equation}
where the prefactors are determined by the free case.

Plugging \eqref{eq:ansatz} into the SD equation \eqref{eq:SDGSigma}, utilizing the relation \eqref{eq:GlobalG} and taking the large $q$ expansion, we find that the $\kc{1/q}^0$ order is automatically solved, and the $(1/q)^1$ order gives rise to the Liouville equation. Because the factor $\sgn(\tau)$ in the SD equation changes its sign at $\tau=0$, we should further divide the triangular domain \eqref{eq:triangle} into two subdomains, as shown in Fig.~\ref{fig:domains}, where the Liouville equations behave differently in respected interiors
\begin{subequations}
\label{eq:Liouville}
\begin{align}
\text{Domain 1: } 0>\tau_1>\tau_2,~~~~~~~
&\pa_{\tau_1}\pa_{\tau_2}g_1(\tau_1,\tau_2)=2\J^2e^{g_1(\tau_1,\tau_2)}
\label{eq:Liouv1},\\
\text{Domain 2: } -\tau_2>\tau_1>0,\quad
&\pa_{\tau_1}\pa_{\tau_2}g_2(\tau_1,\tau_2)=-2\J^2e^{g_2(\tau_1,\tau_2)}
\label{eq:Liouv3}.
\end{align}
\end{subequations}
$g_1(\tau_1,\tau_2)$ is termed uncrossed function because it represents the correlation not crossing the point $\tau=0$, while $g_2(\tau_1,\tau_2)$ is termed crossed function because it represents the correlation crossing the point $\tau=0$.

Next, we discuss the conditions on the boundaries of domains 1 and 2, namely  $\tau_-=0$, $\tau_1=0$ and $\tau_+=0$, as labeled in Fig.~\ref{fig:domains}, where $\tau_\pm\equiv\tau_1\pm\tau_2$ and $\partial_{\tau_\pm}=\frac12(\partial_{\tau_1}\pm\partial_{\tau_2})$:
\begin{enumerate}
    \item[(a)] From the anti-commutation relation, we have $G_{LL}(\tau,\tau)=1$, so
    \begin{equation}
        g_1(\tau,\tau)=0.
    \end{equation}
    
    \item[(b)] When $\tau_1$ passes through $\tau_2$, we should take the additional delta $\nu\delta(\tau_1-\tau_2)$ term in the SD equation \eqref{eq:SDGSigma} into account. The Liouville equation \eqref{eq:Liouv1} acquires an additional delta term,
\begin{equation}\label{eq:LiouvilleDelta}
    \partial_{\tau_1}\partial_{\tau_2}g_1(\tau_1,\tau_2)=2\J^2e^{g_1(\tau_1,\tau_2)}+2\hat{\nu}\delta(\tau_1-\tau_2)
\end{equation}
Integrating $\tau_-$ over an infinitesimally internal around $\tau_-=0$ and using the symmetry \eqref{eq:sym}, we obtain the boundary condition
\begin{equation}
    \lim_{\tau_-\to0^+}\partial_{\tau_-}g_1(\tau_1,\tau_2)=-\hat\nu.
\end{equation}

\item[(c)] The insertion of $e^{-\mu n(0)}$ yields the twisted boundary condition \cite{Qi:2018quantum,Streicher:2019wek}
\begin{equation}
    e^{-\mu n}
    \begin{pmatrix}
	\psi_j^L\\
	i\psi_j^R
    \end{pmatrix}
 =
 \begin{pmatrix}
	\cosh\mu&\sinh\mu\\
	\sinh\mu&\cosh\mu
    \end{pmatrix}
    \begin{pmatrix}
	\psi_j^L\\
	i\psi_j^R
    \end{pmatrix}
    e^{-\mu n}.
\end{equation}
It leads to the following twisted condition for both two-point functions at $\tau_1=0$
\begin{equation}\label{eq:Twist}
    \begin{pmatrix}
	G_{La}(0^-,\tau)\\
	iG_{Rb}(0^-,\tau)
    \end{pmatrix}
    = 
    \begin{pmatrix}
	\cosh\mu&\sinh\mu\\
	\sinh\mu&\cosh\mu
    \end{pmatrix}
     \begin{pmatrix}
	G_{La}(0^+,\tau)\\
	iG_{Rb}(0^+,\tau)
    \end{pmatrix}
    \xrightarrow{q\gg1}
    e^{\hat\mu/q}
    \begin{pmatrix}
	G_{La}(0^+,\tau)\\
	iG_{Rb}(0^+,\tau)
    \end{pmatrix}
\end{equation}
and similar one for $\tau_2=0$.
The two-point function exhibits discontinuity across the $\tau_1=0$ interface and the $\tau_2=0$ interface, as shown in the numerical solution in Fig.~\ref{fig:numSD}.  At large $q$, where $\mu=\hat{\mu}/q$ is small, the twisted boundary conditions decouple as shown at the last step of \eqref{eq:Twist}. It leads to the following boundary condition between the crossed and uncrossed functions at the $\tau_1=0$ interface between domain 1 and domain 2,
\begin{equation}
    \lim_{\tau_1\to0}g_1(\tau_1,\tau_2)
    =\lim_{\tau_1\to0}g_2(\tau_1,\tau_2)+\hat{\mu}.
\end{equation}

\item[(d)] On the boundary $\tau_+=0$ of domain 2, the SD equation \eqref{eq:SDGSigma} does not have any delta source. From \eqref{eq:sym}, we know that the two-point function is symmetric with respect to this line. So we can impose the smoothness condition on this boundary,
\begin{equation}
	\left.\partial_{\tau_+}g_2(\tau_1,\tau_2)\right|_{\tau_+=0}=0.
\end{equation}

\end{enumerate}

\noindent Below we summarize all the boundary conditions derived above:
\begin{subequations}
\label{eq:LiouvBC}
\begin{align}    
{\rm (a)}&~g_1(\tau,\tau)=0.\\
{\rm (b)}&~\lim_{\tau_-\to0^+}\partial_{\tau_-}g_1(\tau_1,\tau_2)=-\hat\nu.\\
{\rm (c)}&~\lim_{\tau_1\to0}g_1(\tau_1,\tau_2)
    =\lim_{\tau_1\to0}g_2(\tau_1,\tau_2)+\hat{\mu}.\\ 
{\rm (d)}&~\left.\partial_{\tau_+}g_2(\tau_1,\tau_2)\right|_{\tau_+=0}=0
\end{align}
\end{subequations}

\subsection{Solution and generating functions}

The Liouville equations \eqref{eq:Liouville} with boundary conditions \eqref{eq:LiouvBC} can determine the solution. Here we find a solution through the following analysis. For the uncrossed function $g_1(\tau_1,\tau_2)$ in domain 1, the operators $e^{t_1\L^\dagger}$ and $e^{-\mu n}$ in the partition function \eqref{eq:PatitionFunction} are eliminated by observing $\bra0 \L^\dagger=\bra0n=0$. So the uncrossed function $g_1(\tau_1,\tau_2)$ should be the same as the translational invariant solution in \cite{Kulkarni:2021gtt}.
Then we can determine the crossed function $g_2(\tau_1,\tau_2)$ in domain 2 by matching the general solution of the Liouville equation in \cite{Gao:2019nyj} to the boundary conditions \eqref{eq:LiouvBC} with the known uncrossed function $g_1(\tau_1,\tau_2)$. The analytical solution we find is
\begin{subequations}\label{eq:asol}
\begin{align}
	&e^{g_1(\tau_1,\tau_2)}=\kc{\frac{\cosh\gamma}{\cosh(\alpha(\tau_1-\tau_2)+\gamma)}}^2,\label{eq:asol1}\\
	&e^{g_2(\tau_1,\tau_2)}= \label{eq:asol2}\\
	&\kc{\frac{2 e^{\hat{\mu}/2}\cosh^2\gamma}{(e^{\hat{\mu}}+1)\cosh\kc{\alpha(\tau_1+\tau_2)}+e^{\hat{\mu}}\cosh\kc{\alpha(\tau_1-\tau_2)+2 \gamma }-\cosh\kc{\alpha(\tau_1-\tau_2)}}}^2\nn,
\end{align}
\end{subequations}
where the parameters $\alpha$ and $\gamma$ are determined by
\begin{equation}
	\alpha=\J \cosh\gamma, \quad \gamma=\arcsinh\frac{\hat\nu}{2\J}.
\end{equation}
In the following two limits of weak and strong dissipation, the two parameters approach
\begin{align}
    \text{weak dissipation}\quad &\hat\nu/\J\ll1,\quad \alpha\approx \J,\quad \gamma\approx \frac{\hat\nu}{2\J},\\
    \text{strong dissipation}\quad  &\hat\nu/\J\gg1,\quad 
    \alpha\approx \frac{\hat\nu}{2},\quad \gamma\approx \ln\frac{\hat\nu}{\J}.
\end{align}

The large $q$ expansion below \eqref{eq:ansatz} is valid when both $e^{g_1(\tau_1,\tau_2)/q}$ and $e^{g_2(\tau_1,\tau_2)/q}$ are ${\cal O}(1)$. When $\hat\nu>0$, they decay as $e^{-2\alpha(\tau_1-\tau_2)/q}$ for large time differences $\tau_1-\tau_2\gg 1/\alpha$. The solution is valid when $\tau_1-\tau_2\ll q/\alpha$. When $\hat\nu<0$, $g_2(\tau_1,\tau_2)$ suffers from a divergence along a line in domain 2 where the denominator in \eqref{eq:asol2} vanishes. The large $q$ solution is not applicable beyond that line, and necessitates regularization due to the finite $N$ effect. Fortunately, the validity region in time is, due to the large $q$ limit, extensive enough for our analysis of operator size.

Plugging \eqref{eq:asol2} into \eqref{eq:GeneratingG} and \eqref{eq:GeneratingG}, we obtain the generating function
\begin{equation}\label{eq:largeqGen}
    \G_\mu(t)=e^{g_2(t,-t)/q}
    =\kc{\frac{2 e^{\hat\mu /2}\cosh^2\gamma } {1-\cosh (2 \alpha  t)+2 e^{\hat\mu } \cosh ^2(\alpha  t+\gamma)}}^{2/q}
\end{equation}
and the double-time generating function
\begin{equation}\label{eq:largeqDGen}
\begin{split}
    \G_\mu(t_1,t_2)
    =&~e^{g_2(t_1,-t_2)/q}\\
    =&~ \kc{\frac{e^{\hat\mu/2}\cosh^2\gamma} {\sinh\left(\alpha  t_1\right) \sinh \left(\alpha  t_2\right) - e^{\hat\mu } \cosh \left(\alpha  t_1+\gamma\right) \cosh \left(\alpha  t_2+\gamma\right)}}^{2/q},
\end{split}
\end{equation}
By taking the derivatives of $\G_\mu(t_1,t_2)$ with respect to $t_1,t_2$ and $-\mu$, we can obtain the expectation values of $\L^\dagger,\, \L$ and $n$ respectively.

\subsection{Loschmidt echo fidelity}

Before calculating observable, we investigate the state $\ket{\psi_1(t)}$ first. 
The generating function \eqref{eq:largeqGen} at $\mu=0$ reduces to the (square root of) Loschmidt echo fidelity \cite{Yan:2019fbg,Schuster:2022bot}
\begin{align}\label{eq:Fidelity}
    \G_0(t)=\bra{\psi_1}e^{t\L^\dagger}e^{t\L}\ket{\psi_1}=\left(\frac{\cosh^2\gamma}{1+\sinh (\gamma )\sinh(2 \alpha  t+\gamma)}\right)^{2/q}.
\end{align}
It measures the return amplitude for the state $\ket{\psi_1}$, which has undergone a forward-backward evolution in the presence of the difference between the two Hamiltonians $i\L+i\L^\dagger=-2i\nu n$. Contrary to the typical Loschmidt echo \cite{Gorin:2006loschmidt,Goussev:2012loschmidt}, the difference in this case is non-Hermitian. The Loschmidt echo fidelity decays exponentially as
\begin{equation}\label{eq:echodecay}
    \G_0(t)\approx \kd{2 e^{-\gamma}\cosh(\gamma)\coth(\gamma )}^{2/q} e^{-4 \alpha  t/q}
\end{equation}
after a time 
\begin{equation}\label{eq:PlateauTime}
    t_p\equiv\frac1{2\alpha}\ln \frac4{e^{2\gamma}-1},
\end{equation}
which indicates the time when the state $\ket{\psi_1(t)}$ becomes stable up to normalization. We call $t_p$ the plateau time since most of the observable will become stable after this time, such as the operator size \cite{Schuster:2022bot} and the Krylov complexity \cite{Bhattacharjee:2022lzy}. At weak dissipation $\hat\nu/\J\ll1$, the plateau time reduces to
\begin{equation}
    t_p\approx \frac1{2\J}\ln\frac{\J}{\hat\nu}.
\end{equation}
This plateau time is similar to the Ehrenfest time in the unitary chaotic evolution \cite{Gorin:2006loschmidt,Goussev:2012loschmidt}.
Recall the scrambling time $t_*\approx \frac1{2\J}\ln N$ in many-body chaotic systems, indicating the time when the local information spread over the system's degree of freedom $N$ \cite{Sekino:2008scramblers,Roberts:2014localized,Shenker:2013yza,Maldacena:2015waa,Maldacena:2016remarks}. The plateau time $t_p$ could be comparable to the scrambling time $t_*$ if $\hat\nu/\J\sim 1/N$.

As predicted by \cite{Jalabert:2001decoherence} and observed in \cite{Schuster:2022bot}, at weak dissipation, the decay rate of the Loschmidt echo \eqref{eq:echodecay} $4\alpha/q\approx 4\J/q$ depends on the pure SYK Hamiltonian coupling but not on the dissipation strength $\hat\nu$. While, at finite dissipation, the decay rate $4\alpha/q$ becomes dependent on $\hat\nu$. Furthermore, at large $q$, due to the $1/q$ power in \eqref{eq:Fidelity} and \eqref{eq:echodecay}, the fidelity $\G_0(t)$ is of order $1$ at the plateau time, decays extremely slowly, and becomes significantly lower than $1$ only after the time $q/\alpha$.

The double-time generating function \eqref{eq:largeqDGen} at $\mu=0$ reduces to the overlap between the states at different times without normalization. We can examine the normalized overlap probability
\begin{equation}\label{eq:overlap}
\begin{split}
    &~\frac{\abs{\avg{\psi_1(t_1)|\psi_1(t_2)}}^2}{\avg{\psi_1(t_1)|\psi_1(t_1)}\avg{\psi_1(t_2)|\psi_1(t_2)}}
    =\frac{\abs{\G_0(t_1,t_2)}^2}{\G_0(t_1)\G_0(t_2)}\\
    =&~\kd{\frac{\left(\cosh ^2\left(\gamma +\alpha  t_1\right)-\sinh ^2\left(\alpha  t_1\right)\right) \left(\cosh ^2\left(\gamma +\alpha  t_2\right)-\sinh ^2\left(\alpha  t_2\right)\right)}{\left(\cosh \left(\gamma +\alpha  t_1\right) \cosh \left(\gamma +\alpha  t_2\right)-\sinh \left(\alpha  t_1\right) \sinh \left(\alpha  t_2\right)\right){}^2}}^{2/q}\\
    =&~\kd{1-\frac1{\gamma ^2}e^{-2 \gamma}\kc{e^{-4 \alpha  t_1}+e^{-4 \alpha  t_2}-2 e^{ -2 \alpha  \left(t_1+t_2\right)}}+\cdots}^{2/q},
\end{split}
\end{equation}
which is close to $1^{2/q}$ when both $t_1,t_2\gg t_p$. It means that the state $\ket{\psi_1(t)}$ remains nearly unchanged up to normalization long after the plateau time $t_p$.

\subsection{Size growth}

\begin{figure}
    \centering
    \includegraphics[height=0.27\linewidth]{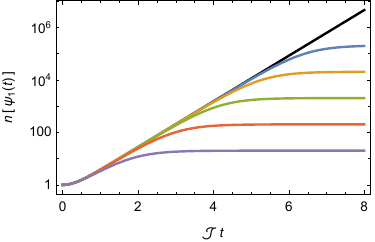}
    \includegraphics[height=0.27\linewidth]{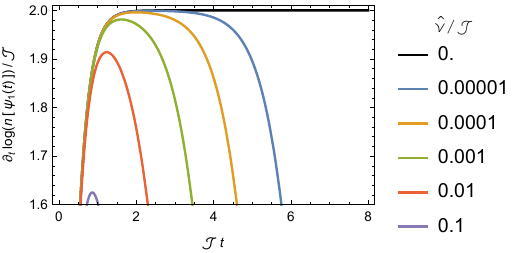}
    \caption{The operator size $n[\psi_1(t)]$ and its ``exponent'' $\partial_t\log n[\psi_1(t)]/\J$ as functions of time. The dimensionless plateau times are $\J t_p\approx\kc{\infty ,5.8,4.6,3.5,2.3,1.2}$ respectively.}
    \label{fig:sizeexp}
\end{figure}

We denote the expectation value of an operator $O$ in state $\ket{\psi_1(t)}$ as \\
$\avg{O}=\bra{\psi_1(t)}O\ket{\psi_1(t)}/\avg{\psi_1(t)|\psi_1(t)}$.
Plugging the generating function \eqref{eq:largeqGen} into \eqref{eq:size}, we obtain the operator size
\begin{equation}\label{eq:largeqsize}
		n[\psi_1(t)]\equiv\avg{n}
  =\frac{\cosh(\gamma)  \cosh (2 \alpha  t+\gamma)}{1+\sinh (\gamma ) \sinh (2 \alpha  t+\gamma)}.
\end{equation}
In Fig.~\ref{fig:SDlargeqSize}, we show the behavior of the size $n[\psi_1(t)]$ in the large $q$ limit.

For vanishing dissipation $\hat\nu=0$, we have $\alpha=\J$ and $\gamma=0$. The size reduces to the pure SYK case $n[\psi_1(t)]=\cosh(2\J t)$ \cite{Roberts:2018operator}.

For dissipation $\hat \nu>0$, we always have $n[\psi_1(t)]\leq \cosh(2\J t)$ and the plateau value is $n[\psi_1(\infty)]=\coth(\gamma)$. For weak dissipation with $\hat\nu/\J\ll1$, the size $n[\psi_1(t)]$ at different time scales behaves as
\begin{equation}\label{eq:sizebehavior}
    n[\psi_1(t)]\approx \begin{cases}
        1+2\J^2 t^2-2\hat \nu\J^2 t^3,&  t\ll \hat\nu/4\J \\
        \frac12e^{2\J t}-\frac12\frac{\hat\nu}{\J}e^{4\J t},& 1/2\J\ll t \ll t_p\\
        2\J/\hat\nu,& t_p\ll t
    \end{cases},
\end{equation}
with the plateau time \eqref{eq:PlateauTime}. So the dissipation reduces the growth rate as well as the plateau value.
Only for weak dissipation $\hat\nu/\J\ll1$, the hierarchy between $1/2\J$ and $t_p$ is huge enough for the emergence of an exponential growth region, as shown in Fig.~\ref{fig:sizeexp}. From \eqref{eq:sizebehavior}, the leading correction resulting from dissipation in the exponential growth region breaks the pure exponential behavior rather than modifying its exponent.
The plateau value $\coth(\gamma)$ approaches to $2\J/\hat\nu$ at weak dissipation, as predicted in \cite{Schuster:2022bot}.
At strong dissipation $\hat\nu/\J\gg1$, the size ceases to grow, $n[\psi_1(t)]\approx1$.

For production $\hat\nu<0$, the early time behavior of the size is still described by \eqref{eq:sizebehavior}, whose growth rate is enhanced. But the size diverges at finite time 
\begin{equation}\label{eq:divergencetime}
    t_{\rm div}=\frac1{2\alpha} \kd{\ln \coth \left(\frac{-\gamma }{2}\right)-\gamma}, 
\end{equation}
where $\gamma<0$. The size divergence should be regularized by finite $N$ effects, as shown in Fig.~\ref{fig:EDSize}. After this time, \eqref{eq:largeqsize} is not reliable.

\subsection{Size distribution}

\begin{figure}[H]
	\centering 
    \includegraphics[height=0.32\textwidth]{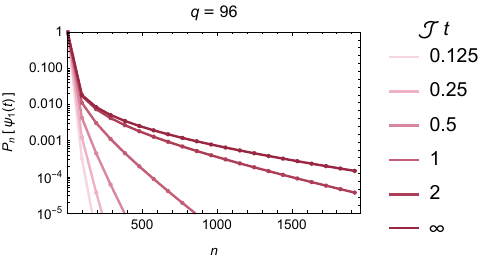}
    \includegraphics[height=0.32\textwidth]{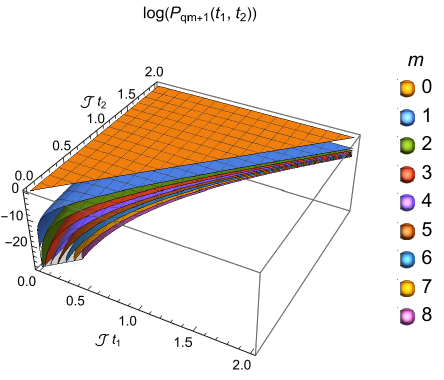}
	\caption{Operator size distribution $P_n[\psi_1(t)]$ (left) and double-time distribution $P_{qm+1}(t_1,t_2)$ (right) at $q=96$ and $\hat{\nu}/\J=0.1$.}
	\label{fig:largeqdistr}
\end{figure}

To study the details of size growth, we calculate the size distribution.
Expanding the generating function \eqref{eq:largeqGen} according to \eqref{eq:Distribution} using the binomial theorem and normalizing the distribution, we obtain the size distribution
\begin{equation}\label{eq:largeqDist}
	P_{qm+1}[\psi_1(t)]\equiv P_{qm+1}(t)=\binom{-2/q}{m}(-1)^m
 \theta(t)^{2m}
 \kc{1-\theta(t)^2}^{2/q},\quad m\in \mathbb N,
\end{equation}
where 
\begin{equation}\label{eq:theta}
    \theta(t)=\frac{\sinh(\alpha t)}{\cosh(\alpha t+\gamma)}\in [0,e^{-\gamma})
\end{equation}
and $P_n[\psi_1(t)]=0$ for $n\neq qm+1$, due to the large-$N$ suppression of non-melonic diagrams and also the large $q$ limit \cite{Maldacena:2016remarks,Gu:2016local,Roberts:2018operator,Qi:2018quantum}. More precisely, the size with nonzero distribution should be $(q-2)m+1$, but the $-2$ term is neglected in our discussion when considering large $q$. We plot the distribution for $n=qm+1$ in the left panel of Fig.~\ref{fig:largeqdistr}. 

When $\hat\nu\geq0$, the distribution in the long time limit converges to 
\begin{equation}
    P_{qm+1}[\psi_1(\infty)]=\binom{-2/q}{m}(-1)^m e^{-2 \gamma  m} \left(1-e^{-2 \gamma }\right)^{2/q}. 
\end{equation}
Unlike the pure SYK case, for which $P_{qm+1}[\psi_1(\infty)]=0$ \cite{Qi:2018quantum}, once we have dissipation $\hat\nu>0$, the distribution $P_{qm+1}[\psi_1(\infty)]$ is finite in the long time limit. The dissipation also leads to the exponential decay $P_{qm+1}[\psi_1(\infty)]\sim e^{-2\gamma m}$ when $m$ is large enough.

Similarly, we can expand the double-time generating function \eqref{eq:largeqDGen} according to 
\begin{equation}
    \G_\mu(t_1,t_2)=\G_0(t_1,t_2)\sum_n e^{-\mu n} P_n(t_1,t_2),\quad 
    P_n(t_1,t_2)\equiv \frac{\bra{\psi_1(t_1)}\pi_n\ket{\psi_1(t_2)}}{\avg{\psi_1(t_1)|\psi_1(t_2)}}
\end{equation}
with normalization $\sum_n P_n(t_1,t_2)=1$ and get the double-time distribution
\begin{equation}\label{eq:largeqDDist}
    P_{qm+1}(t_1,t_2)=
    \binom{-2/q}{n} (-1)^m \kc{\theta(t_1)\theta(t_2)}^m  \kc{1-\theta(t_1)\theta(t_2)}^{2/q},\quad m\in\mathbb N,
\end{equation}
and $P_n(t_1,t_2)=0$ for $n\neq qm+1$. It measures the overlap between $\ket{\psi_1(t_1)}$ and $\ket{\psi_1(t_2)}$ in the size subspace $\H_{qm+1}$ without state normalization. Its behavior is shown in the right panel of Fig.~\ref{fig:largeqdistr}.

\subsection{Operator size concentration}\label{sec:concentration}

In this subsection, we prove the property of operator size concentration at finite dissipation in the large $q$ limit. The operator size concentration, proposed in \cite{Bhattacharjee:2022lzy,Bhattacharjee:2023uwx}, states that the state $\ket{\psi_1(t)}$ can be expanded on a time-independent and orthonormal $\ke{\ket{O_m}|m\in\mathbb N}$, namely
\begin{equation}\label{eq:KrylovWaveFunction}
    \G_0(t)^{-1/2}\ket{\psi_1(t)}=\sum_{m=0} i^m\varphi_m(t) \ket{O_m},
\end{equation}
where the $m$-th basis state $\ket{O_m}$ has size $qm+1$, i.e. $\ket{O_m}\in \H_{qm+1}$ with $\H_{qm+1}$ the size subspace \eqref{eq:sizesubspace}. The statement is nontrivial, because $\H_{qm+1}$ has dimension $\binom{N}{qm+1}$ and $\pi_{qm+1}\ket{\psi_1(t)}$ has to be localized in only one basis state $\ket{O_m}$ in $\H_{qm+1}$ for all $t$, where $\pi_{qm+1}$ is the projection operator of the size subspace $\H_{qm+1}$.

The operator size concentration was proved via constructing the basis from the (bi-)Lanczos algorithm in the large $q$ limit in \cite{Bhattacharjee:2022lzy,Bhattacharjee:2023uwx}. So the basis $\ke{\ket{O_m}}$ is called the Krylov basis and the coefficients $\varphi_m(t)$ is the Krylov wave function. Alternatively, in this work, we will prove the operator size concentration via the generating functions without constructing the Krylov basis from an iterative algorithm.

Because $P_n(t)=0$ for $n\neq qm+1$, we can expand $\ket{\psi_1(t)}=\sum_m \pi_{qm+1}\ket{\psi_1(t)}$. So we expect $\ket{O_m}\propto\pi_{qm+1}\ket{\psi_1(t)}$. As we explained, the nontrivial task is to prove that $\pi_{qm+1}\ket{\psi_1(t)}$ is time-independent up to a normalization coefficient, in other words, to prove that $\pi_{qm+1}\ket{\psi_1(t_1)}$ and $\pi_{qm+1}\ket{\psi_1(t_2)}$ are linearly dependent for any $t_1,t_2$. By using \eqref{eq:overlap}\eqref{eq:largeqDist}\eqref{eq:largeqDDist}, we find that their normalized overlap in the size subspace automatically equals $1$, namely
\begin{equation}\label{eq:projectoverlap}
\begin{split}
    &~\frac{\bra{\psi_1(t_1)}\pi_{qm+1}\ket{\psi_1(t_2)}}{\sqrt{{\bra{\psi_1(t_1)}\pi_{qm+1}\ket{\psi_1(t_1)}}{\bra{\psi_1(t_2)}\pi_{qm+1}\ket{\psi_1(t_2)}}}}\\
    =&~\frac{\G_0(t_1,t_2)P_{qm+1}(t_1,t_2)}{\sqrt{\G_0(t_1)P_{qm+1}(t_1)\G_0(t_2)P_{qm+1}(t_2)}}
    =1.
\end{split}
\end{equation}
So we can simply choose the orthonormal basis state as
\begin{equation}\label{eq:concentrationbasis}
    \ket{O_m}=\frac1{i^m}\frac{\pi_{qm+1}\ket{\psi_1(t)}}{\sqrt{\bra{\psi_1(t)}\pi_{qm+1}\ket{\psi_1(t)}}},
\end{equation}
which is time-independent in the large $q$ limit. Thus, the basis $\ke{\ket{O_m}|m\in\mathbb N}$ with \eqref{eq:concentrationbasis} satisfies all the above conditions of operator size concentration: time-independence, orthonormality, the ability to express $\ket{\psi_1(t)}$, and each state belongs to a different size subspace, which concludes our proof. 

From this point of view, the operator size distribution $P_{qm+1}(t)$ is actually the transition probability from $\ket{\psi_1}$ to the orthonormal basis state $\ket{O_m}$ under the Lindbladian evolution for time $t$. So the Krylov wave function is
\begin{equation}
    \varphi_m(t)=\sqrt{P_{qm+1}(t)}.
\end{equation}
We further find that the wave function satisfies the discrete Schrödinger equation in the bi-Lanczos algorithm \cite{Bhattacharya:2023zqt,Bhattacharjee:2023uwx}
\begin{equation}
    \partial_t \varphi_m(t)= b_m \varphi_{m-1}(t)-\abs{a_m} \varphi_{m}(t) - b_{m+1} \varphi_{m+1}(t)
\end{equation}
to the order of $1/q$ with the Lanczos coefficients
\begin{equation}\label{eq:Lanczos}
    a_m=i\hat\nu m,\quad 
    b_0=0,\quad 
    b_1=\J \sqrt{2/q},\quad
    b_m=\J \sqrt{m(m-1)+\O(1/q)}, \ m\geq 2,
\end{equation}
except for $m=0$, where the overall factor $(1-\theta(t)^2)^{2/q}$ in \eqref{eq:largeqDist} needs a $\O(1/q)$ correction to fulfill the equation for $m=0$ at $1/q$ order.
Thus, $\ke{\ket{O_m}}$ is the Krylov basis in the large-$q$ SYK model in \cite{Bhattacharjee:2022lzy}. Due to the operator size concentration, the operator size $n[\psi_1(t)]$ is related to the Krylov complexity $K(t)$ by
\begin{equation}
    q K(t)+1= \sum_m (qm+1)P_{qm+1}(t)= n[\psi_1(t)]
\end{equation}
in the large $q$ limit.
Also, we note that the operator size $n[\psi_1(t)]$ is proportional to the connected part of the OTOC or the square of anti-commutator $\mathcal F(t,0,t,0)$ normalized by the two-point function \cite{Roberts:2018operator,Qi:2018quantum}. It is derived via the double-time generating function \eqref{eq:largeqDGen} as
\begin{align}
    \frac12\mathcal F(t_1,0,t_2,0)\equiv&~\frac14\sum_j\frac{\Tr[\ke{\psi_j,\psi_1(t_1)}\ke{\psi_j,\psi_1(t_2)}^\dagger]}{\Tr[\psi_1(t_1)\psi_1(t_2)^\dagger]}
    =-\partial_\mu \ln\G_\mu(t_1,t_2)|_{\mu=0} \nn\\
    =&~\frac{\cosh (\gamma ) \cosh \left(\alpha  \left(t_1+t_2\right)+\gamma \right)}{\cosh \left(\alpha  \left(t_1-t_2\right)\right)+\sinh (\gamma ) \sinh \left(\alpha  \left(t_1+t_2\right)+\gamma \right)}.
\end{align}
It can not be simply factorized into a form of $e^{\lambda_L(t_1+t_2)}f(t_1-t_2)$ at finite dissipation even $t_1+t_2\gg1/\alpha$ and $\abs{t_1-t_2}\lesssim1/\alpha$, in contrast to the Hermitian case \cite{Gu:2018jsv,Gu:2021xaj}
\footnote{We thank the authors of \cite{Garcia-Garcia:2024tbd} for helpful discussion.}.

The Krylov complexity and the
normalized square of anti-commutator were calculated in the same Lindbladian SYK model as ours in the first-order perturbation theory with respect to $\hat\nu/\J$ before, see (6.15) in \cite{Bhattacharjee:2022lzy} and (7.21) in \cite{Bhattacharjee:2023uwx}. As expected, their expressions coincide with our operator size in \eqref{eq:largeqsize} at the leading-order in  $\hat\nu/\J$. Moreover, the square of the wave function in the Krylov space in (6.10) in \cite{Bhattacharjee:2022lzy} also coincides with our size distribution \eqref{eq:largeqDist} at the leading-order
\footnote{
We thank the authors of \cite{Bhattacharjee:2023uwx}, especially Pratik Nandy, for helpful discussion and clarification.
}.

\subsection{Variances}

We further investigate the quantum fluctuations of the size operator $n$ and two terms $\P,\X$ in the Liouvillian $\L$ in \eqref{eq:Liouvillian} by calculating their variances.
We define the deviation of operator $O$ from its expectation value as $\Delta O=O-\avg{O}$.
Using the generating function \eqref{eq:largeqGen}, we can calculate the size variance
\begin{equation}\label{eq:SizeVariance}
    \Delta n^2[\psi_1(t)]\equiv\avg{\Delta n^2}=\left.\partial_\mu^2\ln\G_\mu(t)\right |_{\mu=0}
    =\frac{q}{2}\xi(t)^2\csch^2\gamma,
\end{equation}
where
\begin{equation}\label{eq:xi}
    \xi(t)=1-\G_0(t)^{q/2}=1-\frac{\cosh ^2(\gamma )}{1+\sinh (\gamma ) \sinh (2 \alpha  t+\gamma)}
    \approx\begin{cases}
        0 ,& t \ll \gamma/2\alpha \\
        1 ,& t\gg t_p \\
    \end{cases}.
\end{equation}
When $\hat\nu=0$, $\xi(t)$ vanishes. 
The size variance is very big compared to $n[\psi_1(t)]^2$ because of the factor $q$ in \eqref{eq:SizeVariance} due to fluctuations generated by $q$-body interactions. When $\tau\gg1/2\alpha$, the relative variance approaches the long time limit \\$\Delta n^2[\psi_1(\infty)]/n[\psi_1(\infty)]^2=(q/2)\sech^2\gamma$.

Using the double-time generating function \eqref{eq:largeqDGen}, we can calculate the expectation values of some combinations between $\L^\dagger,\L,\P,\X$. The expectation values are given by
\begin{align}
    \avg{\L}=&~\partial_{t_2}\ln \G_0(t_1,t_2)|_{t_1=t_2=t}
    =-\nu\avg{n},\\
    \avg{\L^\dagger}=&~\partial_{t_1}\ln \G_0(t_1,t_2)|_{t_1=t_2=t}=\avg{\L},
    \\
    \avg{\P}=&~i\frac12\avg{\L^\dagger-\L}=0,\\
    \avg{\X}=&~-\nu\avg{n}=\avg{\L}.
\end{align}
The variances of the Liouvilian are
\begin{align}\label{eq:LiouvillianVariance}
    \avg{\Delta\L^\dagger\Delta\L}
    =&~\left.\partial_{t_1}\partial_{t_2}\ln\G_0(t_1,t_2)\right|_{t_1=t_2=t} 
    =\frac2q (1-\xi(t))^2\J^2,\\
    \avg{\Delta\L^2}=&~\left.\partial_{t_2}^2\ln\G_0(t_1,t_2)\right|_{t_1=t_2=t}=\frac2q(\xi(t)^2-1)\J^2,\\
    \avg{(\Delta\L^\dagger)^2}=&~\left.\partial_{t_1}^2\ln\G_0(t_1,t_2)\right|_{t_1=t_2=t}=\frac2q(\xi(t)^2-1)\J^2,
\end{align}
Notice that $\avg{\Delta\L^\dagger\Delta\L}>0$ and $\avg{(\Delta\L^\dagger)^2}=\avg{\Delta\L^2}<0$ always. 
When $\hat\nu=0$, $\avg{\Delta\L^\dagger\Delta\L},\avg{\Delta\L^2}$ and $\avg{(\Delta\L^\dagger)^2}$ reduce to $(2/q)\J^2,-(2/q)\J^2$ and $-(2/q)\J^2$ respectively. When $\hat\nu/\J>0$, they decay exponentially after the plateau time $t_p$. The relative variance of Liouvillian is huge $\avg{\Delta\L^2}/\avg{\L}^2\sim \O(q^1)$. The moment of Liouvillian can be constructed as 
$\avg{\L^2}=\avg{\Delta \L^2}+\avg{\L}^2=\avg{(\L^\dagger)^2},\, 
\avg{\L^\dagger \L}=\avg{\Delta \L^\dagger\Delta\L}+\avg{\L^\dagger}\avg{\L}$.

Finally, we list some useful results for later convenience. 
\begin{align}
    &\avg{\Delta \P^2}=\avg{\P^2}=-\frac12\avg{(\L^\dagger)^2+\L^2-2\X^2}=\frac2q\J^2 \label{eq:P2},\\
    &\avg{\ke{\Delta \X,i\Delta \P}}=\avg{\ke{\X,i\P}}=\frac12\avg{\L^2-(\L^\dagger)^2}=0,\\
    &\avg{\kd{\Delta \X,i\Delta \P}}=\avg{\kd{\X,i\P}}=\avg{\L^\dagger \L-\P^2-\X^2}
    =-\frac{4}{q}\xi(t) \J^2. \label{eq:CXiP}
\end{align}
It is surprising that $\avg{\Delta \P^2}$ is independent of dissipation. Actually, we recognize $\sqrt{\avg{\Delta \P^2}}$ as the $b_1$ Lanczos coefficient in \eqref{eq:Lanczos}. The balance between these quantum fluctuations is crucial for understanding the emergence of classical dynamics in the next section.

\section{Emergence of classical size growth}\label{sec:Epidemic}

In \cite{Qi:2018quantum,Schuster:2022bot}, the authors constructed classical equations to describe the dynamics of the operator size growth phenomenologically. Since we have nearly all the information about the operator size growth in this Lindbladian SYK at large $q$, we can show that the emergence of classical dynamics of operator size growth is due to the saturation of the uncertainty relation of size growth in quantum mechanics. 

Following \cite{Schuster:2022bot}, the growth rate of $\avg{n}$ can be written as
\begin{equation}\label{eq:unitarydissipationterms}
    \partial_t \avg{n}
    =\avg{\L^\dagger n+n \L}-\avg{n}\avg{\L^\dagger+\L}
    =i\avg{[n,\P]}-2\nu\avg{\Delta n^2}.
\end{equation}
The first term is the unitary term, and the second term is the dissipation term. Since both $n$ and $\P$ are Hermitian, they have the uncertainty relation
\begin{equation}
    \abs{\avg{\kd{n,\P}}}
    =\abs{\avg{\kd{\Delta n,\Delta \P}}}\leq
    2\sqrt{\avg{\Delta n^2}\avg{\Delta \P^2}}.
\end{equation}
Similar uncertainty relations are applied in \cite{Hornedal:2022pkc,Bhattacharya:2024uxx}. 
Since the size \eqref{eq:largeqsize} never decreases, the unitary term in \eqref{eq:unitarydissipationterms} should be non-negative. This yields a limit on the growth rate
\begin{equation}\label{eq:uncertainty}
    \partial_t \avg{n}\leq 2\sqrt{\avg{\Delta n^2}\avg{\Delta \P^2}}-2\nu\avg{\Delta n^2}.
\end{equation}
The uncertainty relation for the size operator and the Liouvillian is discussed in App.~\ref{sec:AlterUncertainty}.

The inequality \eqref{eq:uncertainty} holds generically for the Liouvillian in the form of \eqref{eq:Liouvillian}, independent of the microscopic details of the Hamiltonian $H$. However, thanks to the large $q,N$ limit of the Lindbladian SYK, we can check this inequality by comparing \eqref{eq:SizeVariance}, \eqref{eq:P2} and \eqref{eq:CXiP}. We find that the uncertainty relation is actually saturated, namely
\begin{equation}
    \avg{[\X,i\P]}=\avg{[\Delta\X,i\Delta\P]} =-2\sqrt{\avg{\Delta \X^2}\avg{\Delta \P^2}},
\end{equation}
where $\X=-\nu n$.
The saturation of the Cauchy-Schwarz inequality means that states $\Delta\X\ket{\psi_1(t)}$ and $\Delta\P\ket{\psi_1(t)}$ are linearly dependent. By calculating $\avg{\Delta\X\Delta\P}$ from the generating functions, we find that the linear coefficient is $\xi(t)$ in \eqref{eq:xi}, namely
\begin{equation}
    \Delta\X\ket{\psi_1(t)}=-\xi(t) i\Delta\P\ket{\psi_1(t)}
    \quad \text{or}    \quad 
    (\X+i\xi(t)\P)\ket{\psi_1(t)}=\avg{\X}\ket{\psi_1(t)}.
\end{equation}
So $\ket{\psi_1(t)}$ behaves like a coherent state of the time-dependent ``annihilation operator'' $(\xi(t)^{-1}\X+i\P)$, which consists of the ``position'' operator $\xi(t)^{-1}\X$ and the ``momentum'' operator $\P$. The two operators follow the commutation relation in the bracket  $\avg{[\xi(t)^{-1}\X,\P]}=i(4/q)\J^2$, and have the time-independent variances $\avg{\xi(t)^{-2}\Delta\X^2}=\avg{\Delta\P^2}=(2/q)\J^2$. 
From the asymptotic behavior of $\xi(t)$ in \eqref{eq:xi}, the operator $(\xi(t)^{-1}\X+i\P)$ interpolates between the size operator $-\nu n\xi(t)^{-1}$ at early times and the Liouvillian $\L$ at late times.

Coming back to the growth rate of the operator size, we find
\begin{align}
    \partial_t \avg{n}=&~ 2\J\sqrt{2/q}\sqrt{\avg{\Delta n^2}}-2\nu\avg{\Delta n^2} \label{eq:SizeQM}\\
    =&~2\J r \avg{n}\kc{1 - \frac{\hat\nu r}{2\J} \avg{n}} \label{eq:RIS}
\end{align}
where 
\begin{equation}\begin{split}
    r=&~\sqrt{\frac{2\avg{\Delta n^2}}{q \avg{n}^2}}
    =\tanh (\alpha  t) \kd{\sech(\gamma )+\sech(2 \alpha  t+\gamma)} \\ \approx&~
    \begin{cases}
        2\J t, & t\ll 1/\J\\
        1, & t\gg1/\J
    \end{cases},\quad \text{ when } \hat\nu/\J\ll1.
    \end{split}
\end{equation}
The first line \eqref{eq:SizeQM} indicates that the size's growth rate is controlled by its variance in quantum mechanics. Specifically, the larger the size variance, the more the operator will fail to commute with the Hamiltonian, but the dissipation also becomes larger in the meantime. In total, the growth rate of the size \eqref{eq:SizeQM} is determined by this interplay. If we recognize the factor $\J \sqrt{2/q}$ as the $b_1$ Lanczos coefficient in \eqref{eq:Lanczos}, then the first term in \eqref{eq:SizeQM} could be identified as the saturated growth rate of Krylov complexity in \cite{Hornedal:2022pkc,Bhattacharya:2024uxx}.

In the second line \eqref{eq:RIS}, we derive the differential equation of operator size in \cite{Qi:2018quantum,Schuster:2022bot}. Following \cite{Qi:2018quantum}, we could interpret \eqref{eq:RIS} as the Susceptible-Infectious (SI) epidemic model with a stock of infected individuals $\avg{n}$, a contact rate $2\J r$, the total population $2\J/\hat\nu r$, and no recovery. The total population $2\J/\hat\nu r$ in the SI model corresponds to an effective number of qubits in the Lindbladian SYK, controlling the steady state maximum operator size in \eqref{eq:RIS}, and taking the place of $N/2$ in the pure SYK case. At weak dissipation $\hat\nu/\J\ll1$, the ratio $r$ converges to $1$ much after $1/\J$, which is still much early than the plateau time $t_p\sim \frac1\J \ln \frac{\J}{\hat \nu}$. So we can take the approximation $r\approx 1$ in the differential equation \eqref{eq:RIS}, solve it with the initial condition $n(0)=1$, and find the solution valid in the $t\gg 1/\J$ region 
\begin{equation}
    n(t)\approx \frac{e^{2 \J t}}{1+\frac{\hat\nu}{2 \J}\left(e^{2 \J t}-1\right)}
    \approx \begin{cases}
        e^{2\J t}   ,& \frac1\J\ll t\ll \frac1{\J} \ln \frac{\J}{\hat\nu}  \\
        2\J/\hat\nu, & \frac1{\J} \ln \frac{\J}{\hat\nu} \ll t \\
    \end{cases}
\end{equation}
It is close to the exact expression \eqref{eq:largeqsize} at weak dissipation, including the time scales  \eqref{eq:sizebehavior}.

\section{Conclusion and outlook}\label{sec:conclusion}

\subsection{Conclusion}

We comprehensively studied the operator size growth of a single Majorana fermion in the Lindbladian SYK model with a linear jump operator for both finite dissipation and finite production. The operator size and distribution can be derived from the two-point function with the insertion of $e^{-\nu n}$ in the path integral. The symmetries of the two-point function and the properties of the maximally entangled state greatly simplify the problem.

First, we used exact diagonalization to solve the model at finite $N$ and numerically solved the Schwinger-Dyson equation at infinite $N$. We observed the slowdown (acceleration) of the operator size growth and the suppression (enhancement) of the size plateau due to the dissipation (production) introduced by the Lindblad terms with $\nu>0$ ($\nu<0$). Second, we analytically solved the Liouville equations in the large $q$ limit and obtained the expression for the Loschmidt echo fidelity, operator size, size distribution, and the variances of size and the Liouvillian. The plateau time is extracted from the analytical results given. The operator size exhibits a quadratic-exponential-plateau behavior at weak dissipation and the size distribution is localized at finite size. Third, we construct the time-independent orthogonal basis exhibiting operator size concentration. Fourth, we derive a growth rate limit on the operator size from an uncertainty relation and find that it is saturated at large $q$, which gives rise to classical dynamics of operator growth with dissipation. Five, we study the size of operators at finite temperature, including $n[e^{t\L_H}[e^{-\beta H/2}]]$ and $n[e^{t\L_H}[\psi_1e^{-\beta H/2}]]$, at finite $N$ in App.~\ref{sec:Temperature}.

We formulated a self-consistent path integral for studying operator size in Lindblad-ian dynamics. Our approach does not depend on the weak dissipation limit and can be readily extended to other fermionic models and Lindblad operators. We analytically derived the operator size and distribution of a single Majorana fermion, a feature that was previously explored only at the leading-order perturbation in the dissipation strength in \cite{Schuster:2022bot,Bhattacharjee:2022lzy,Bhattacharjee:2023uwx}. Additionally, we elucidated the reasons behind the emergence of the classical size growth in quantum mechanics.

\subsection{Outlook}

In this paper, we investigate a specific Lindbladian SYK model with single jump operators across different parameter regions, rather than deriving it from a microscopic model of an SYK system coupled with a bath. The operator growth in a SYK-plus-bath model is also an important question, and it can be compared with our results from the Lindbladian SYK model.

In Sec.~\ref{sec:concentration}, we constructed the orthogonal basis and proved the operator size concentr-ation via the path integral formalism rather than apply the (bi-)Lanczos algorithm \cite{Bhattacharjee:2022lzy,Bhattacharjee:2023uwx}. The square root of the size distribution coincides with the wave function in the Krylov basis and satisfies the same discrete Schrödinger equation. This implies that the Liouville equation \eqref{eq:Liouville} with the boundary conditions \eqref{eq:LiouvBC} is equivalent to the  (bi-)Lanczos algorithm even at finite dissipation in the large $q$ limit. The detailed connection and realization of their equivalence deserves further investigation. For example, how to derive the Lanczos coefficients from the Liouville equation with boundary conditions? How to understand the saturated uncertainty relation from the Krylov space? We leave these important questions for future work.

Furthermore, we could study the sizes $n[e^{t\L_H}[e^{-\beta H/2}]]$ and $n[e^{t\L_H}[\psi_1e^{-\beta H/2}]]$ in the large $N$ limit, by writing down the path integral for the partition function at finite temperature $\bra0 e^{-\beta H^L/2} e^{t\L^\dagger}e^{-\mu n}e^{t\L}e^{-\beta H^L/2}\ket0$ and solving the SD equation at finite $q$ or the Liouville equation at large $q$. Now the whole time window $[0,\beta+2t]$ is divided into $4$ analytical domains $(0,\beta/2),\,(\beta/2,\beta/2+t),\,(\beta/2+t,\beta/2+2t),\,(\beta/2+2t,\beta+2t)$. In this case, the simplification \eqref{eq:GlobalG} does not work anymore. At large $q$, one has to solve the Liouville equations for $4$ components in $4\times4\times2=32$ double time $(\tau_1,\tau_2)$ domains separately and glue them together with some boundary conditions at $\tau_{1,2}=0,\beta/2,\beta/2+t,\beta/2+2t,\beta+2t$ and $\tau_1=\tau_2$. By using symmetries, one may reduce the number of independent components and domains. But the task is still complicated and will be left for future work.

We only considered the Lindblad operator with the linear jump operator, which introduces a size gap in the spectrum of the Lindbladian. One may consider other kinds of Lindblad operators such as the random $p$-body jump operators $L_a=\sum_{j_1\cdots j_p} K^a_{j_1\cdots j_p}\psi_{j_1}\cdots\psi_{j_p}$ \cite{Sa:2021tdr,Kulkarni:2021gtt}. We expect a spectrum in a lemon-like shape as predicted in \cite{Denisov:2018nif}. These Lindblad operators are transformed into product terms of the left and right Lindblad operators in the Liouvillian, which play the role of the non-local and non-Hermitian double-trace deformation on the SYK model in the double-copy Hilbert space \cite{Jian:2017tzg}. We expect to solve the resulting deformed Liouville equations at large $q$ and $p$.

It is also interesting to compare the regular SYK dynamics and the Brownian SYK dynamics in the presence of the same Lindblad operator. A significant effect of the time-dependent disorder interaction in the Brownian SYK is the breaking of energy conservation. However, such an effect seems to be not so important in the Lindbladian, since the Lindblad dynamics already break the energy conservation even in the regular SYK model. Also, the time-dependent disorder in the Brownian SYK Hamiltonian will usually benefit the solvability of the operator size \cite{Zhang:2023BrownianSize}.

Finally, the operator size in the SYK model is important in the SL(2,R) generator and the microscopic description of the volume of the AdS$_2$ space \cite{Lin:2019symmetries,Lensky:2020size,Lin:2022rbf}. The Lindbladian SYK model was proposed to be dual to a Keldysh wormhole \cite{Garcia-Garcia:2022adg}. The operator size calculated in this paper probes the correlation in the dual spacetime. Here we alternatively suggest the investigation of the gravity duality of the dissipating thermofield-double state $e^{t\L}\ket{e^{-\beta H/2}}$ in the $N\gg q^2\gg \J\beta\gg \nu\beta\gg1$ and $\beta\gg t$ limit. We expect the gravity duality to be the AdS$_2$ wormhole deformed by double-trace terms between the two boundaries \cite{Maldacena:2001eternal,Jian:2017tzg,Maldacena:2018lmt,Xian:2019qmt,He:2021dhr}, but with the imaginary sources \cite{Arean:2019pom,Xian:2023zgu,Chen:2023hra}. From our numerical result, the decrease of size $n[e^{t\L_H}[e^{-\beta H/2}]]$ implies the decrease of the spacetime distance between the two boundary trajectories traveling along the boost time. This will become more clear if the same Liouville equations \eqref{eq:LiouvilleDelta} could be derived from the gravity side, similar to \cite{Maldacena:2018lmt,Milekhin:2022bzx}.

\section*{Acknowledgments}
We are grateful to Budhaditya Bhattacharjee, Yu Chen, Antonio M. García-García, Hyun-Sik Jeong, Shao-Kai Jian, Changan Li, Pratik Nandy, Cheng Peng, Tanay Pathak, Jacobus J. M. Verbaarschot, Zhenbin Yang, Pengfei Zhang, and Jie-ping Zheng for helpful discussions. Ren\'e Meyer and ZYX acknowledge funding by DFG through the Collaborative Research Center SFB 1170 ToCoTronics, Project-ID 258499086-SFB 1170, as well as by Germany's Excellence Strategy through the W\"urzburg‐Dresden Cluster of Excellence on Complexity and Topology in Quantum Matter ‐ ct.qmat (EXC 2147, project‐id 390858490). ZYX also acknowledges support from the National Natural Science Foundation of China under Grant No.~12075298.

\appendix

\section{Keldysh contour}\label{sec:Keldysh}

Alternatively, we can write the partition function \eqref{eq:PatitionFunction} in the path integral along the Keldysh contour with $s\in[0,2t_1+2t_2)$ \cite{Kulkarni:2021gtt,Garcia-Garcia:2022adg}, as shown in Fig.~\ref{fig:Contour}. Here we take $t_1=t_2=t$ for conciseness and introduce a real Grassmann variables $\chi_j(s)$ along the contour to unify $\psi_j^L(\tau)$ and $\psi_j^R(\tau)$
\begin{equation}
    \chi_j(s)=\begin{cases}
        \psi_j^L(s-t),& 0\leq s<2t\\
        i\psi_j^R(3t-s),& 2t\leq s<4t
    \end{cases}.
\end{equation}
Now the boundary condition \eqref{eq:BCpsi} becomes the ordinary continuious condition $\chi_j(2t^-)=\chi_j(2t^+)$ and anti-periodic condition $\chi_j(0^+)=-\chi_j(4t^-)$ in the Keldysh contour.
Replacing $\psi_j^a(\tau)$ with $\chi_j(s)$ in the action \eqref{eq:Actionpsi}, we obtain
\begin{equation}\label{eq:Actionchi}
\begin{split}
    -S=&~\int_{0}^{4t}\dd s \sum_j \Big[
    -\frac14\chi_j(s)\partial_s\chi_j(s)  
    -\frac12\kc{\nu+\mu \delta(s-t)} (\chi_j(s)\chi_j(4t-s)+1) \Big]     \\  
    &~ -\frac{i^q J^2}{2qN^{q-1}}\int_0^{2t}\dd s_1\dd s_2  \sum_{j_1\cdots j_N} \tsq\kc{\frac{s_1}{2t}}\tsq\kc{\frac{s_2}{2t}} \chi_{j_1}(s_1)\cdots\chi_{j_q}(s_1)\chi_{j_1}(s_2)\cdots\chi_{j_q}(s_2) ,
\end{split}
\end{equation}
where $\tsq(x)$ is the square wave function of period $1$. 
Similarly, we introduce the bi-local field
\begin{equation}
    F(s_1,s_2)=\frac1N\sum_j \chi_j(s_1)\chi_j(s_2)
\end{equation}
and $S(s_1,s_2)$ via the Lagrange multiplier method similar to \eqref{eq:multiplier}. They unify the two $4$-component fields $G_{ab}$ and $\Sigma_{ab}$ on the time domains $[-t,t]$ into two single-component fields $F$ and $S$ on the time domain $[0,4t]$ via
\begin{equation}
\begin{aligned}
&F(s_1,s_2)=\begin{cases}
    G_{LL}(s_1-t,s_2-t), & 0<s_1<2t,\ 0<s_2<2t\\
    iG_{RL}(3t-s_1,s_2-t), & 2t<s_1<4t,\ 0<s_2<2t\\
    iG_{LR}(s_1-t,3t-s_2), & 0<s_1<2t,\ 2t<s_2<4t\\
    -G_{RR}(3t-s_1,3t-s_2), & 2t<s_1<4t,\ 2t<s_2<4t\\
\end{cases}, \\
&S(s_1,s_2)=\begin{cases}
    \Sigma_{LL}(s_1-t,s_2-t), & 0<s_1<2t,\ 0<s_2<2t\\
    -i\Sigma_{RL}(3t-s_1,s_2-t), & 2t<s_1<4t,\ 0<s_2<2t\\
    -i\Sigma_{LR}(s_1-t,3t-s_2), & 0<s_1<2t,\ 2t<s_2<4t\\
    -\Sigma_{RR}(3t-s_1,3t-s_2), & 2t<s_1<4t,\ 2t<s_2<4t\\
\end{cases}. 
\end{aligned}
\end{equation}
Then the boundary condition \eqref{eq:BCG} for $G_{ab}$ becomes the continuous condition and anti-periodic condition for $F$
\begin{equation}
\begin{split}
    &F(2t^-,s_2)=F(2t^+,s_2),\quad 
    F(s_1,2t^-)=F(s_1,2t^+),\\
    &F(0^+,s_2)=-F(4t^-,s_2),\quad
    F(s_1,0^+)=-F(s_1,4t^-).
\end{split}
\end{equation}
We further integrate out $\chi_j(s)$ and obtain the effective action
\begin{align}
    -S/N=&\frac12\log\det\kc{\partial-2S}  \label{eq:effactionFS}\\
  &~-\frac{1}{2}\int_{0}^{4t}\dd s_1\dd s_2 \kd{S(s_1,s_2)F(s_1,s_2)
  +\tsq\kc{\frac{s_1}{2t}}\tsq\kc{\frac{s_2}{2t}}\frac{J^2}{q}F(s_1,s_2)^q} \nn\\
		&-\frac{1}{4}\int_{0}^{4t}\dd s\kc{2\nu+\mu\delta(s-t)}\kd{\sgn(2t-s)\kc{F(s,4t-s)-F(4t-s,s)}+2}, \nn
\end{align}
So the $4$-component SD equation are unified into a single-component SD equation
\begin{subequations}\label{eq:SDFS}
    \begin{align}
        \delta(s_1-s_2)=&~\frac12\partial_{s_1}F(s_1,s_2)-\int_0^{4t}\dd s_3 S(s_1,s_3)F(s_3,s_2),\\
        S(s_1,s_2)=&~-\tsq\kc{\frac{s_1}{2t}}\tsq\kc{\frac{s_2}{2t}}J^2F(s_1,s_2)^{q-1} \nn \\
        &~+\sgn(s_1-s_2)\kc{2\nu+\mu\delta(\abs{s_1-s_2}-2t)}\delta(s_1+s_2-4t)
    \end{align}
\end{subequations}
The unified SD equation is real, so we expect a real solution. One can solve this SD equation either numerically or analytically, following the same approach as outlined in the main text. The generating function \eqref{eq:Generatring} under the disorder average is equal to the specific two-point function
\begin{equation}
    \G_\mu(t)=Z_\mu(t)F(2t,0).
\end{equation}

\section{Numerical result at finite temperature}\label{sec:Temperature}

\begin{figure}
    \centering
     \includegraphics[height=0.25\textwidth]{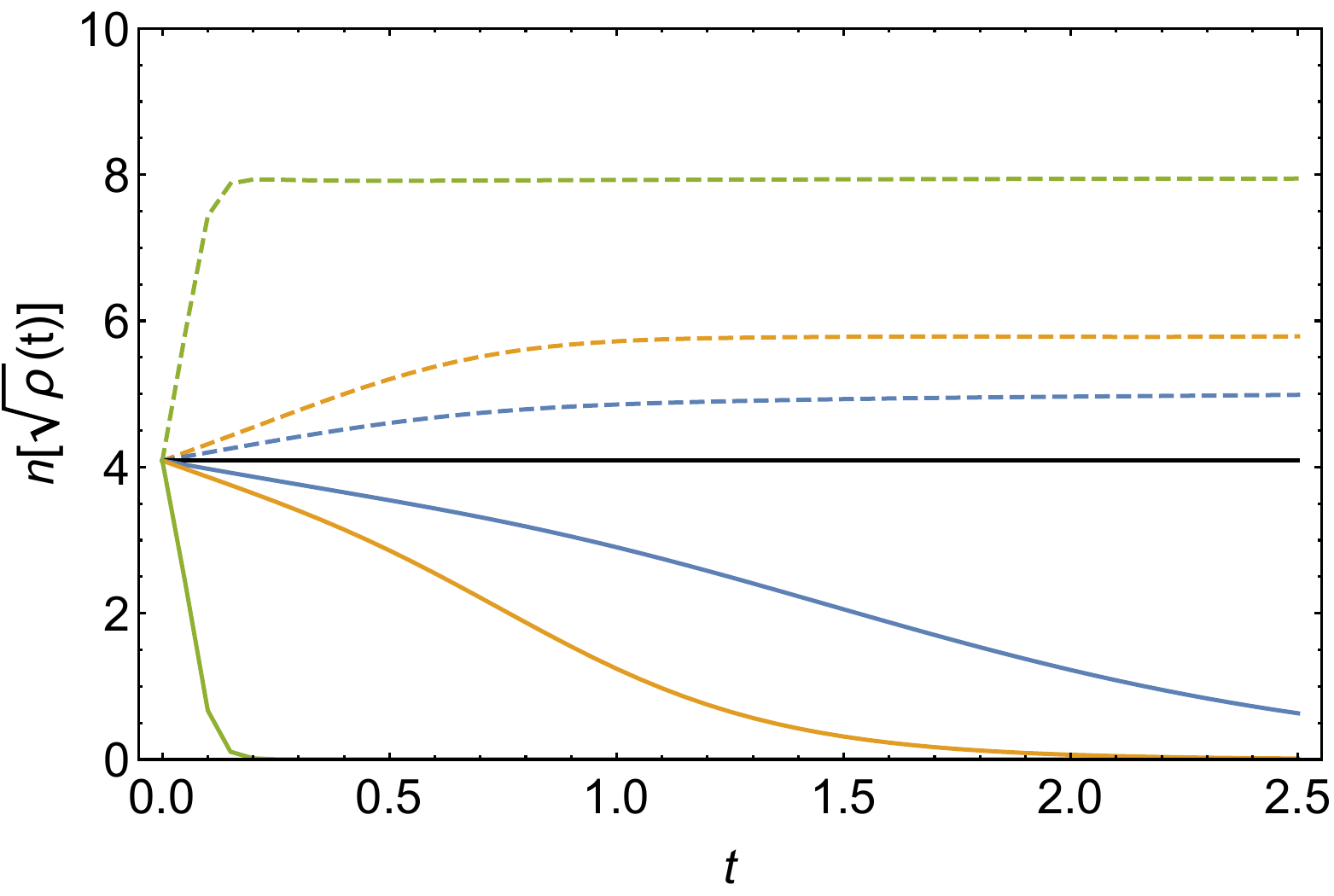}
     \includegraphics[height=0.25\textwidth]{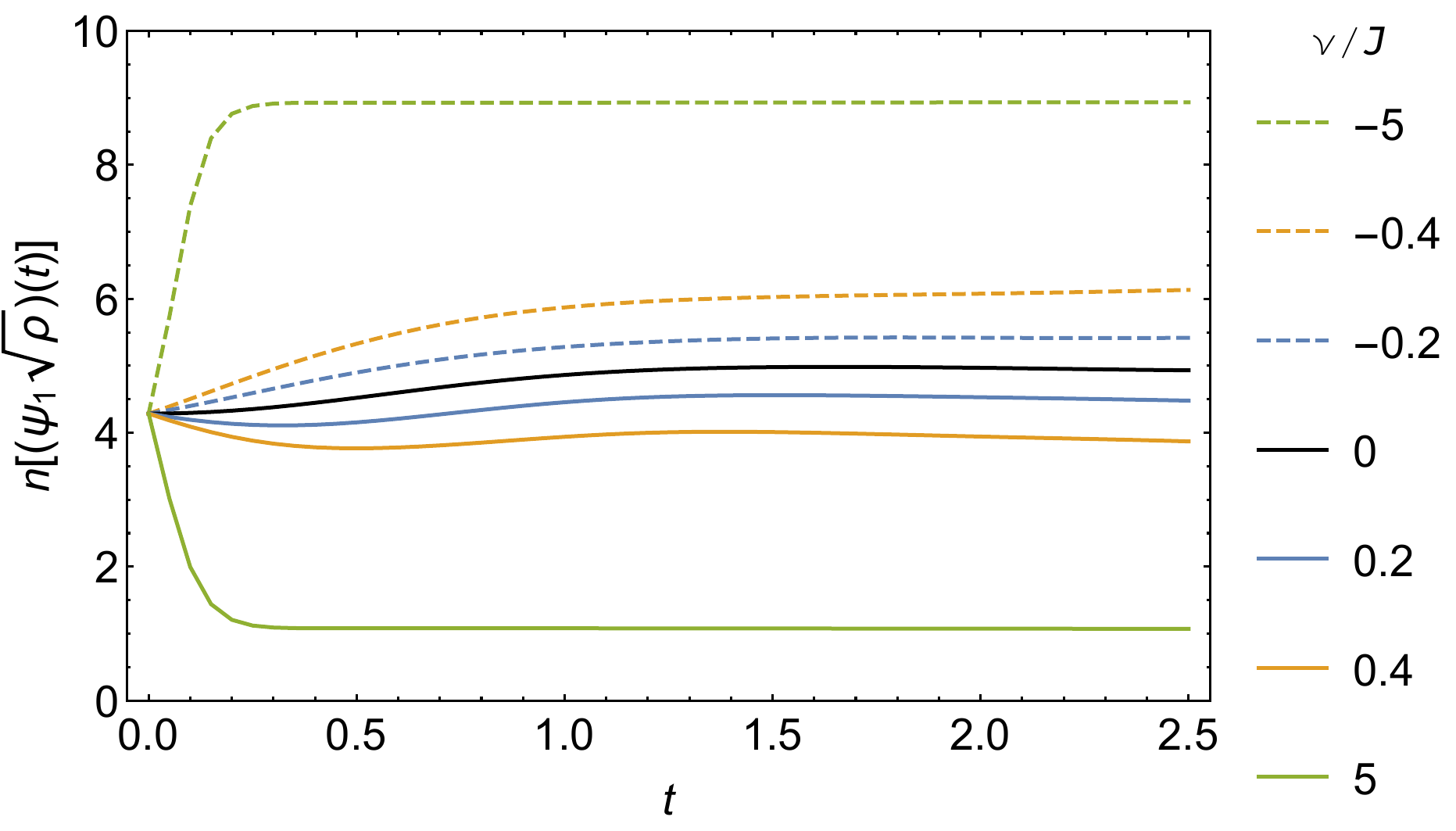}
    \caption{Operator sizes $n[\sqrt{\rho}(t)]$ and $n[(\psi_1\sqrt{\rho})(t)]$ as functions of time. The parameters in the current and subsequent figures are all $q=4$, $N=10$, and $\beta=10$.}
    \label{fig:opsizetherma}
    \centering
      \includegraphics[height=0.25\textwidth]{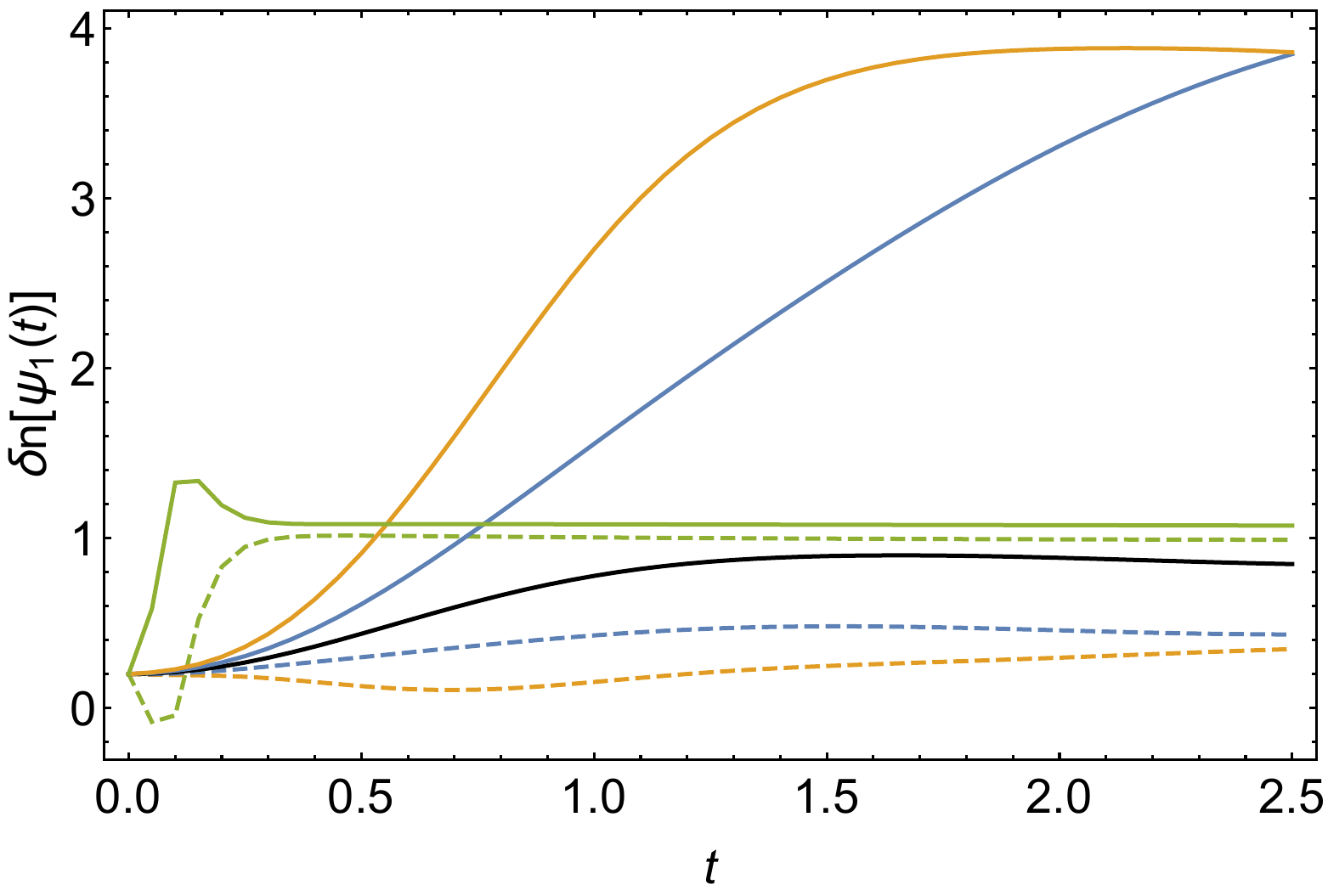}
      \includegraphics[height=0.25\textwidth]{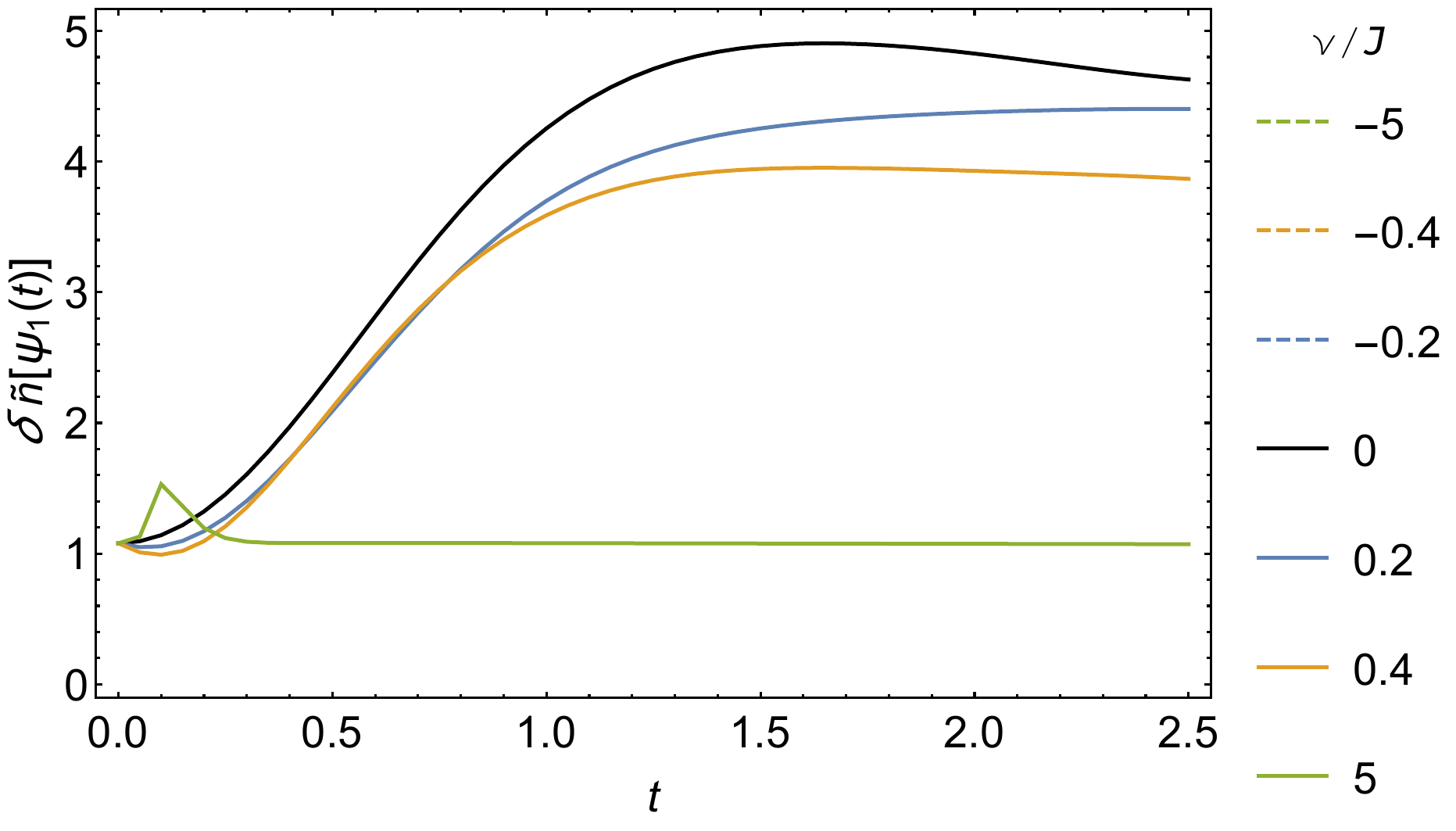}
    \caption{The size difference $\delta n[\psi_1(t)]$ and normalized size difference $\delta \tilde n[\psi_1(t)]$ with $\nu>0$ at finite temperature.}
    \label{fig:opsizethermb}
\end{figure}

In this appendix, we study the thermal operator growth in Lindbladian SYK model. We consider the initial operators $\sqrt{\rho}$ and $\psi_1\sqrt{\rho}$ with $\rho=e^{-\beta H}$. Following the same numerical method in Subsec.~\ref{sec:ED}, we construct their time evolution $\sqrt{\rho}(t)=e^{t\L_H}[\sqrt{\rho}]$ and $(\psi_1\sqrt{\rho})(t)=e^{t\L_H}[\psi_1\sqrt{\rho}]$ under the Lindbladian \eqref{eq:LindbOperator} and calculate their sizes and distributions. The size growth $n[\sqrt{\rho}(t)]$ and $n[(\psi_1\sqrt{\rho})(t)]$ are shown in Fig.~\ref{fig:opsizetherma}. 

We further define the size difference and the normalized size difference
\begin{align}
    &\delta n[\psi_1(t)]\equiv n[(\psi_1\sqrt{\rho})(t)]-n[\sqrt{\rho}(t)],\\
    &\delta \tilde n[\psi_1(t)]\equiv \frac{\delta n[\psi_1(t)]}{1-\frac2N n[\sqrt{\rho}(t)]}\label{eq:normthermalsize}
\end{align}
with a normalization factor generalized from \cite{Qi:2018quantum}, which could be interpreted as the effective operator size of a thermal Majorana fermion under the Lindbladian evolution. We plot both size differences in Fig.~\ref{fig:opsizethermb}.

When $\nu>0$, $n[\sqrt{\rho}(t)]$ decreases to zero at the late times.  $n[(\psi_1\sqrt{\rho})(t)]$ decreases first and then grows to a plateau lower than $N/2$ at late times. This behavior could be understood as the competition between the dissipation on thermal state $\rho$ and the scrambling of fermion $\psi_1$. Such competition is visualized in the size difference $\delta n[\psi_1(t)]$, which grows and reaches a plateau. After the normalization, the scale of the thermal fermion size $\delta \tilde n[\psi_1(t)]$ in the right panel of Fig.~\ref{fig:opsizethermb} is similar to the scale of the size of single fermion $n[\psi_1(t)]$ in Fig.~\ref{fig:EDSize}, which aligns with the pure SYK result in \cite{Qi:2018quantum}. 

When $\nu<0$, both $n[\sqrt{\rho}(t)]$ and $n[(\psi_1\sqrt{\rho})(t)]$ grow to a finite plateau greater than $N/2$. The size difference $\delta n[\psi_1(t)]$ fluctuates at early times and becomes stable at late times. The normalization in $\delta \tilde n[\psi_1(t)]$ fails since $n[\sqrt{\rho}(t)]$ can surpass $N/2$, resulting in a singularity in \eqref{eq:normthermalsize}.

\begin{figure}
	\centering 
		\includegraphics[height=0.29\textwidth]{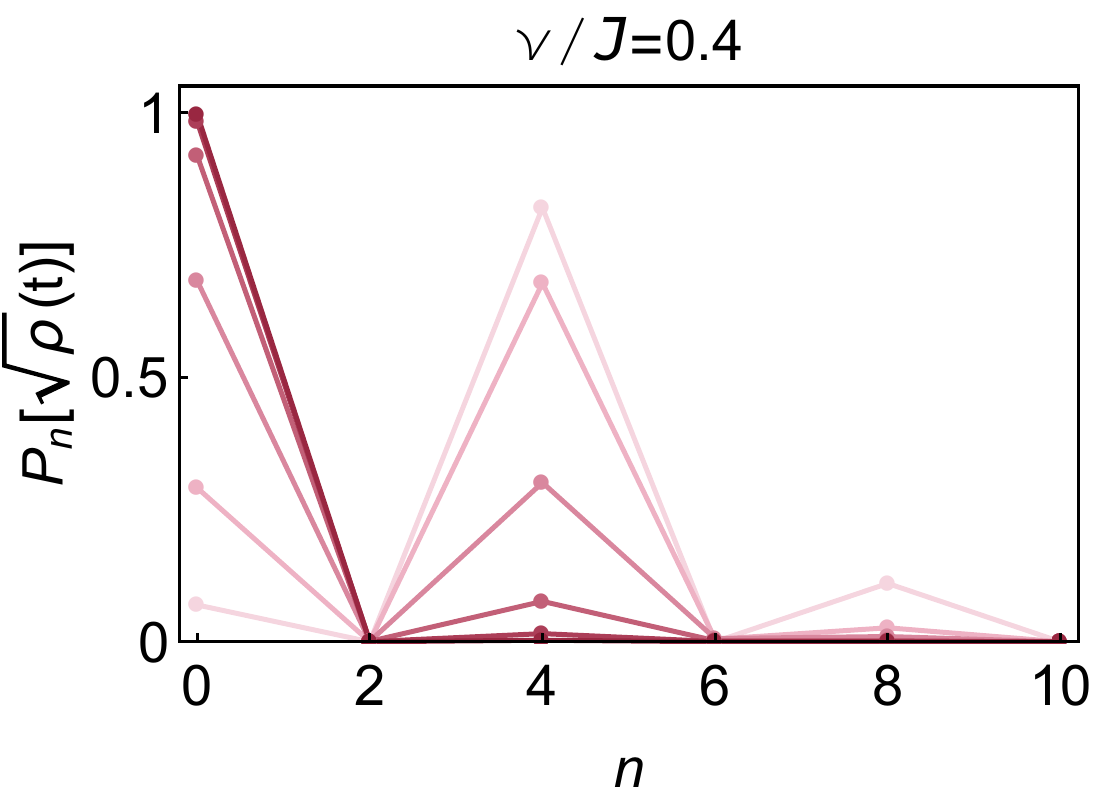}
		\includegraphics[height=0.29\textwidth]{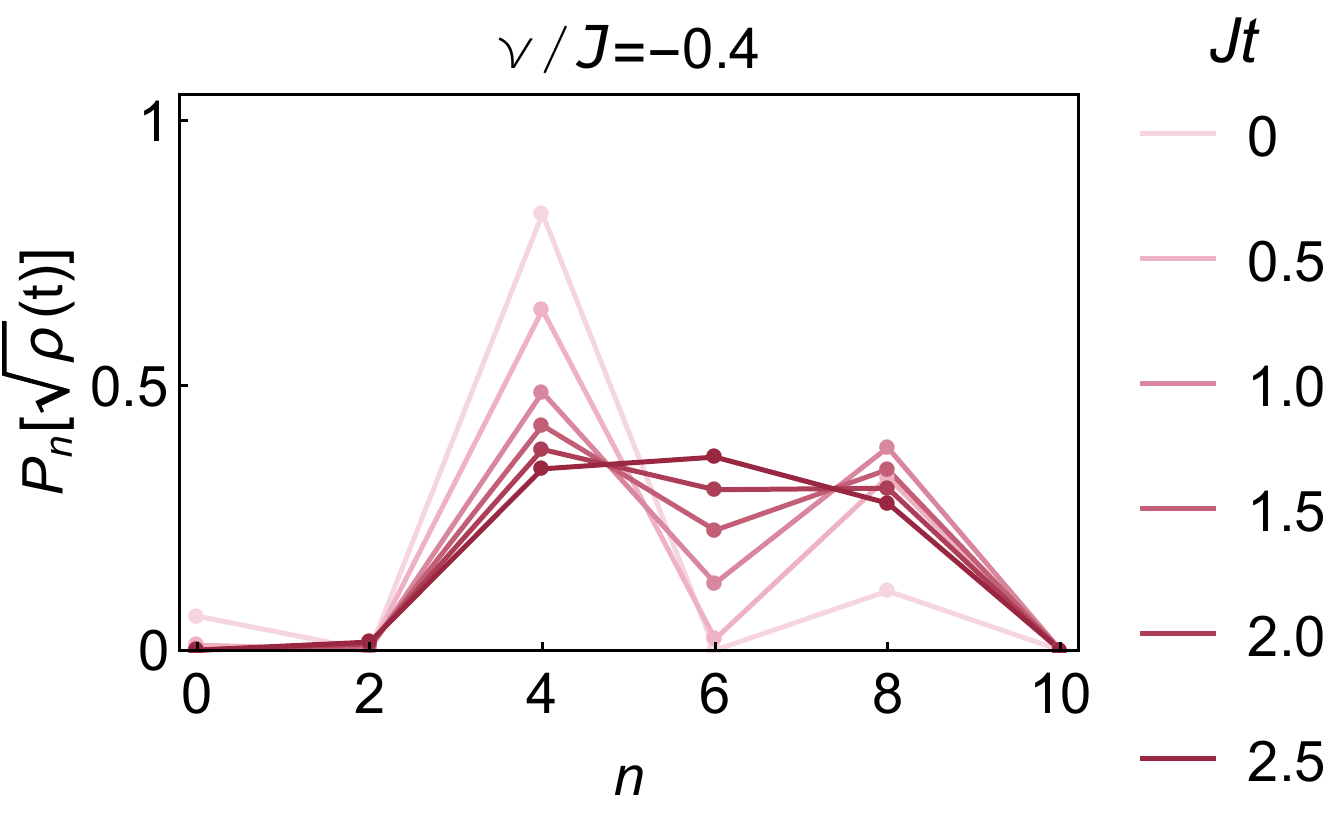}
	\caption{Some snapshots of size distribution $P_n[\sqrt{\rho}(t)]$.}
	\label{fig:vactherm}
	\centering 
		\includegraphics[height=0.29\textwidth]{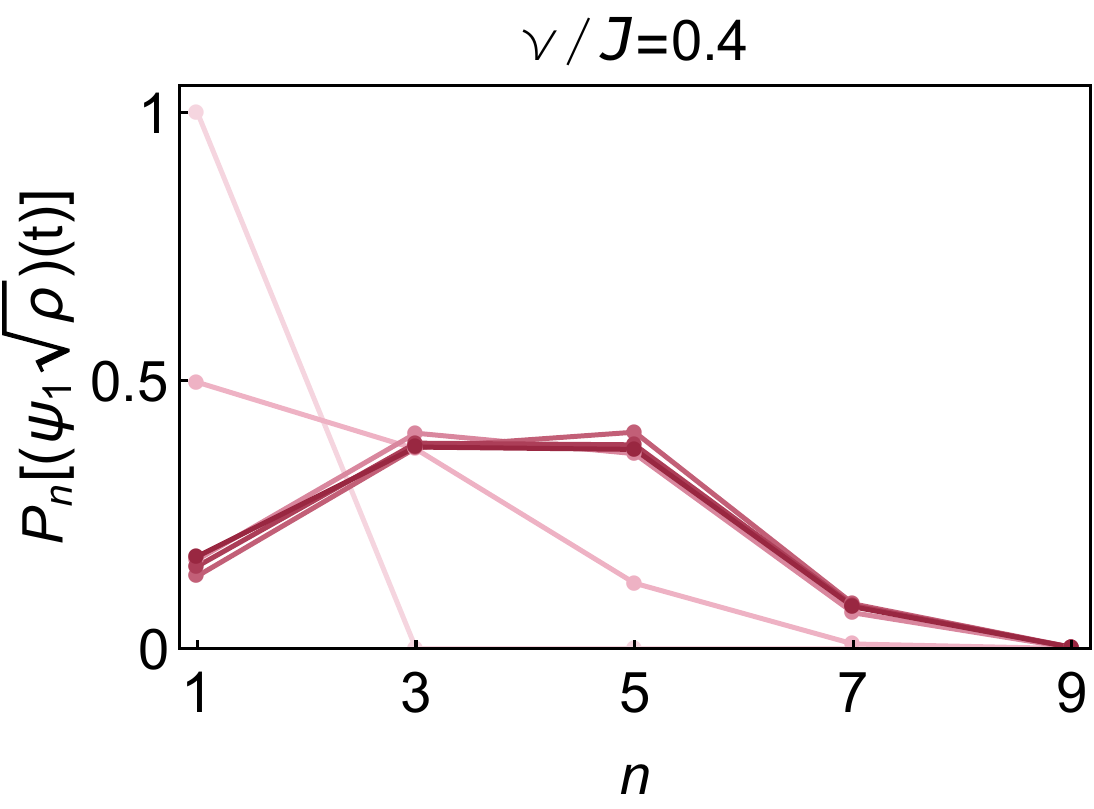}
		\includegraphics[height=0.29\textwidth]{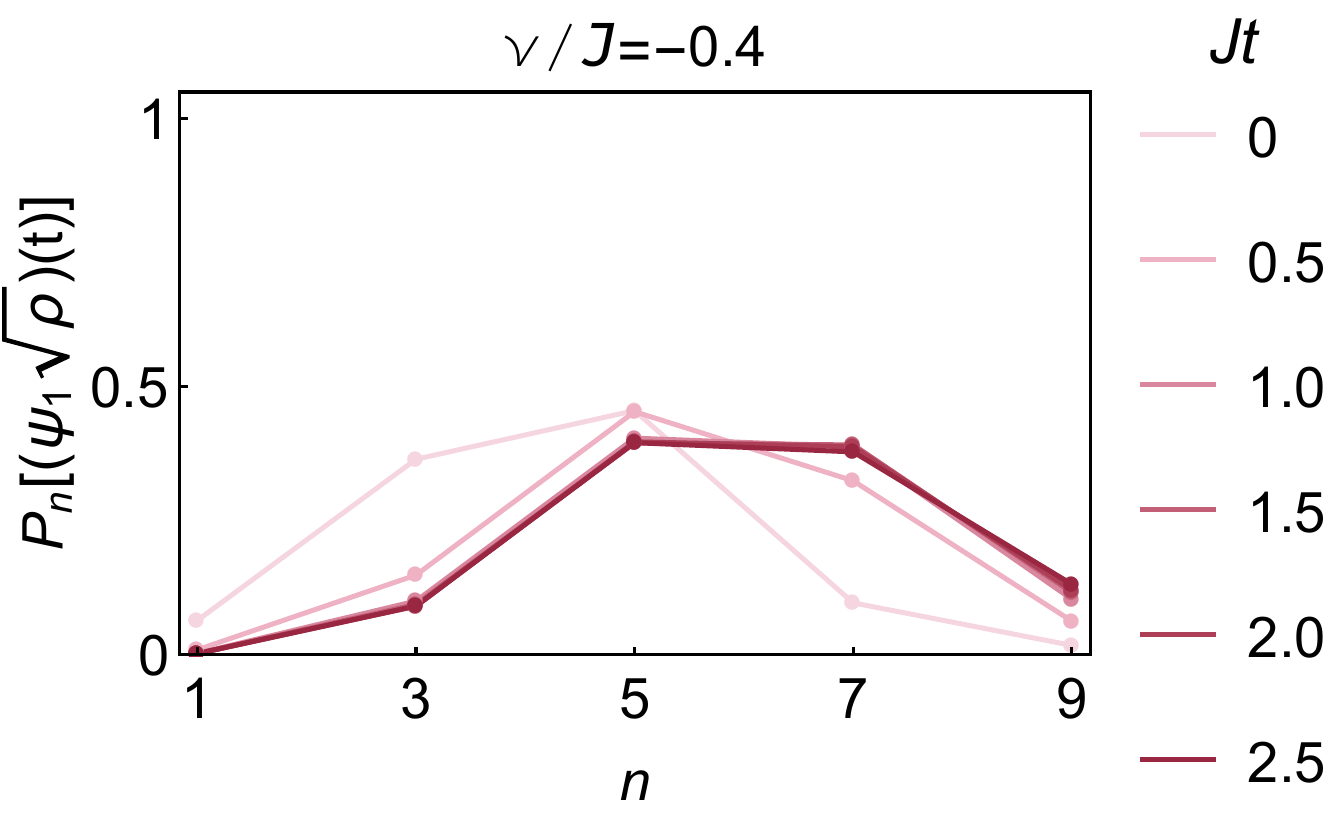}
	\caption{Some snapshots of size distribution $P_n[(\psi_1\sqrt{\rho})(t)]$.}
	\label{fig:fermitherm}
\end{figure}

Based on the fermion parity, $P_n[\sqrt{\rho}(t)]$ and $P_n[(\psi_1\sqrt{\rho})(t)]$ are nonzero only for even $n$s and odd $n$s respectively. Their size distributions are exhibited in Fig.~\ref{fig:vactherm} and Fig.~\ref{fig:fermitherm}. 

When $\nu>0$, both $P_n[\sqrt{\rho}(t)]$ and $P_n[(\psi_1\sqrt{\rho})(t)]$ tend to concentrate on their lowest possible value after a long evolution. But due to the scrambling of fermion $\psi_1$ in  $(\psi_1\sqrt{\rho})(t)$, it has nonzero distribution at $n\geq3$ even at late times. Moreover, the distribution $P_n[(\psi_1\sqrt{\rho})(t)]$ in Fig.~\ref{fig:fermitherm} presents the similar late time behavior with the distribution $P_n[\psi_1(t)]$ in Fig~\ref{fig:EDDist}.
When $\nu<0$, both $P_n[\sqrt{\rho}(t)]$ and $P_n[(\psi_1\sqrt{\rho})(t)]$ are pushed to the region of large size at late times.

\section{Alternative saturated uncertainty relation}\label{sec:AlterUncertainty}

By using the Cauchy-Schwarz inequality and the hermicity of $\X,\P$, we write the uncertainty relation in another form
\begin{equation}
    \avg{\Delta \X^2}\avg{\Delta \L^\dagger\Delta \L}
    \geq 
    \abs{\avg{\Delta \X\Delta \L}}^2
    \geq \kc{\frac12\avg{[\X,i\P]+\Delta \X^2}}^2 
    =\kc{\frac12\partial_t\avg{\X}}^2.
\end{equation}
The inequality connects the operator size growth to the standard variances of the Liouvillian and the operator size
\begin{equation}\label{eq:uncertaintyL}
    \partial_t \avg{n}
    \leq 2\sqrt{\avg{\Delta \L^\dagger \Delta \L}\avg{\Delta n^2}}.
\end{equation}
Comparing the expressions of each term on the two sides \eqref{eq:largeqsize} \eqref{eq:LiouvillianVariance} \eqref{eq:SizeVariance} in the large $q$ limit, we find that the inequality \eqref{eq:uncertaintyL} is saturated. 
Actually, this condition stems from the property of the large $q$ solution at finite $\hat\mu$
\begin{equation}
    \left.(\partial_{\hat\mu}\partial_{\tau_-}g_2(\tau_1,\tau_2))^2 +(\partial_{\tau_1}\partial_{\tau_2}g_2(\tau_1,\tau_2))
    (\partial_{\hat\mu}^2 g_2(\tau_1,\tau_2))\right|_{\tau_+=0}=0.
\end{equation}

\bibliographystyle{JHEP}
\bibliography{refs}

\providecommand{\href}[2]{#2}\begingroup\raggedright\begin{thebibliography}{100}

\bibitem{Lieb2004}
E.H.~Lieb and D.W.~Robinson, \emph{The finite group velocity of quantum spin
  systems},  in \emph{Statistical Mechanics: Selecta of Elliott H. Lieb},
  B.~Nachtergaele, J.P.~Solovej and J.~Yngvason, eds., pp.~425--431, Springer
  Berlin Heidelberg (2004),
  \href{https://doi.org/10.1007/978-3-662-10018-9_25}{DOI}.

\bibitem{Hayden:2007cs}
P.~Hayden and J.~Preskill, \emph{{Black holes as mirrors: Quantum information
  in random subsystems}},
  \href{https://doi.org/10.1088/1126-6708/2007/09/120}{\emph{JHEP} {\bfseries
  09} (2007) 120} [\href{https://arxiv.org/abs/0708.4025}{{\ttfamily
  0708.4025}}].

\bibitem{Sekino:2008scramblers}
Y.~Sekino and L.~Susskind, \emph{Fast scramblers},
  \href{https://doi.org/10.1088/1126-6708/2008/10/065}{\emph{JHEP} {\bfseries
  10} (2008) 065} [\href{https://arxiv.org/abs/0808.2096}{{\ttfamily
  0808.2096}}].

\bibitem{larkin:1969quasiclassical}
A.I.~Larkin and Y.N.~Ovchinnikov, \emph{Quasiclassical method in the theory of
  superconductivity}, {\emph{Sov Phys JETP} {\bfseries 28} (1969) 1200}.

\bibitem{Shenker:2013yza}
S.H.~Shenker and D.~Stanford, \emph{Multiple shocks},
  \href{https://doi.org/10.1007/JHEP12(2014)046}{\emph{JHEP} {\bfseries 12}
  (2014) 046} [\href{https://arxiv.org/abs/1312.3296}{{\ttfamily 1312.3296}}].

\bibitem{Roberts:2014localized}
D.A.~Roberts, D.~Stanford and L.~Susskind, \emph{Localized shocks},
  \href{https://doi.org/10.1007/JHEP03(2015)051}{\emph{JHEP} {\bfseries 03}
  (2015) 051} [\href{https://arxiv.org/abs/1409.8180}{{\ttfamily 1409.8180}}].

\bibitem{PRXQuantum.5.010201}
S.~Xu and B.~Swingle, \emph{Scrambling dynamics and out-of-time-ordered
  correlators in quantum many-body systems},
  \href{https://doi.org/10.1103/PRXQuantum.5.010201}{\emph{PRX Quantum}
  {\bfseries 5} (2024) 010201}.

\bibitem{Roberts:2016wdl}
D.A.~Roberts and B.~Swingle, \emph{{Lieb-Robinson Bound and the Butterfly
  Effect in Quantum Field Theories}},
  \href{https://doi.org/10.1103/PhysRevLett.117.091602}{\emph{Phys. Rev. Lett.}
  {\bfseries 117} (2016) 091602}
  [\href{https://arxiv.org/abs/1603.09298}{{\ttfamily 1603.09298}}].

\bibitem{Shenker:2013black}
S.H.~Shenker and D.~Stanford, \emph{Black holes and the butterfly effect},
  \href{https://doi.org/10.1007/JHEP03(2014)067}{\emph{JHEP} {\bfseries 03}
  (2014) 067} [\href{https://arxiv.org/abs/1306.0622}{{\ttfamily 1306.0622}}].

\bibitem{Hosur:2015channels}
P.~Hosur, X.-L.~Qi, D.A.~Roberts and B.~Yoshida, \emph{Chaos in quantum
  channels}, \href{https://doi.org/10.1007/JHEP02(2016)004}{\emph{JHEP}
  {\bfseries 02} (2016) 004}
  [\href{https://arxiv.org/abs/1511.04021}{{\ttfamily 1511.04021}}].

\bibitem{Roberts:2014ifa}
D.A.~Roberts and D.~Stanford, \emph{{Two-dimensional conformal field theory and
  the butterfly effect}},
  \href{https://doi.org/10.1103/PhysRevLett.115.131603}{\emph{Phys. Rev. Lett.}
  {\bfseries 115} (2015) 131603}
  [\href{https://arxiv.org/abs/1412.5123}{{\ttfamily 1412.5123}}].

\bibitem{Stanford:2015owe}
D.~Stanford, \emph{{Many-body chaos at weak coupling}},
  \href{https://doi.org/10.1007/JHEP10(2016)009}{\emph{JHEP} {\bfseries 10}
  (2016) 009} [\href{https://arxiv.org/abs/1512.07687}{{\ttfamily
  1512.07687}}].

\bibitem{Roberts:2018operator}
D.A.~Roberts, D.~Stanford and A.~Streicher, \emph{Operator growth in the syk
  model}, \href{https://doi.org/10.1007/JHEP06(2018)122}{\emph{JHEP} {\bfseries
  06} (2018) 122} [\href{https://arxiv.org/abs/1802.02633}{{\ttfamily
  1802.02633}}].

\bibitem{Qi:2018quantum}
X.-L.~Qi and A.~Streicher, \emph{Quantum epidemiology: Operator growth, thermal
  effects, and syk}, \href{https://doi.org/10.1007/JHEP08(2019)012}{\emph{JHEP}
  {\bfseries 08} (2019) 012}
  [\href{https://arxiv.org/abs/1810.11958}{{\ttfamily 1810.11958}}].

\bibitem{Qi:2019rpi}
X.-L.~Qi, E.J.~Davis, A.~Periwal and M.~Schleier-Smith, \emph{{Measuring
  operator size growth in quantum quench experiments}},
  \href{https://arxiv.org/abs/1906.00524}{{\ttfamily 1906.00524}}.

\bibitem{Parker:2018a}
D.E.~Parker, X.~Cao, A.~Avdoshkin, T.~Scaffidi and E.~Altman, \emph{A universal
  operator growth hypothesis},
  \href{https://doi.org/10.1103/PhysRevX.9.041017}{\emph{Phys. Rev. X}
  {\bfseries 9} (2019) 041017}
  [\href{https://arxiv.org/abs/1812.08657}{{\ttfamily 1812.08657}}].

\bibitem{Barbon:2019on}
J.~Barbón, E.~Rabinovici, R.~Shir and R.~Sinha, \emph{On the evolution of
  operator complexity beyond scrambling},
  \href{https://doi.org/10.1007/JHEP10(2019)264}{\emph{JHEP} {\bfseries 10}
  (2019) 264} [\href{https://arxiv.org/abs/1907.05393}{{\ttfamily
  1907.05393}}].

\bibitem{Avdoshkin:2019trj}
A.~Avdoshkin and A.~Dymarsky, \emph{Euclidean operator growth and quantum
  chaos}, \href{https://doi.org/10.1103/PhysRevResearch.2.043234}{\emph{Phys.
  Rev. Res.} {\bfseries 2} (2020) 043234}
  [\href{https://arxiv.org/abs/1911.09672}{{\ttfamily 1911.09672}}].

\bibitem{Jian:2020qpp}
S.-K.~Jian, B.~Swingle and Z.-Y.~Xian, \emph{Complexity growth of operators in
  the syk model and in jt gravity},
  \href{https://doi.org/10.1007/JHEP03(2021)014}{\emph{JHEP} {\bfseries 03}
  (2021) 014} [\href{https://arxiv.org/abs/2008.12274}{{\ttfamily
  2008.12274}}].

\bibitem{Rabinovici:2020operator}
E.~Rabinovici, A.~S\'anchez-Garrido, R.~Shir and J.~Sonner, \emph{Operator
  complexity: a journey to the edge of krylov space},
  \href{https://arxiv.org/abs/2009.01862}{{\ttfamily 2009.01862}}.

\bibitem{Dymarsky:2019quantum}
A.~Dymarsky and A.~Gorsky, \emph{Quantum chaos as delocalization in krylov
  space}, \href{https://doi.org/10.1103/PhysRevB.102.085137}{\emph{Phys. Rev.
  B} {\bfseries 102} (2020) 085137}
  [\href{https://arxiv.org/abs/1912.12227}{{\ttfamily 1912.12227}}].

\bibitem{Carrega:2020unveiling}
M.~Carrega, J.~Kim and D.~Rosa, \emph{Unveiling operator growth in syk quench
  dynamics},  \href{https://arxiv.org/abs/2007.03551}{{\ttfamily 2007.03551}}.

\bibitem{Kar:2021nbm}
A.~Kar, L.~Lamprou, M.~Rozali and J.~Sully, \emph{Random matrix theory for
  complexity growth and black hole interiors},
  \href{https://doi.org/10.1007/JHEP01(2022)016}{\emph{JHEP} {\bfseries 01}
  (2022) 016} [\href{https://arxiv.org/abs/2106.02046}{{\ttfamily
  2106.02046}}].

\bibitem{Caputa:2021sib}
P.~Caputa, J.M.~Magan and D.~Patramanis, \emph{Geometry of krylov complexity},
  \href{https://doi.org/10.1103/PhysRevResearch.4.013041}{\emph{Phys. Rev.
  Res.} {\bfseries 4} (2022) 013041}
  [\href{https://arxiv.org/abs/2109.03824}{{\ttfamily 2109.03824}}].

\bibitem{Dymarsky:2021bjq}
A.~Dymarsky and M.~Smolkin, \emph{Krylov complexity in conformal field theory},
  \href{https://doi.org/10.1103/PhysRevD.104.L081702}{\emph{Phys. Rev. D}
  {\bfseries 104} (2021) L081702}
  [\href{https://arxiv.org/abs/2104.09514}{{\ttfamily 2104.09514}}].

\bibitem{Rabinovici:2021qqt}
E.~Rabinovici, A.~S\'anchez-Garrido, R.~Shir and J.~Sonner, \emph{Krylov
  localization and suppression of complexity},
  \href{https://doi.org/10.1007/JHEP03(2022)211}{\emph{JHEP} {\bfseries 03}
  (2022) 211} [\href{https://arxiv.org/abs/2112.12128}{{\ttfamily
  2112.12128}}].

\bibitem{Muck:2022xfc}
W.~M\"uck and Y.~Yang, \emph{{Krylov complexity and orthogonal polynomials}},
  \href{https://doi.org/10.1016/j.nuclphysb.2022.115948}{\emph{Nucl. Phys. B}
  {\bfseries 984} (2022) 115948}
  [\href{https://arxiv.org/abs/2205.12815}{{\ttfamily 2205.12815}}].

\bibitem{Rabinovici:2022beu}
E.~Rabinovici, A.~S\'anchez-Garrido, R.~Shir and J.~Sonner, \emph{Krylov
  complexity from integrability to chaos},
  \href{https://doi.org/10.1007/JHEP07(2022)151}{\emph{JHEP} {\bfseries 07}
  (2022) 151} [\href{https://arxiv.org/abs/2207.07701}{{\ttfamily
  2207.07701}}].

\bibitem{Bhattacharjee:2022vlt}
B.~Bhattacharjee, X.~Cao, P.~Nandy and T.~Pathak, \emph{Krylov complexity in
  saddle-dominated scrambling},
  \href{https://doi.org/10.1007/JHEP05(2022)174}{\emph{JHEP} {\bfseries 05}
  (2022) 174} [\href{https://arxiv.org/abs/2203.03534}{{\ttfamily
  2203.03534}}].

\bibitem{Bhattacharjee:2022ave}
B.~Bhattacharjee, P.~Nandy and T.~Pathak, \emph{{Krylov complexity in large q
  and double-scaled SYK model}},
  \href{https://doi.org/10.1007/JHEP08(2023)099}{\emph{JHEP} {\bfseries 08}
  (2023) 099} [\href{https://arxiv.org/abs/2210.02474}{{\ttfamily
  2210.02474}}].

\bibitem{Avdoshkin:2022xuw}
A.~Avdoshkin, A.~Dymarsky and M.~Smolkin, \emph{Krylov complexity in quantum
  field theory, and beyond},
  \href{https://arxiv.org/abs/2212.14429}{{\ttfamily 2212.14429}}.

\bibitem{Alishahiha:2022anw}
M.~Alishahiha and S.~Banerjee, \emph{{A universal approach to Krylov state and
  operator complexities}},
  \href{https://doi.org/10.21468/SciPostPhys.15.3.080}{\emph{SciPost Phys.}
  {\bfseries 15} (2023) 080}
  [\href{https://arxiv.org/abs/2212.10583}{{\ttfamily 2212.10583}}].

\bibitem{Huh:2023jxt}
K.-B.~Huh, H.-S.~Jeong and J.F.~Pedraza, \emph{{Spread complexity in
  saddle-dominated scrambling}},
  \href{https://arxiv.org/abs/2312.12593}{{\ttfamily 2312.12593}}.

\bibitem{Tang:2023ocr}
H.~Tang, \emph{{Operator Krylov complexity in random matrix theory}},
  \href{https://arxiv.org/abs/2312.17416}{{\ttfamily 2312.17416}}.

\bibitem{Aguilar-Gutierrez:2023nyk}
S.E.~Aguilar-Gutierrez and A.~Rolph, \emph{{Krylov complexity is not a measure
  of distance between states or operators}},
  \href{https://arxiv.org/abs/2311.04093}{{\ttfamily 2311.04093}}.

\bibitem{Kundu:2023hbk}
A.~Kundu, V.~Malvimat and R.~Sinha, \emph{State dependence of krylov complexity
  in $2d$ cfts},  \href{https://arxiv.org/abs/2303.03426}{{\ttfamily
  2303.03426}}.

\bibitem{Beetar:2023mfn}
C.~Beetar, N.~Gupta, S.S.~Haque, J.~Murugan and H.J.R.~Van~Zyl,
  \emph{{Complexity and Operator Growth for Quantum Systems in Dynamic
  Equilibrium}},  \href{https://arxiv.org/abs/2312.15790}{{\ttfamily
  2312.15790}}.

\bibitem{Bhattacharyya:2023dhp}
A.~Bhattacharyya, D.~Ghosh and P.~Nandi, \emph{{Operator growth and Krylov
  complexity in Bose-Hubbard model}},
  \href{https://doi.org/10.1007/JHEP12(2023)112}{\emph{JHEP} {\bfseries 12}
  (2023) 112} [\href{https://arxiv.org/abs/2306.05542}{{\ttfamily
  2306.05542}}].

\bibitem{Loc:2024oen}
T.Q.~Loc, \emph{{Lanczos spectrum for random operator growth}},
  \href{https://arxiv.org/abs/2402.07980}{{\ttfamily 2402.07980}}.

\bibitem{Malvimat:2024vhr}
V.~Malvimat, S.~Porey and B.~Roy, \emph{{Krylov Complexity in $2d$ CFTs with
  SL$(2,\mathbb{R})$ deformed Hamiltonians}},
  \href{https://arxiv.org/abs/2402.15835}{{\ttfamily 2402.15835}}.

\bibitem{Bhattacharya:2024uxx}
A.~Bhattacharya, P.P.~Nath and H.~Sahu, \emph{{Speed limits to the growth of
  Krylov complexity in open quantum systems}},
  \href{https://arxiv.org/abs/2403.03584}{{\ttfamily 2403.03584}}.

\bibitem{Balasubramanian:2022tpr}
V.~Balasubramanian, P.~Caputa, J.M.~Magan and Q.~Wu, \emph{Quantum chaos and
  the complexity of spread of states},
  \href{https://doi.org/10.1103/PhysRevD.106.046007}{\emph{Phys. Rev. D}
  {\bfseries 106} (2022) 046007}
  [\href{https://arxiv.org/abs/2202.06957}{{\ttfamily 2202.06957}}].

\bibitem{Balasubramanian:2022dnj}
V.~Balasubramanian, J.M.~Magan and Q.~Wu, \emph{{Tridiagonalizing random
  matrices}}, \href{https://doi.org/10.1103/PhysRevD.107.126001}{\emph{Phys.
  Rev. D} {\bfseries 107} (2023) 126001}
  [\href{https://arxiv.org/abs/2208.08452}{{\ttfamily 2208.08452}}].

\bibitem{Erdmenger:2023wjg}
J.~Erdmenger, S.-K.~Jian and Z.-Y.~Xian, \emph{{Universal chaotic dynamics from
  Krylov space}}, \href{https://doi.org/10.1007/JHEP08(2023)176}{\emph{JHEP}
  {\bfseries 08} (2023) 176}
  [\href{https://arxiv.org/abs/2303.12151}{{\ttfamily 2303.12151}}].

\bibitem{Balasubramanian:2023kwd}
V.~Balasubramanian, J.M.~Magan and Q.~Wu, \emph{{Quantum chaos, integrability,
  and late times in the Krylov basis}},
  \href{https://arxiv.org/abs/2312.03848}{{\ttfamily 2312.03848}}.

\bibitem{Camargo:2023eev}
H.A.~Camargo, V.~Jahnke, H.-S.~Jeong, K.-Y.~Kim and M.~Nishida, \emph{{Spectral
  and Krylov complexity in billiard systems}},
  \href{https://doi.org/10.1103/PhysRevD.109.046017}{\emph{Phys. Rev. D}
  {\bfseries 109} (2024) 046017}
  [\href{https://arxiv.org/abs/2306.11632}{{\ttfamily 2306.11632}}].

\bibitem{Bhattacharyya:2023grv}
A.~Bhattacharyya, S.S.~Haque, G.~Jafari, J.~Murugan and D.~Rapotu,
  \emph{{Krylov complexity and spectral form factor for noisy random matrix
  models}}, \href{https://doi.org/10.1007/JHEP10(2023)157}{\emph{JHEP}
  {\bfseries 10} (2023) 157}
  [\href{https://arxiv.org/abs/2307.15495}{{\ttfamily 2307.15495}}].

\bibitem{Caputa:2024vrn}
P.~Caputa, H.-S.~Jeong, S.~Liu, J.F.~Pedraza and L.-C.~Qu, \emph{{Krylov
  complexity of density matrix operators}},
  \href{https://arxiv.org/abs/2402.09522}{{\ttfamily 2402.09522}}.

\bibitem{Kitaev:2015a}
A.~Kitaev, ``A simple model of quantum holography.''
  \url{http://online.kitp.ucsb.edu/online/entangled15/kitaev/}
  \url{http://online.kitp.ucsb.edu/online/entangled15/kitaev2/}, 2015.

\bibitem{Maldacena:2016remarks}
J.~Maldacena and D.~Stanford, \emph{Remarks on the sachdev-ye-kitaev model},
  \href{https://doi.org/10.1103/PhysRevD.94.106002}{\emph{Phys. Rev. D}
  {\bfseries 94} (2016) 106002}
  [\href{https://arxiv.org/abs/1604.07818}{{\ttfamily 1604.07818}}].

\bibitem{Lin:2023trc}
H.W.~Lin and D.~Stanford, \emph{{A symmetry algebra in double-scaled SYK}},
  \href{https://doi.org/10.21468/SciPostPhys.15.6.234}{\emph{SciPost Phys.}
  {\bfseries 15} (2023) 234}
  [\href{https://arxiv.org/abs/2307.15725}{{\ttfamily 2307.15725}}].

\bibitem{Roberts:2016design}
D.A.~Roberts and B.~Yoshida, \emph{Chaos and complexity by design},
  \href{https://doi.org/10.1007/JHEP04(2017)121}{\emph{JHEP} {\bfseries 04}
  (2017) 121} [\href{https://arxiv.org/abs/1610.04903}{{\ttfamily
  1610.04903}}].

\bibitem{Cotler:2017jue}
J.~Cotler, N.~Hunter-Jones, J.~Liu and B.~Yoshida, \emph{Chaos, complexity, and
  random matrices}, \href{https://doi.org/10.1007/JHEP11(2017)048}{\emph{JHEP}
  {\bfseries 11} (2017) 048}
  [\href{https://arxiv.org/abs/1706.05400}{{\ttfamily 1706.05400}}].

\bibitem{Nahum:2017operator}
A.~Nahum, S.~Vijay and J.~Haah, \emph{Operator spreading in random unitary
  circuits}, \href{https://doi.org/10.1103/PhysRevX.8.021014}{\emph{Phys. Rev.
  X} {\bfseries 8} (2018) 021014}
  [\href{https://arxiv.org/abs/1705.08975}{{\ttfamily 1705.08975}}].

\bibitem{vonKeyserlingk:2017operator}
C.~von Keyserlingk, T.~Rakovszky, F.~Pollmann and S.~Sondhi, \emph{Operator
  hydrodynamics, otocs, and entanglement growth in systems without conservation
  laws}, \href{https://doi.org/10.1103/PhysRevX.8.021013}{\emph{Phys. Rev. X}
  {\bfseries 8} (2018) 021013}
  [\href{https://arxiv.org/abs/1705.08910}{{\ttfamily 1705.08910}}].

\bibitem{Xu:2018locality}
S.~Xu and B.~Swingle, \emph{Locality, quantum fluctuations, and scrambling},
  \href{https://doi.org/10.1103/PhysRevX.9.031048}{\emph{Phys. Rev. X}
  {\bfseries 9} (2019) 031048}
  [\href{https://arxiv.org/abs/1805.05376}{{\ttfamily 1805.05376}}].

\bibitem{Maldacena:2016conformal}
J.~Maldacena, D.~Stanford and Z.~Yang, \emph{Conformal symmetry and its
  breaking in two dimensional nearly anti-de-sitter space},
  \href{https://doi.org/10.1093/ptep/ptw124}{\emph{PTEP} {\bfseries 2016}
  (2016) 12C104} [\href{https://arxiv.org/abs/1606.01857}{{\ttfamily
  1606.01857}}].

\bibitem{Lensky:2020size}
Y.D.~Lensky, X.-L.~Qi and P.~Zhang, \emph{Size of bulk fermions in the syk
  model},  \href{https://arxiv.org/abs/2002.01961}{{\ttfamily 2002.01961}}.

\bibitem{Mousatov:2019operator}
A.~Mousatov, \emph{Operator size for holographic field theories},
  \href{https://arxiv.org/abs/1911.05089}{{\ttfamily 1911.05089}}.

\bibitem{Lin:2019symmetries}
H.W.~Lin, J.~Maldacena and Y.~Zhao, \emph{Symmetries near the horizon},
  \href{https://doi.org/10.1007/JHEP08(2019)049}{\emph{JHEP} {\bfseries 08}
  (2019) 049} [\href{https://arxiv.org/abs/1904.12820}{{\ttfamily
  1904.12820}}].

\bibitem{Maldacena:2015waa}
J.~Maldacena, S.H.~Shenker and D.~Stanford, \emph{A bound on chaos},
  \href{https://doi.org/10.1007/JHEP08(2016)106}{\emph{JHEP} {\bfseries 08}
  (2016) 106} [\href{https://arxiv.org/abs/1503.01409}{{\ttfamily
  1503.01409}}].

\bibitem{Fan:2016ean}
R.~Fan, P.~Zhang, H.~Shen and H.~Zhai, \emph{{Out-of-Time-Order Correlation for
  Many-Body Localization}},
  \href{https://doi.org/10.1016/j.scib.2017.04.011}{\emph{Sci. Bull.}
  {\bfseries 62} (2017) 707}
  [\href{https://arxiv.org/abs/1608.01914}{{\ttfamily 1608.01914}}].

\bibitem{Chen:2017dbb}
Y.~Chen, H.~Zhai and P.~Zhang, \emph{{Tunable Quantum Chaos in the
  Sachdev-Ye-Kitaev Model Coupled to a Thermal Bath}},
  \href{https://doi.org/10.1007/JHEP07(2017)150}{\emph{JHEP} {\bfseries 07}
  (2017) 150} [\href{https://arxiv.org/abs/1705.09818}{{\ttfamily
  1705.09818}}].

\bibitem{Zhang:2018oop}
Y.-L.~Zhang, Y.~Huang and X.~Chen, \emph{{Information scrambling in chaotic
  systems with dissipation}},
  \href{https://doi.org/10.1103/PhysRevB.99.014303}{\emph{Phys. Rev. B}
  {\bfseries 99} (2019) 014303}
  [\href{https://arxiv.org/abs/1802.04492}{{\ttfamily 1802.04492}}].

\bibitem{Li:2016xhw}
J.~Li, R.~Fan, H.~Wang, B.~Ye, B.~Zeng, H.~Zhai et~al., \emph{{Measuring
  Out-of-Time-Order Correlators on a Nuclear Magnetic Resonance Quantum
  Simulator}}, \href{https://doi.org/10.1103/PhysRevX.7.031011}{\emph{Phys.
  Rev. X} {\bfseries 7} (2017) 031011}
  [\href{https://arxiv.org/abs/1609.01246}{{\ttfamily 1609.01246}}].

\bibitem{Garttner:2016mqj}
M.~G\"arttner, J.G.~Bohnet, A.~Safavi-Naini, M.L.~Wall, J.J.~Bollinger and
  A.M.~Rey, \emph{{Measuring out-of-time-order correlations and multiple
  quantum spectra in a trapped ion quantum magnet}},
  \href{https://doi.org/10.1038/nphys4119}{\emph{Nature Phys.} {\bfseries 13}
  (2017) 781} [\href{https://arxiv.org/abs/1608.08938}{{\ttfamily
  1608.08938}}].

\bibitem{Sanchez:2020LoschmidtEcho}
C.M.~S\'anchez, A.K.~Chattah, K.X.~Wei, L.~Buljubasich, P.~Cappellaro and
  H.M.~Pastawski, \emph{Perturbation independent decay of the loschmidt echo in
  a many-body system},
  \href{https://doi.org/10.1103/PhysRevLett.124.030601}{\emph{Phys. Rev. Lett.}
  {\bfseries 124} (2020) 030601}.

\bibitem{Sanchez:2022LoschmidtEchoNER}
C.M.~S\'anchez, A.K.~Chattah and H.M.~Pastawski, \emph{Emergent decoherence
  induced by quantum chaos in a many-body system: A loschmidt echo observation
  through nmr}, \href{https://doi.org/10.1103/PhysRevA.105.052232}{\emph{Phys.
  Rev. A} {\bfseries 105} (2022) 052232}.

\bibitem{Dominguez:2021decoherence}
F.D.~Dom\'{\i}nguez, M.C.~Rodr\'{\i}guez, R.~Kaiser, D.~Suter and
  G.A.~\'Alvarez, \emph{Decoherence scaling transition in the dynamics of
  quantum information scrambling},
  \href{https://doi.org/10.1103/PhysRevA.104.012402}{\emph{Phys. Rev. A}
  {\bfseries 104} (2021) 012402}.

\bibitem{Dominguez:2021keq}
F.D.~Dom\'\i{}nguez and G.A.~\'Alvarez, \emph{{Dynamics of quantum information
  scrambling under decoherence effects measured via active spin clusters}},
  \href{https://doi.org/10.1103/PhysRevA.104.062406}{\emph{Phys. Rev. A}
  {\bfseries 104} (2021) 062406}
  [\href{https://arxiv.org/abs/2107.03870}{{\ttfamily 2107.03870}}].

\bibitem{Mi:2021gdf}
X.~Mi et~al., \emph{{Information scrambling in quantum circuits}},
  \href{https://doi.org/10.1126/science.abg5029}{\emph{Science} {\bfseries 374}
  (2021) abg5029} [\href{https://arxiv.org/abs/2101.08870}{{\ttfamily
  2101.08870}}].

\bibitem{Cotler:2022fin}
J.~Cotler, T.~Schuster and M.~Mohseni, \emph{{Information-theoretic hardness of
  out-of-time-order correlators}},
  \href{https://doi.org/10.1103/PhysRevA.108.062608}{\emph{Phys. Rev. A}
  {\bfseries 108} (2023) 062608}
  [\href{https://arxiv.org/abs/2208.02256}{{\ttfamily 2208.02256}}].

\bibitem{Swingle:2016var}
B.~Swingle, G.~Bentsen, M.~Schleier-Smith and P.~Hayden, \emph{{Measuring the
  scrambling of quantum information}},
  \href{https://doi.org/10.1103/PhysRevA.94.040302}{\emph{Phys. Rev. A}
  {\bfseries 94} (2016) 040302}
  [\href{https://arxiv.org/abs/1602.06271}{{\ttfamily 1602.06271}}].

\bibitem{Islam:2015mom}
R.~Islam, R.~Ma, P.M.~Preiss, M.E.~Tai, A.~Lukin, M.~Rispoli et~al.,
  \emph{{Measuring entanglement entropy through the interference of quantum
  many-body twins}},  \href{https://arxiv.org/abs/1509.01160}{{\ttfamily
  1509.01160}}.

\bibitem{Landsman:2018jpm}
K.A.~Landsman, C.~Figgatt, T.~Schuster, N.M.~Linke, B.~Yoshida, N.Y.~Yao
  et~al., \emph{{Verified Quantum Information Scrambling}},
  \href{https://doi.org/10.1038/s41586-019-0952-6}{\emph{Nature} {\bfseries
  567} (2019) 61} [\href{https://arxiv.org/abs/1806.02807}{{\ttfamily
  1806.02807}}].

\bibitem{Blok:2020may}
M.S.~Blok, V.V.~Ramasesh, T.~Schuster, K.~O'Brien, J.M.~Kreikebaum, D.~Dahlen
  et~al., \emph{{Quantum Information Scrambling on a Superconducting Qutrit
  Processor}}, \href{https://doi.org/10.1103/PhysRevX.11.021010}{\emph{Phys.
  Rev. X} {\bfseries 11} (2021) 021010}
  [\href{https://arxiv.org/abs/2003.03307}{{\ttfamily 2003.03307}}].

\bibitem{Brydges:2019probing}
T.~Brydges, A.~Elben, P.~Jurcevic, B.~Vermersch, C.~Maier, B.P.~Lanyon et~al.,
  \emph{Probing r{\'e}nyi entanglement entropy via randomized measurements},
  {\emph{Science} {\bfseries 364} (2019) 260}.

\bibitem{Joshi:2020PRL}
M.K.~Joshi, A.~Elben, B.~Vermersch, T.~Brydges, C.~Maier, P.~Zoller et~al.,
  \emph{Quantum information scrambling in a trapped-ion quantum simulator with
  tunable range interactions},
  \href{https://doi.org/10.1103/PhysRevLett.124.240505}{\emph{Phys. Rev. Lett.}
  {\bfseries 124} (2020) 240505}.

\bibitem{Blocher:2023hvk}
P.D.~Blocher, K.~Chinni, S.~Omanakuttan and P.M.~Poggi, \emph{{Probing
  scrambling and operator size distributions using random mixed states and
  local measurements}},  \href{https://arxiv.org/abs/2305.16992}{{\ttfamily
  2305.16992}}.

\bibitem{breuer2002theory}
H.-P.~Breuer and F.~Petruccione, \emph{The theory of open quantum systems},
  Oxford University Press, USA (2002).

\bibitem{lidar2019lecture}
D.A.~Lidar, \emph{Lecture notes on the theory of open quantum systems},
  {\emph{arXiv preprint arXiv:1902.00967} (2019) }.

\bibitem{manzano2020short}
D.~Manzano, \emph{A short introduction to the lindblad master equation},
  {\emph{Aip Advances} {\bfseries 10} (2020) }.

\bibitem{Denisov:2018nif}
S.~Denisov, T.~Laptyeva, W.~Tarnowski, D.~Chru\'sci\'nski and K.~\.Zyczkowski,
  \emph{{Universal spectra of random Lindblad operators}},
  \href{https://doi.org/10.1103/PhysRevLett.123.140403}{\emph{Phys. Rev. Lett.}
  {\bfseries 123} (2019) 140403}
  [\href{https://arxiv.org/abs/1811.12282}{{\ttfamily 1811.12282}}].

\bibitem{Schuster:2022bot}
T.~Schuster and N.Y.~Yao, \emph{{Operator Growth in Open Quantum Systems}},
  \href{https://doi.org/10.1103/PhysRevLett.131.160402}{\emph{Phys. Rev. Lett.}
  {\bfseries 131} (2023) 160402}
  [\href{https://arxiv.org/abs/2208.12272}{{\ttfamily 2208.12272}}].

\bibitem{Kulkarni:2021gtt}
A.~Kulkarni, T.~Numasawa and S.~Ryu, \emph{{Lindbladian dynamics of the
  Sachdev-Ye-Kitaev model}},
  \href{https://doi.org/10.1103/PhysRevB.106.075138}{\emph{Phys. Rev. B}
  {\bfseries 106} (2022) 075138}
  [\href{https://arxiv.org/abs/2112.13489}{{\ttfamily 2112.13489}}].

\bibitem{Kawabata:2022osw}
K.~Kawabata, A.~Kulkarni, J.~Li, T.~Numasawa and S.~Ryu, \emph{{Dynamical
  quantum phase transitions in Sachdev-Ye-Kitaev Lindbladians}},
  \href{https://doi.org/10.1103/PhysRevB.108.075110}{\emph{Phys. Rev. B}
  {\bfseries 108} (2023) 075110}
  [\href{https://arxiv.org/abs/2210.04093}{{\ttfamily 2210.04093}}].

\bibitem{Garcia-Garcia:2022adg}
A.M.~Garc\'\i{}a-Garc\'\i{}a, L.~S\'a, J.J.M.~Verbaarschot and J.P.~Zheng,
  \emph{{Keldysh wormholes and anomalous relaxation in the dissipative
  Sachdev-Ye-Kitaev model}},
  \href{https://doi.org/10.1103/PhysRevD.107.106006}{\emph{Phys. Rev. D}
  {\bfseries 107} (2023) 106006}
  [\href{https://arxiv.org/abs/2210.01695}{{\ttfamily 2210.01695}}].

\bibitem{Sa:2021tdr}
L.~S\'a, P.~Ribeiro and T.~Prosen, \emph{{Lindbladian dissipation of
  strongly-correlated quantum matter}},
  \href{https://doi.org/10.1103/PhysRevResearch.4.L022068}{\emph{Phys. Rev.
  Res.} {\bfseries 4} (2022) L022068}
  [\href{https://arxiv.org/abs/2112.12109}{{\ttfamily 2112.12109}}].

\bibitem{Bhattacharjee:2023uwx}
B.~Bhattacharjee, P.~Nandy and T.~Pathak, \emph{{Operator dynamics in
  Lindbladian SYK: a Krylov complexity perspective}},
  \href{https://doi.org/10.1007/JHEP01(2024)094}{\emph{JHEP} {\bfseries 01}
  (2024) 094} [\href{https://arxiv.org/abs/2311.00753}{{\ttfamily
  2311.00753}}].

\bibitem{Liu:2022god}
C.~Liu, H.~Tang and H.~Zhai, \emph{Krylov complexity in open quantum systems},
  \href{https://arxiv.org/abs/2207.13603}{{\ttfamily 2207.13603}}.

\bibitem{Bhattacharjee:2022lzy}
B.~Bhattacharjee, X.~Cao, P.~Nandy and T.~Pathak, \emph{Operator growth in open
  quantum systems: lessons from the dissipative syk},
  \href{https://arxiv.org/abs/2212.06180}{{\ttfamily 2212.06180}}.

\bibitem{Bhattacharya:2022gbz}
A.~Bhattacharya, P.~Nandy, P.P.~Nath and H.~Sahu, \emph{Operator growth and
  krylov construction in dissipative open quantum systems},
  \href{https://doi.org/10.1007/JHEP12(2022)081}{\emph{JHEP} {\bfseries 12}
  (2022) 081} [\href{https://arxiv.org/abs/2207.05347}{{\ttfamily
  2207.05347}}].

\bibitem{Bhattacharya:2023zqt}
A.~Bhattacharya, P.~Nandy, P.P.~Nath and H.~Sahu, \emph{On krylov complexity in
  open systems: an approach via bi-lanczos algorithm},
  \href{https://arxiv.org/abs/2303.04175}{{\ttfamily 2303.04175}}.

\bibitem{Bhattacharya:2023yec}
A.~Bhattacharya, R.N.~Das, B.~Dey and J.~Erdmenger, \emph{{Spread complexity
  for measurement-induced non-unitary dynamics and Zeno effect}},
  \href{https://arxiv.org/abs/2312.11635}{{\ttfamily 2312.11635}}.

\bibitem{Zhang:2023BrownianSize}
P.~Zhang and Z.~Yu, \emph{Dynamical transition of operator size growth in
  quantum systems embedded in an environment},
  \href{https://doi.org/10.1103/PhysRevLett.130.250401}{\emph{Phys. Rev. Lett.}
  {\bfseries 130} (2023) 250401}.

\bibitem{Gao:2016bin}
P.~Gao, D.L.~Jafferis and A.C.~Wall, \emph{{Traversable Wormholes via a Double
  Trace Deformation}},
  \href{https://doi.org/10.1007/JHEP12(2017)151}{\emph{JHEP} {\bfseries 12}
  (2017) 151} [\href{https://arxiv.org/abs/1608.05687}{{\ttfamily
  1608.05687}}].

\bibitem{Maldacena:2018lmt}
J.~Maldacena and X.-L.~Qi, \emph{{Eternal traversable wormhole}},
  {\emph{arXiv:1804.00491} (2018) }.

\bibitem{Milekhin:2022bzx}
A.~Milekhin and F.K.~Popov, \emph{{Measurement-induced phase transition in
  teleportation and wormholes}},
  \href{https://arxiv.org/abs/2210.03083}{{\ttfamily 2210.03083}}.

\bibitem{Prosen:2012sn}
T.~Prosen, \emph{{PT-Symmetric Quantum Liouvillean Dynamics}},
  \href{https://doi.org/10.1103/PhysRevLett.109.090404}{\emph{Phys. Rev. Lett.}
  {\bfseries 109} (2012) 090404}
  [\href{https://arxiv.org/abs/1207.4395}{{\ttfamily 1207.4395}}].

\bibitem{Garcia-Garcia:2023yet}
A.M.~Garc\'\i{}a-Garc\'\i{}a, L.~S\'a, J.J.M.~Verbaarschot and C.~Yin,
  \emph{{Towards a classification of PT-symmetric quantum systems: from
  dissipative dynamics to topology and wormholes}},
  \href{https://arxiv.org/abs/2311.15677}{{\ttfamily 2311.15677}}.

\bibitem{Bender:1998PT}
C.M.~Bender and S.~Boettcher, \emph{Real spectra in non-hermitian hamiltonians
  having $\mathcal{P}\mathcal{T}$ symmetry},
  \href{https://doi.org/10.1103/PhysRevLett.80.5243}{\emph{Phys. Rev. Lett.}
  {\bfseries 80} (1998) 5243}.

\bibitem{Mostafazadeh:2001jk}
A.~Mostafazadeh, \emph{{Pseudo-Hermiticity versus PT symmetry: The necessary
  condition for the reality of the spectrum}},
  \href{https://doi.org/10.1063/1.1418246}{\emph{J. Math. Phys.} {\bfseries 43}
  (2002) 205}.

\bibitem{Mostafazadeh:2001nr}
A.~Mostafazadeh, \emph{{PseudoHermiticity versus PT symmetry 2. A Complete
  characterization of nonHermitian Hamiltonians with a real spectrum}},
  \href{https://doi.org/10.1063/1.1461427}{\emph{J. Math. Phys.} {\bfseries 43}
  (2002) 2814}.

\bibitem{Mostafazadeh:2002id}
A.~Mostafazadeh, \emph{{PseudoHermiticity versus PT symmetry 3: Equivalence of
  pseudoHermiticity and the presence of antilinear symmetries}},
  \href{https://doi.org/10.1063/1.1489072}{\emph{J. Math. Phys.} {\bfseries 43}
  (2002) 3944}.

\bibitem{Zhang:2019gyc}
R.~Zhang, H.~Qin and J.~Xiao, \emph{{PT-symmetry entails pseudo-Hermiticity
  regardless of diagonalizability}},
  \href{https://doi.org/10.1063/1.5117211}{\emph{J. Math. Phys.} {\bfseries 61}
  (2020) 012101}.

\bibitem{Cotler:2016fpe}
J.S.~Cotler, G.~Gur-Ari, M.~Hanada, J.~Polchinski, P.~Saad, S.H.~Shenker
  et~al., \emph{Black holes and random matrices},
  \href{https://doi.org/10.1007/JHEP05(2017)118}{\emph{JHEP} {\bfseries 05}
  (2017) 118} [\href{https://arxiv.org/abs/1611.04650}{{\ttfamily
  1611.04650}}].

\bibitem{Zhou:2021yyw}
Y.-N.~Zhou, L.~Mao and H.~Zhai, \emph{{R\'enyi entropy dynamics and Lindblad
  spectrum for open quantum systems}},
  \href{https://doi.org/10.1103/PhysRevResearch.3.043060}{\emph{Phys. Rev.
  Res.} {\bfseries 3} (2021) 043060}
  [\href{https://arxiv.org/abs/2101.11236}{{\ttfamily 2101.11236}}].

\bibitem{Garcia-Garcia:2016mno}
A.M.~Garc\'\i{}a-Garc\'\i{}a and J.J.M.~Verbaarschot, \emph{Spectral and
  thermodynamic properties of the sachdev-ye-kitaev model},
  \href{https://doi.org/10.1103/PhysRevD.94.126010}{\emph{Phys. Rev. D}
  {\bfseries 94} (2016) 126010}
  [\href{https://arxiv.org/abs/1610.03816}{{\ttfamily 1610.03816}}].

\bibitem{you2017sachdev}
Y.-Z.~You, A.W.~Ludwig and C.~Xu, \emph{Sachdev-ye-kitaev model and
  thermalization on the boundary of many-body localized fermionic
  symmetry-protected topological states}, {\emph{Physical Review B} {\bfseries
  95} (2017) 115150}.

\bibitem{Brody2013BiorthogonalQM}
D.C.~Brody, \emph{Biorthogonal quantum mechanics},
  \href{https://doi.org/10.1088/1751-8113/47/3/035305}{\emph{J. Phys. A}
  {\bfseries 47} (2013) 035305}.

\bibitem{Streicher:2019wek}
A.~Streicher, \emph{{SYK Correlators for All Energies}},
  \href{https://doi.org/10.1007/JHEP02(2020)048}{\emph{JHEP} {\bfseries 02}
  (2020) 048} [\href{https://arxiv.org/abs/1911.10171}{{\ttfamily
  1911.10171}}].

\bibitem{Gao:2019nyj}
P.~Gao and D.L.~Jafferis, \emph{{A traversable wormhole teleportation protocol
  in the SYK model}},
  \href{https://doi.org/10.1007/JHEP07(2021)097}{\emph{JHEP} {\bfseries 07}
  (2021) 097} [\href{https://arxiv.org/abs/1911.07416}{{\ttfamily
  1911.07416}}].

\bibitem{Yan:2019fbg}
B.~Yan, L.~Cincio and W.H.~Zurek, \emph{{Information Scrambling and Loschmidt
  Echo}}, \href{https://doi.org/10.1103/PhysRevLett.124.160603}{\emph{Phys.
  Rev. Lett.} {\bfseries 124} (2020) 160603}
  [\href{https://arxiv.org/abs/1903.02651}{{\ttfamily 1903.02651}}].

\bibitem{Gorin:2006loschmidt}
T.~Gorin, T.~Prosen, T.H.~Seligman and M.~Žnidarič, \emph{Dynamics of
  loschmidt echoes and fidelity decay},
  \href{https://doi.org/https://doi.org/10.1016/j.physrep.2006.09.003}{\emph{Physics
  Reports} {\bfseries 435} (2006) 33}.

\bibitem{Goussev:2012loschmidt}
A.~Goussev, R.A.~Jalabert, H.M.~Pastawski and D.~Wisniacki, \emph{Loschmidt
  echo}, {\emph{arXiv preprint arXiv:1206.6348} (2012) }.

\bibitem{Jalabert:2001decoherence}
R.A.~Jalabert and H.M.~Pastawski, \emph{Environment-independent decoherence
  rate in classically chaotic systems},
  \href{https://doi.org/10.1103/PhysRevLett.86.2490}{\emph{Phys. Rev. Lett.}
  {\bfseries 86} (2001) 2490}.

\bibitem{Gu:2016local}
Y.~Gu, X.-L.~Qi and D.~Stanford, \emph{Local criticality, diffusion and chaos
  in generalized sachdev-ye-kitaev models},
  \href{https://doi.org/10.1007/JHEP05(2017)125}{\emph{JHEP} {\bfseries 05}
  (2017) 125} [\href{https://arxiv.org/abs/1609.07832}{{\ttfamily
  1609.07832}}].

\bibitem{Gu:2018jsv}
Y.~Gu and A.~Kitaev, \emph{{On the relation between the magnitude and exponent
  of OTOCs}}, \href{https://doi.org/10.1007/JHEP02(2019)075}{\emph{JHEP}
  {\bfseries 02} (2019) 075}
  [\href{https://arxiv.org/abs/1812.00120}{{\ttfamily 1812.00120}}].

\bibitem{Gu:2021xaj}
Y.~Gu, A.~Kitaev and P.~Zhang, \emph{{A two-way approach to out-of-time-order
  correlators}}, \href{https://doi.org/10.1007/JHEP03(2022)133}{\emph{JHEP}
  {\bfseries 03} (2022) 133}
  [\href{https://arxiv.org/abs/2111.12007}{{\ttfamily 2111.12007}}].

\bibitem{Garcia-Garcia:2024tbd}
A.M.~Garc\'\i{}a-Garc\'\i{}a, J.J.M.~Verbaarschot and J.-p.~Zheng, \emph{{The
  Lyapunov exponent as a signature of dissipative many-body quantum chaos}},
  \href{https://arxiv.org/abs/2403.12359}{{\ttfamily 2403.12359}}.

\bibitem{Hornedal:2022pkc}
N.~H\"ornedal, N.~Carabba, A.S.~Matsoukas-Roubeas and A.~del Campo,
  \emph{Ultimate speed limits to the growth of operator complexity},
  \href{https://doi.org/10.1038/s42005-022-00985-1}{\emph{Commun. Phys.}
  {\bfseries 5} (2022) 207} [\href{https://arxiv.org/abs/2202.05006}{{\ttfamily
  2202.05006}}].

\bibitem{Jian:2017tzg}
S.-K.~Jian, Z.-Y.~Xian and H.~Yao, \emph{{Quantum criticality and duality in
  the Sachdev-Ye-Kitaev/AdS$_2$ chain}},
  \href{https://doi.org/10.1103/PhysRevB.97.205141}{\emph{Phys. Rev. B}
  {\bfseries 97} (2018) 205141}
  [\href{https://arxiv.org/abs/1709.02810}{{\ttfamily 1709.02810}}].

\bibitem{Lin:2022rbf}
H.W.~Lin, \emph{The bulk hilbert space of double scaled syk},
  \href{https://doi.org/10.1007/JHEP11(2022)060}{\emph{JHEP} {\bfseries 11}
  (2022) 060} [\href{https://arxiv.org/abs/2208.07032}{{\ttfamily
  2208.07032}}].

\bibitem{Maldacena:2001eternal}
J.M.~Maldacena, \emph{Eternal black holes in anti-de sitter},
  \href{https://doi.org/10.1088/1126-6708/2003/04/021}{\emph{JHEP} {\bfseries
  04} (2003) 021} [\href{https://arxiv.org/abs/hep-th/0106112}{{\ttfamily
  hep-th/0106112}}].

\bibitem{Xian:2019qmt}
Z.-Y.~Xian and L.~Zhao, \emph{{Wormholes and the Thermodynamic Arrow of Time}},
  \href{https://doi.org/10.1103/PhysRevResearch.2.043095}{\emph{Phys. Rev.
  Res.} {\bfseries 2} (2020) 043095}
  [\href{https://arxiv.org/abs/1911.03021}{{\ttfamily 1911.03021}}].

\bibitem{He:2021dhr}
S.~He and Z.-Y.~Xian, \emph{{TT\textasciimacron{} deformation on multiquantum
  mechanics and regenesis}},
  \href{https://doi.org/10.1103/PhysRevD.106.046002}{\emph{Phys. Rev. D}
  {\bfseries 106} (2022) 046002}
  [\href{https://arxiv.org/abs/2104.03852}{{\ttfamily 2104.03852}}].

\bibitem{Arean:2019pom}
D.~Are\'an, K.~Landsteiner and I.~Salazar~Landea, \emph{{Non-hermitian
  holography}},
  \href{https://doi.org/10.21468/SciPostPhys.9.3.032}{\emph{SciPost Phys.}
  {\bfseries 9} (2020) 032} [\href{https://arxiv.org/abs/1912.06647}{{\ttfamily
  1912.06647}}].

\bibitem{Xian:2023zgu}
Z.-Y.~Xian, D.~Rodr\'\i{}guez~Fern\'andez, Z.~Chen, Y.~Liu and R.~Meyer,
  \emph{{Electric conductivity in non-Hermitian holography}},
  \href{https://doi.org/10.21468/SciPostPhys.16.1.004}{\emph{SciPost Phys.}
  {\bfseries 16} (2024) 004}
  [\href{https://arxiv.org/abs/2304.11183}{{\ttfamily 2304.11183}}].

\bibitem{Chen:2023hra}
Y.~Chen, V.~Ivo and J.~Maldacena, \emph{{Comments on the double cone
  wormhole}},  \href{https://arxiv.org/abs/2310.11617}{{\ttfamily 2310.11617}}.

\end{thebibliography}\endgroup

\end{document}